%% file: main.tex
\documentclass[11pt,a4paper]{article}
\usepackage{jheppub}

\usepackage{amssymb,amsmath}
\usepackage{color}
\usepackage{hyperref}
\usepackage{graphicx}
\usepackage{mathrsfs}
\usepackage{MnSymbol}
\usepackage{cleveref}
\usepackage{bbold}

\usepackage{tikz}
\usetikzlibrary{arrows.meta}

\numberwithin{equation}{section}

\crefformat{footnote}{#2\footnotemark[#1]#3}

\title{Gravitational Hilbert spaces:\\ invariant and co-invariant states, inner products, gauge-fixing, and BRST}
\author[a]{Jesse Held, }
\author[b]{Henry Maxfield}

\affiliation[a]{Department of Physics, University of California
Santa Barbara, USA}
\affiliation[b]{Leinweber Institute for Theoretical Physics at Stanford, 382 Via Pueblo, Stanford, CA 94305, USA}

\emailAdd{jheld@ucsb.edu}
\emailAdd{henrym@stanford.edu}

\abstract{
     Hilbert spaces in theories of gravity are notoriously subtle due to the Hamiltonian constraints, particularly regarding the inner product. To demystify this subject, we review and extend a collection of ideas in canonical gravity, and connect to the sum-over-histories approach by clarifying the Hilbert space interpretation of various gravitational path integrals. We use one-dimensional (or mini-superspace) models as the simplest context to exemplify the conceptual ideas. We emphasise that a physical Hilbert space can be defined either by requiring states to be annihilated by constraint operators (e.g., the Wheeler-DeWitt equation) or by equivalence relations between wavefunctions, and explain that these two approaches are related by an inner product. We advocate that the group averaging procedure constructs the correct physical inner product. The Klein-Gordon inner product is not positive-definite, which we explain as arising from a bad gauge choice; nonetheless, it agrees with group averaging when such a problem is absent. These concepts are all embedded in the BRST/BFV formalism, which provides a systematic way to construct these and other physically equivalent inner products (e.g., from maximal-volume gauge and Gaussian averaged gauges). Finally we discuss the application of these ideas in the semi-classical approximation, including non-perturbative gravitational effects.
}

\newcommand{\hilb}{\mathcal{H}}	
\newcommand{\hdens}{\mathscr{H}} 
\newcommand{\pdens}{\mathscr{P}} 

\newcommand{\RR}{\mathbb{R}}		
\newcommand{\CC}{\mathbb{C}}		

\newcommand{\hinv}{\hilb_\mathrm{inv}}
\newcommand{\hco}{\hilb_\mathrm{co}}
\newcommand{\Vinv}{V_\mathrm{inv}}
\newcommand{\Vco}{V_\mathrm{co}}

\newcommand{\target}{\mathcal{M}}

\newcommand{\li}{\lsem}
\newcommand{\ri}{\rsem}
\newcommand{\lc}{\llangle}
\newcommand{\rc}{\rrangle}

\DeclareMathOperator{\im}{im}
\DeclareMathOperator{\Ai}{Ai}
\DeclareMathOperator{\sgn}{sgn}

\newcommand{\scri}{\mathcal{I}}

\renewcommand{\Im}{\operatorname{Im}}

\DeclareMathOperator{\Tr}{Tr}

\graphicspath{{./Figures/}}

\begin{document}

\maketitle

\section{Introduction and detailed summary}

Among numerous approaches to quantum theories of gravity, the oldest and most straightforward is a direct quantisation of a dynamical spacetime metric and matter fields. In modern terms this is best understood as an effective quantum theory, valid for sufficiently low energy scales and weakly curved spacetimes. It has recently become apparent that this effective theory is far more powerful than previously envisaged, yielding several surprising results on quantum aspects of black holes. Highlights include calculations of the entropy of Hawking radiation \cite{Almheiri:2019qdq,Penington:2019kki}, statistics of energy eigenvalues \cite{Saad:2019lba}, and state counting of BPS black holes \cite{Iliesiu:2022kny}. In many cases, these results rely on non-perturbative effects involving fluctuations in the topology of spacetime.

These results have largely been understood in the formalism of the gravitational path integral. While this has proven a very powerful tool, it has several drawbacks relative to the canonical formalism. A Hilbert space interpretation in bulk variables is not manifest, since the usual way to obtain intermediate states by cutting the path integral on a Cauchy surface is not straightforward due to diffeomorphism invariance. Relatedly, a local (or approximately local) Lorentzian description of the dynamics is obscured, particularly when using a saddle-point approximation dominated by spacetimes with Euclidean or complex metrics. These features make it challenging to apply recent results to questions inherently about the bulk dynamics (rather than quantities defined asymptotically), most notably involving the black hole interior and quantum cosmology.

This motivates us to recast recent developments in the canonical formalism, and in a language which makes the bulk physics manifest. In particular, in the `semi-classical' $G_N\to 0$ limit we aim for the closest possible approach to a standard Hilbert space description of local quantum field theory (LQFT). Explicitly, our aim is to formulate a gravitational Hilbert space satisfying the following criteria:
\begin{enumerate}
    \item To leading order in the $G_N\to 0$ limit, the Hilbert space reduces to that of LQFT (including free gravitons) on a classical background spacetime.\footnote{If the background in question possesses isometries, then we should recover only the sector of the QFT Hilbert space invariant under the symmetries. We comment on this further in the discussion section \ref{sec:disc}.} In particular, this means we can define `Schr\"odinger picture' states defined by wavefunctions of field configurations on a specific Cauchy surface.
    \item Perturbative corrections can (at least in principle) be calculated order-by-order in $G_N$ perturbation theory (up to the usual ambiguities from renormalisation of coupling constants).
    \item Contributions from non-perturbative gravitational effects (e.g., saddle point geometries which are not close to the original background) can be incorporated.
    \item The Hilbert space is equipped with a positive-definite inner product.
\end{enumerate}
Furthermore, we would like to be able to construct gauge-invariant bulk observables acting on this Hilbert space, which reduce to local QFT operators in the $G_N\to 0$ limit (but require `gravitational dressing' once gravity is present). We emphasise that we only attempt to describe physics in the regime of validity of gravitational effective field theory, so in this way our aims are more modest than much of the older literature on canonical gravity.

As already noted, the main challenge to such a description comes from diffeomorphism invariance and the associated constraints. For example, the common description of the Hilbert space as solutions to the Wheeler-DeWitt (WDW) equation does not (by itself) achieve our aims, because this description of states is not local in time (failing point (1) above). Point (4) emphasises that a Hilbert space requires not only a space of states, but also a positive-definite inner product; once again this is not straightforward because of the constraints (and is not satisfied by the commonly used Klein-Gordon (KG) inner product). This motivates us to explore alternatives which overcome these issues, and we will subsequently explain the relation to the better-known WDW/KG constructions.

The canonical methods we describe should not be viewed as a completely separate alternative to the path integral, but as a complementary language. In particular, the canonical language will be useful to provide robust Hilbert space interpretations of various path integrals, especially those with boundary conditions at finite time (rather than asymptotically). It will also be helpful for identifying the correct measure associated with such path integrals, including path integrals which compute matrix elements of gauge-fixed bulk observables.

A specific goal of this program is to shed light on the idea of `null states', which has become prominent in recent developments. This refers to a non-zero wavefunction of gravitational variables which (due to non-perturbative effects) has zero norm, leading to linear relations between apparently distinct states. From the perspective of the path integral, the existence of such states (but none with negative norm) appears to require a highly-tuned conspiracy, which demands some explanation. We advocate an explanation in terms of gauge symmetry, specifically non-perturbative diffeomorphism invariance (or perhaps a mild extension thereof). Null states  can then result as the difference between two states related by a gauge transformation. One of our central motivations is to provide a language in which this idea can be made as precise and concrete as possible.

In this paper we describe a strategy for achieving these goals. We draw heavily on previous ideas in the literature on canonical quantum gravity, many of which are not as well-known as they deserve. Indeed, many of the technical ingredients used below appear already in the classic treatment of constrained Hamiltonian systems and BRST quantisation by Henneaux and Teitelboim \cite{Henneaux:1994lbw}. As such, much of the paper will be dedicated to a review of this older work, emphasising the conceptual points which are most relevant to our aims, though in many places we extend these ideas and illustrate them with novel examples. We sketch the main ideas in the remainder of the introduction, while leaving details to the main text. To provide the simplest possible illustration and avoid excessive technical distractions, in the bulk of the paper we mostly flesh out the details in theories of one-dimensional (or mini-superspace) theories of gravity, which is sufficient to illustrate the salient points. Nonetheless, we concentrate on ideas which readily generalise to  more interesting theories in higher dimensions (both closed universes and those with spatial asymptotic boundaries such as AdS).

The reader may find a specific solvable example useful: many of the main ideas in this paper  are implemented and illustrated in detail for de Sitter JT gravity in \cite{Held:2024rmg}.

\subsection{States in quantum gravity: not (only) Wheeler-DeWitt}

Gravity is exceptional among physical theories in that it does not have a conventional description of dynamics, due to the fact that time evolution is a gauge transformation (except at boundaries of spacetime). The Hamiltonian density $\hdens$ does not generate time translations relating different states and observables, but rather imposes the Hamiltonian constraints $\hdens=0$.\footnote{Diffeomorphisms of a given Cauchy surface also give rise to momentum constraints $\pdens=0$, though these are less challenging (both conceptually and technically) and we do not emphasise their role.} This raises both conceptual and technical issues, particularly if we wish to describe non-perturbative gravitational effects. These include defining the space of physical states, the inner product on those states, and observables which describe (approximately) local physics.

Starting with an unconstrained quantum theory with a Hilbert space  $\hilb_0$ and constraint operators $\hdens$, we would like to construct a physical Hilbert space of the constrained theory with $\hdens=0$ (i.e., we quantise first and subsequently impose constraints).\footnote{This is \emph{not} equivalent to the alternative approach of first imposing constraints classically and then quantising. We will see some specific examples in section \ref{ssec:nonpert} where `constrain first' misses some interesting and physically important non-perturbative effects.} For gravity, $\hilb_0$ is the space of wavefunctionals $\psi[\gamma_{ij}(x)]$ of spatial metrics (as well as spatial configurations of matter fields). We first discuss the space of states, addressing the inner product later. In particular, we will not worry about normalisability of wavefunctions (and associated technical issues like completions to form a Hilbert space) since the relevant norm follows later from our choice of inner product. 

 The oldest and perhaps most well-known approach goes back to Dirac \cite{Dirac:1950pj}, and in the context of gravity the foundational work of DeWitt \cite{DeWitt:1967yk}. In this Dirac-DeWitt approach, the constraints are imposed as equations restricting the physical states, namely the Wheeler-DeWitt equation:
\begin{equation}\label{eq:introInv}
    \hdens|\psi\rangle = 0 \qquad (\text{invariant states $|\psi\ri$} ).
\end{equation}
We call such states \emph{invariants} since this condition expresses invariance of the wavefunction under the infinitesimal gauge transformations generated by the constraints, in this case changes of coordinates that change the initial $t=0$ Cauchy surface. We use the notation $|\psi\ri\in \hinv$ for an invariant state to distinguish from arbitrary states in $\hilb_0$.

However, solving the Wheeler-DeWitt equation is not the only way to construct a space of states. An alternative is to put no restriction on the physical wavefunctions, but instead to declare equivalences between wavefunctions related by action of the constraints:
\begin{equation}\label{eq:introCo}
    |\psi\rangle \sim |\psi\rangle+ \hdens|\zeta\rangle \qquad (\text{co-invariant states $|\psi\rc$}).
\end{equation}
In this equation, $\hdens$ can represent any linear combination of constraints, and $|\zeta\rangle$ an arbitrary state of the unconstrained theory. The physical states are cosets, equivalence classes of wavefunctions related by gauge transformations. We denote such an equivalence class as $|\psi\rc\in\hco$ (often using $|\psi\rc$ to mean the equivalence class containing a state $|\psi\rangle\in\hilb_0$), and call these states \emph{co-invariants} (following terminology in \cite{Chandrasekaran:2022cip}). This name reflects the duality (in the sense of linear algebra) between the co-invariants and invariants: the inner product on the unconstrained Hilbert space induces a well-defined sesquilinear pairing between invariants and co-invariants (though not an inner product on either space).
\begin{equation}\label{eq:invcodual}
    \text{$\hinv$ \& $\hco$ are dual:}\qquad  \li \cdot | \cdot \rc \: \text{ is a non-degenerate sesqulinear form } \: \overline{\hinv} \otimes \hco \to \CC.
\end{equation}

Standard approaches in other contexts (such as gauge theories and the string worldsheet) typically use a mixture of these two prescriptions for different constraints. For example, Gupta-Bleuler quantisation (or old covariant quantisation of the string) imposes invariance under positive frequency constraints and equivalence (co-invariance) under negative frequency constraints. BRST quantisation encompasses both depending on the state of ghosts. We expect such a mixture to be most practical for many applications to realistic theories of gravity (discussed more in section \ref{sec:disc}), but for now we describe states purely in terms of one or the other.

Much of the literature which we follow tends to regard the `Dirac' invariant Hilbert space \eqref{eq:introInv} as primary, and the unconstrained Hilbert space (and its co-invariant cosets \eqref{eq:introCo}) as an auxiliary or kinematic structure used to define the correct Dirac Hilbert space and inner product as described below. While our presentation is technically equivalent, we shift emphasis to put the invariants and co-invariants on an equal footing, regarding them as two special cases among the more general mixtures described in the preceding paragraph.

\subsection{Co-invariants, gauge-fixing, and null states}\label{ssec:GFintro}

We advocate that for the aims outlined above --- a description of states which is local in time in perturbation theory, and which leaves room to account for non-perturbative effects --- it is often most useful to describe states as co-invariants. For this, we use a quantum analogue of `fixing a gauge'. Specifically, the gauge transformations in question are the redundancies in the description of a state (analogous to the classical freedom that a single spacetime can be defined by many sets of initial data on different Cauchy surfaces); this is distinct from choosing a gauge (e.g., de Donder) in the description of the dynamics or action.

The basic idea is to use the redundancy inherent in co-invariants, namely that a single state can be represented in many ways by choosing a representative of the equivalence class under \eqref{eq:introCo}. We can `gauge-fix' by allowing only states of a particular chosen form. For example, we might impose the gauge condition $\chi(q)=0$ by choosing a wavefunction of the form $\psi(q) = \delta(\chi(q))\psi_\perp(q_\perp)$, and the physical content of the state is contained in a wavefunction $\psi_\perp$ depending on fewer transverse variables $q_\perp$ (chosen so that $\chi$ and $q_\perp$ together constitute good coordinates for configuration space near $\chi=0$).\footnote{In fact we will often need something slightly more complicated for gravity, such as an additional term proportional to $\delta'(\chi)$ because the Hamiltonian is quadratic in momentum. Additionally, replacing the delta function with a narrow Gaussian is useful for a nice classical limit, since this allows for coherent states which are well-localised in phase space. These points will be addressed in the main text.} In a semi-classical limit, this can give us a description of states which is (at least approximately) local in time. Dynamics can emerge by changing the choice of gauge-fixing $\chi$ (perhaps a different value of some `clock' field), and identifying how the wavefunction $\psi_\perp$ changes to keep the state in the same co-invariant coset.

For a good gauge choice, we would like each physical state coset to have a unique representative of the chosen form. This is often difficult or impossible to arrange (a Gribov problem), but it is much easier to find a good gauge in perturbation theory around a given semi-classical background. In this case, the gauge choice eliminates redundancy under infinitesimal diffeomorphisms, but some redundancy under a discrete set of finite non-perturbative (and state-dependent) diffeomorphisms may remain. We propose that such residual diffeomorphisms may be the underlying mechanism for the existence of null states alluded to above (see further comments in section \ref{ssec:null}).

\subsection{Gravitational inner products}

So far, we have only examined the definition of the vector space of physical states. For either $\hinv$ or $\hco$ to be a Hilbert space we need additional structure, namely an inner product. Before discussing specific proposals, we first comment on the general mathematical relationship between invariants and co-invariants and inner products on these spaces.

Because of the duality \eqref{eq:invcodual} between $\hinv$ and $\hco$, four different structures become equivalent: an inner product on invariants $\hinv$, an inner product on co-invariants $\hco$, a map from co-invariants to invariants $\eta:\hco\to \hinv$, or a map from invariants to co-invariants $\eta^{-1}:\hinv\to \hco$. More precisely, a linear isomorphism of vector spaces $\eta$ is equivalent to a non-degenerate sesquilinear form on either space; an inner product should also be Hermitian ($\eta^\dag=\eta$) and positive-definite (or at least semi-definite, $\eta\geq 0$). A consequence of this general structure is that an inner product also allows us to pass between the two ways of defining physical states \eqref{eq:introInv} and \eqref{eq:introCo}. Thus, these are no longer truly different Hilbert spaces (once we have identified an inner product), though one language or  the other may be more useful depending on the context.

We can give more physical interpretations to $\eta$ and its inverse. First, we can think of $\eta$ as acting on an arbitrary wavefunction in the unconstrained Hilbert space $\hilb_0$ (regarded as a representative of a co-invariant coset), mapping it to a solution of the Wheeler-DeWitt equation. So $\eta$ is roughly like a projection onto gauge-invariant states (though this is not literally true for a non-compact space of gauge transformations like diffeomorphisms). For $\eta$ to be well-defined on $\hco$ and to  map into $\hinv$, for each constraint $\hdens$ we should have
\begin{equation}
    \eta \hdens = \hdens \eta = 0.
\end{equation}
The approach of defining an inner product though such a map goes under the name of `refined algebraic quantisation' \cite{Ashtekar:1995zh,Giulini:1998rk,Marolf:2000iq}, and $\eta$ is known as a `rigging map'.

The inverse map $\eta^{-1}$ takes a solution of the Wheeler-DeWitt equation to a co-invariant equivalence class. To make this concrete and practical we would like to represent the outcome as a wavefunction in $\hilb_0$, which means that for each invariant we choose a specific representative of the corresponding coset in $\hco$. We denote such a concrete representative of $\eta^{-1}$ by $\kappa:\hinv\to\hilb_0$. This leaves a great deal of freedom in choosing $\kappa$, which amounts to a choice of gauge in representing the state. For example, we might be able to choose $\kappa$ so that it always outputs states satisfying a gauge condition $\chi=0$ as discussed in section \ref{ssec:GFintro}. We would expect such a map to look something like $\delta(\chi)$ times a Faddeev-Popov determinant factor associated with the gauge-fixing. The details of this construction are fleshed out in the body of the paper. When we use $\kappa$ to define an inner product on invariants $\li\phi'|\phi\ri = \langle \phi'|\kappa|\phi\rangle$, its role is to remove the overcounting from summing over gauge-equivalent wavefunctions.  So, we think of $\kappa$ as a `gauge-fixing' map.

For a given $\kappa$ to be compatible with a chosen rigging map $\eta$, we need it to be a representative of the inverse $\eta^{-1}$ when thought of as a map into $\hco$. The condition for this is that $\kappa$ is a generalised inverse, $\eta\kappa\eta=\eta$:
\begin{equation}\label{eq:etakappaetaintro}
    \eta \text{ compatible with }\kappa \iff \eta\kappa\eta=\eta.
\end{equation}
One way to read this equation is that $\eta\kappa$ is the identity when acting on invariants, namely on the image of $\eta$. Another interpretation is that $\kappa \eta$  acts as a gauge transformation mapping any wavefunction to a different member of the same coset, so $\kappa \eta-\mathbf{1}$ should always yield a pure gauge `null state', which is annihilated by $\eta$.

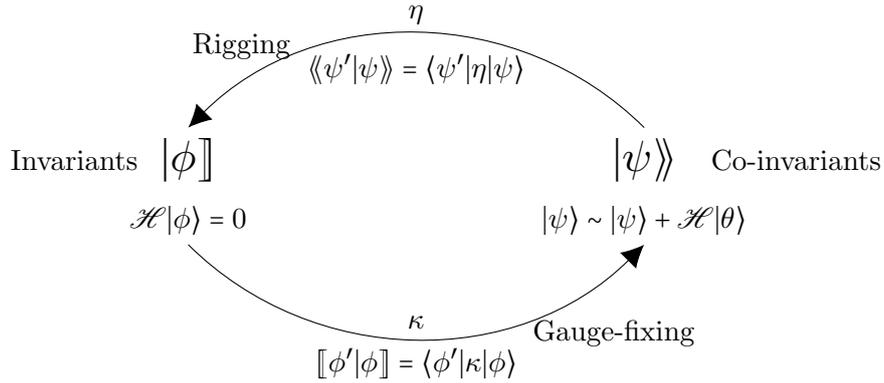
\begin{figure}
    \centering
    \begin{tikzpicture}
    \node (inv1) at (-3,.8) {  \scalebox{1.5}{$|\phi\ri$}};
    \node (inv2) at (-3,0) {
        $\hdens|\phi\rangle=0$};
    \node (coinv1) at (3,.8) {
         \scalebox{1.5}{$|\psi\rc$}};
    \node (coinv2) at (3,0) {
        $|\psi\rangle\sim |\psi\rangle + \hdens|\theta\rangle$};
    \draw[-{Latex[length=3mm,width=3mm]}] (coinv1.north) to[bend right=45] node[above] {$\eta$}  (inv1.north);
    \draw[-{Latex[length=3mm,width=3mm]}] (inv2.south) to[bend right=45] node[above] {$\kappa$} node[below]{$\li\phi'|\phi\ri = \langle \phi'|\kappa|\phi\rangle$} (coinv2.south);
    \node at (0,2) {
        $\lc\psi'|\psi\rc = \langle \psi'|\eta|\psi\rangle$};
    \node at (-4.5,.8) {Invariants};
    \node at (5,.8) {Co-invariants};
    \node at (2.6,-1.5) {Gauge-fixing};
    \node at (-2.3,2.3) {Rigging};
\end{tikzpicture}
    \caption{A summary of the relationships between the spaces of invariant and co-invariant states ($\hinv$ and $\hco$), inner products on these spaces, and maps between them.}
    \label{fig:introcoinv}
\end{figure}

\subsection{Group averaging}\label{ssec:introGA}

Having understood the general structure of inner products, we would like a specific candidate for the maps $\eta$ and/or $\kappa$. Perhaps the most well-known proposal is that of DeWitt \cite{DeWitt:1967yk}, to use a `Klein-Gordon' inner product on the space of invariants (discussed by Wald \cite{Wald:1993kj} for mini-superspace models). This is motivated by the fact that the Hamiltonian constraints $\hdens$ are quadratic in canonical momenta, so act as second-order differential operators on wavefunctions of metrics analogous to a Klein-Gordon operator; the Klein-Gordon form is then defined in analogy with a similar inner product on the single-particle states of free QFT, defined on solutions to the Klein-Gordon equation. But this has a fatal drawback: it is not positive-definite! The typical remedy is to restrict to a subspace of `positive frequency' states on which the Klein-Gordon form is positive definite, but in general there is no canonical choice for this subspace. In QFT this corresponds to the ambiguity in the choice of a Gaussian vacuum state and the corresponding notion of particle, but for gravity this is not a useful interpretation. Additionally, this restriction typically eliminates states which correspond to sensible, physical classical solutions in the semi-classical limit (for example,  cosmologies which expand and re-collapse \cite{Marolf:1994wh}). With this in mind, we would like a better proposal.

Fortunately, there is a much more satisfactory and well-motivated (but less well-known) suggestion for the physical inner product. This is the `group averaging' procedure \cite{Higuchi:1991tk,Ashtekar:1995zh}, which defines the inner product (on the space of co-invariants \eqref{eq:introCo}) by averaging over all possible gauge transformations. In the simplest case of a single constraint (which we call $H$, with 1D gravity in mind) generating the non-compact gauge group $\RR$, we construct a physical inner product $\lc\cdot|\cdot\rc$ as an integral over the gauge group,
\begin{equation}\label{eq:GAIP}
    \lc \psi'|\psi\rc = \int_{-\infty}^\infty dt \langle \psi'|e^{-i H t}|\psi\rangle = 2\pi\langle \psi'|\delta(H)|\psi\rangle.
\end{equation}
In the language above, $\eta= 2\pi\delta(H)$ is a specific natural choice of rigging map, which generalises the projection onto gauge-invariant states to the case of non-compact gauge groups. It is a particularly nice candidate for an inner product because it is manifestly positive-definite ($\delta$ is a positive distribution). Additionally, we will see that it arises naturally both from the path integral and also from the Hamiltonian BRST/BFV formalism, including for constraint algebras with field-dependent structure constants such as gravity. We will also develop a simple understanding of the relationship of group averaging with other candidate inner products (such as Klein-Gordon), and in particular an explanation for their shortcomings (most notably, failure of positivity). For these reasons we advocate for group averaging as the best procedure for defining the inner product on the physical Hilbert space.

\subsection{Klein-Gordon and more inner products}\label{ssec:introKG}

Group averaging defines a good physical inner product, and defines an equivalent inner product on a Hilbert space of invariants through the inverse map $\eta^{-1}$. However, this inverse is only defined abstractly and is not very practical due to the freedom of choosing a representative wavefunction in a given co-invariant coset. How do we concretely find a representative wavefunction which maps to a given invariant under $\eta$? And how can we compute the inner product between a pair of invariant states directly from the wavefunction? We can answer these questions by finding a concrete $\kappa$ map, which takes a solution to the Wheeler-DeWitt equation to a specific wavefunction representing the corresponding co-invariant coset.

Unfortunately, we do not expect that there exists any simple $\kappa$ which inverts $\eta$ (i.e., satisfying  \eqref{eq:etakappaetaintro}). This is essentially due to a Gribov problem: no nice (e.g., local) gauge-fixing condition simultaneously works for all states. Nonetheless, we will be able to find maps $\kappa$ which invert $\eta$ on a restricted set of states, in particular to all orders in perturbation theory around a given classical background (or family of backgrounds). For this reason, we will study maps $\kappa$ which need not satisfy the compatibility condition $\eta\kappa\eta=\eta$ with the group averaging rigging map $\eta$: instead, only certain matrix elements of this  relation will be satisfied (and even then, perhaps only in perturbation theory). Perhaps the central point of the paper is to explain the circumstances under which this compatibility holds, and the manner in which it is violated.

The Klein-Gordon inner product gives us the prominent example of such a $\kappa$ map. It arises from a gauge-fixing condition $\chi=0$ where $\chi(\gamma)$ is a function of spatial metrics $\gamma$ (independent of their conjugate momenta) at each point of space. As familiar from gauge-fixing procedures in path integrals, it takes the schematic form $\kappa_\chi = \delta(\chi)\Delta_\mathrm{FP}$, where $\Delta_\mathrm{FP}$ is a Faddeev-Popov determinant. The $\chi=0$ condition chooses a  hypersurface  in superspace (which corresponds in QFT to a Cauchy surface in the target spacetime), and can be written as an integral over that hypersurface. The result is independent of deformations of the chosen hypersurface, which we can understand as a consequence of gauge invariance (and the job of $\Delta_\mathrm{FP}$).  In the simple case of single Hamiltonian constraint $H$ generating time evolution $\frac{d}{dt} = i[H,\cdot]$, the Klein-Gordon form is given by  $\kappa_\chi = \frac{d\chi}{dt} \delta(\chi)$, up to issues of operator ordering (addressed in the main text), so the Faddeev-Popov factor is $\Delta_\mathrm{FP}=\frac{d\chi}{dt}$. For gravity, the Hamiltonian is quadratic in momentum, so $\Delta_\mathrm{FP}$ is linear in momentum.  This means that the Klein-Gordon form involves first order in derivatives with respect to superspace coordinates $\gamma$. Full details will be given in the body of the paper (see also \cite{Witten:2022xxp}).

In this form, it is clear that we can generalise the Klein-Gordon form by making a different choice of $\chi$ which depends not just on superspace variables, but also on their conjugate momenta, and/or on matter fields. An example is the `maximal volume gauge', where $\chi$ is the trace of the extrinsic curvature \cite{Witten:2022xxp}. Another generalisation uses `Gaussian averaged' gauges, where the sharp $\delta(\chi)$ is replaced by a smooth (but narrow) Gaussian. This is a practical choice for semi-classical calculations, since it allows for states which are well-localised in phase space (both $\chi$ and its canonically conjugate momentum), corresponding to classical initial data. Some details of these generalisations will be explored in the main text.

Having seen how different maps $\kappa_\chi$ arise from a gauge-fixing choice $\chi=0$, we can understand the failure of positivity of the Klein-Gordon form (or one of its generalisations). This arises from a bad choice of measure, coming from a possible sign in the Faddeev-Popov determinant $\Delta_\mathrm{FP} = \frac{d\chi}{dt}$. From the above, it is clear that positivity of $\kappa$ depends on whether a given trajectory crosses the gauge-fixing surface $\chi=0$ in an increasing or decreasing direction: positive-frequency solutions (with positive norm) correspond to increasing $\chi$, and negative-frequency (negative norm) solutions decreasing $\chi$. This is the usual issue that a good measure corresponds to the absolute value of the Faddeev-Popov determinant. An additional potential problem is that a given spacetime may have no Cauchy surface satisfying the gauge condition, or several, and it is hard to find a local gauge-fixing function $\chi$ which does not have this drawback. An example is a bang/crunch spacetime which crosses the surface $\chi=0$ once when expanding and once (in the opposite direction) when re-collapsing; this gives zero Klein-Gordon norm.

This issue is also responsible for $\kappa_\chi$ being incompatible with $\eta$. The upshot is that $\eta\kappa\eta=\eta$ is satisfied only for matrix elements where all contributing spacetimes cross the slice $\chi=0$ once, where we count the number of crossings with signs depending on whether $\chi$ is increasing or decreasing. In other words, $\chi$ constitutes a `good gauge' for spacetimes which have $\chi<0$ in the distant past and $\chi>0$ in the future. We can arrange for this in perturbation theory by a judicious choice of $\chi$, so that $\chi=0$ selects a unique Cauchy surface $\Sigma$ in the class of classical background spacetimes of interest, and also $\chi$ is increasing on $\Sigma$.

Once this issue is taken into account, the compatibility of $\kappa$ and $\eta$ is intuitively clearest from a path integral perspective described in the next subsection. From a more algebraic and canonical point of view, the compatibility can be understood by seeing how both $\eta$ and $\kappa$ maps emerge from the Hamiltonian BRST/BFV formalism (which is also useful, if not essential, for correctly identifying the appropriate measures).

We have emphasised a perspective where $\kappa$ is  thought of  as a map from invariants to co-invariants, not only as a sesquilinear form on invariants. The reason is that a $\kappa$ map $\hinv\to\hco$ is itself useful, since it gives us a way to construct  gauge-fixed wavefunctions as described in section \ref{ssec:GFintro}. The idea is that co-invariant states defined by `initial data' on a Cauchy surface selected by the gauge condition $\chi=0$ are constructed as the image of the map $\kappa_\chi$. Note that we can construct states in this way even if $\chi$ is not a globally good gauge condition (especially in perturbation theory), treating $\kappa$ as a tool for generating an ansatz while relying on the group averaging $\eta$ to provide the correct physical inner product.


\subsection{Gravitational path integrals}

Many recent successes in quantum gravity have been cast in the formalism of the gravitational path integral. This approach is not at odds with the canonical formalism we are using here: on the contrary, they are complementary approaches which are useful for emphasising different aspects of the physics. In particular, a good understanding of the ideas of canonical quantum gravity is invaluable for providing well-justified Hilbert space interpretations of path integral calculations. Conversely, the path integral provides useful intuition for the sort of canonical constructions we have been describing, as well as practical methods for semi-classical calculations.

\subsubsection*{States and group averaging}

Consider first a path integral over spacetimes bounded by a pair of Cauchy surfaces $\Sigma_1$ and $\Sigma_2$, with fixed field configurations on those surfaces: that is, fixed spatial metrics $\gamma_1,\gamma_2$ along with matter fields. In a non-gravitational theory, such a path integral (with fixed spacetime geometry) computes the matrix elements of a time-evolution operator, relating wavefunctionals depending on fields on different Cauchy surfaces. Most simply, if spacetime is a product of space $\Sigma$ and an interval of time, the path integral computes matrix elements of $e^{-i H t}$, where $H$ is the Hamiltonian of the theory on $\Sigma$. But for gravity, the metric is a dynamical field so we also integrate over all possible spacetime metrics $g$ between $\Sigma_{1,2}$ (with induced metrics $\gamma_{1,2}$ on the boundaries $\Sigma_{1,2}$), modulo diffeomorphisms which leave $\Sigma_{1,2}$ invariant. Since time-evolution (as well as changes of coordinates on a fixed spatial slice) is part of the gauge symmetry of diffeomorphisms, the integral over such geometries is precisely an integral over the gauge group.\footnote{More precisely, this integrates over diffeomorphisms of spacetime, modulo the diffs which leave the spatial slice $\Sigma_1$ invariant.} So, the path integral is computing the matrix elements $\langle \gamma_2|\eta|\gamma_1\rangle$ of the group-average rigging map introduced in section \ref{ssec:introGA}.\footnote{In a situation with fixed spatial boundary conditions (such as asymptotically AdS), we are referring to an integral where $\Sigma_1$ and $\Sigma_2$ are required to coincide at the boundary. If we instead require the initial and final Cauchy surfaces go to the boundary at different times, we are not simply computing an inner product, but also inserting a physical time-evolution operator.} For a one-dimensional (or mini-superspace) theory, the integral over metrics becomes an integral over the proper time $t$; since the integral over all other fields gives the time-evolution operator $e^{-i H t}$, we recover precisely \eqref{eq:GAIP}. We also learn that a fixed spatial metric $\gamma_1$ (or more general finite time boundary conditions) should be interpreted as states $|\gamma_1\rangle$ with delta-function wavefunctions $\langle \gamma|\gamma_1\rangle = \delta(\gamma-\gamma_1)$ in the \emph{unconstrained} Hilbert space, which we can regard as representatives of a co-invariant equivalence class $|\gamma_1\rc$. This path integral computes the physical (group-averaging) inner product $\lc \gamma_2 |\gamma_1\rc$ on co-invariant states.

Note that this interpretation as a gauge-invariant inner product on co-invariants only makes sense if the path integral constructs solutions to the constraints \cite{Halliwell:1990qr}. Concretely, $\Psi(\gamma) = \langle \gamma|\eta|\gamma_1\rangle$ should solve the Wheeler-DeWitt equation, or more abstractly  $|\eta \gamma_1\ri = \eta |\gamma_1\rc$ should be in the Hilbert space of invariants. This requires us to integrate over spacetimes where $\Sigma_1$ can lie to the future of $\Sigma_2$ or to its past (or a mixture of the two): in ADM language, we integrate over both positive and negative lapse. Integrating  only over positive lapse would result in a `propagator' for the constraints (solving $\mathsf{H}\Psi(\gamma) =\delta(\gamma-\gamma_1) $ instead of $\mathsf{H}\Psi(\gamma)=0$). See further comments in \cite{Araujo-Regado:2022gvw}.

This illustrates that any boundary condition on a fixed finite-time Cauchy surface can be interpreted as a co-invariant state, and also shows how path integrals can define invariant wavefunctions as a function of boundary conditions. The latter is not limited only to the group-averaging we just saw, and many familiar classes of states are prepared by different path integrals. An important example are states defined by asymptotic boundary conditions in the distant past,  essentially the same  as scattering `in' states (or similarly `out' states defined in the future). These can be prepared by path integrals over only forward time evolution (positive lapse), except in a limit for which only very large times contribute, so the WDW equation is still satisfied (the would-be source which appeared above is pushed off to infinity). An example would be out states defined at future infinity in de Sitter space (discussed for JT gravity in \cite{Held:2024rmg}). Such constructions could also be generalised to allow some Euclidean evolution, for example to prepare a vacuum state in asymptotically AdS spacetimes, or the Hartle-Hawking state. The wavefunction $\Psi(\gamma)$ of such a state is interpreted in our language as the canonical pairing $\lc \gamma|\psi\ri$ between co-invariants and invariants.

\subsubsection*{Cutting path integrals, gauge-fixing, and $\kappa$ maps}

Let's now consider how we recover a Hilbert space interpretation from a path integral. In ordinary QFT, we do this by cutting spacetime into two along a Cauchy surface $\Sigma$, producing a Hilbert space $\hilb_\Sigma$ which can be informally thought of as wavefunctions $\psi(q)$ where $q$ denotes the field configuration on $\Sigma$. The integral over one half of spacetime (with fixed field configuration $q$ on $\Sigma$) defines a wavefunction $\psi_1(q)$ of a `ket' state $|\psi_1\rangle$, the integral over the other half similarly defines a `bra' state $\langle \psi_2|$, and integrating over fields $q$ on $\Sigma$  then sews the two halves together. This last integral is interpreted in Hilbert space language as a completeness relation, the inverse to the inner product on $\Sigma$. Arguably,  such a factorisation  of path integrals is the main purpose of a Hilbert space: it embodies locality (since the states are functions of local data on $\Sigma$) and positivity (since the inner product is positive-definite). This allows us to define local operators on $\Sigma$ and assign sensible probabilistic interpretations to their expectation values.  For gravity it is not so simple: diffeomorphism invariance means we cannot simply cut along a `$t=0$' Cauchy surface $\Sigma$, because gauge transformations move the physical location of $\Sigma$.

To illustrate the difficulty, let's consider cutting the above path integral between surfaces $\Sigma_{1,2}$ computing the group-average $\lc \gamma_2|\gamma_1\rc =\langle \gamma_2|\eta|\gamma_1\rangle$. Suppose we na\"ively follow the usual procedure of integrating over all field configurations (spatial metrics $\gamma$) on an intermediate surface $\Sigma$, as well as all metrics $g_1$ between  $\Sigma_1$ and $\Sigma$ (with fixed boundary metrics $\gamma_1,\gamma$, and modding out by diffeomorphisms which are trivial on these boundaries) and similarly $g_2$ between $\Sigma$ and $\Sigma_2$.  The result is not the same as the integral over all metrics which gave us matrix elements of $\eta$: in fact, the result additionally integrates over all possible embeddings of the Cauchy surface $\Sigma$ in a given spacetime geometry (including all possible choices of coordinates on $\Sigma$ itself). We have overcounted by the volume of a residual gauge group:  diffeomorphisms modulo those which are trivial on $\Sigma$ (see (2.39) of \cite{Witten:2022xxp} for a similar discussion). This construction computes matrix elements of $\eta^2$, which does not make sense for non-compact (and field-dependent) groups of gauge transformations, even for $\eta=2\pi\delta(H)$ in the simplest 1D example given above.

\begin{figure}
    \centering
    \input{Figures/nkn.tex}
    \caption{The generalised inverse relation $\eta\kappa\eta=\eta$ between group averaging and gauge-fixing maps has a geometric interpretation in terms of cutting and gluing spacetimes in the path integral. On the left, we integrate over all geometries $g_1$ and $g_2$, glued together with an operator $\kappa$ which implements a gauge condition fixing the red cutting surface, giving matrix elements of $\eta\kappa\eta$. If this is a good gauge (selecting a unique intermediate Cauchy surface in every geometry), then the result is equivalent to the integral over all geometries $g$ which computes $\eta$.}
    \label{fig:nkn}
\end{figure}
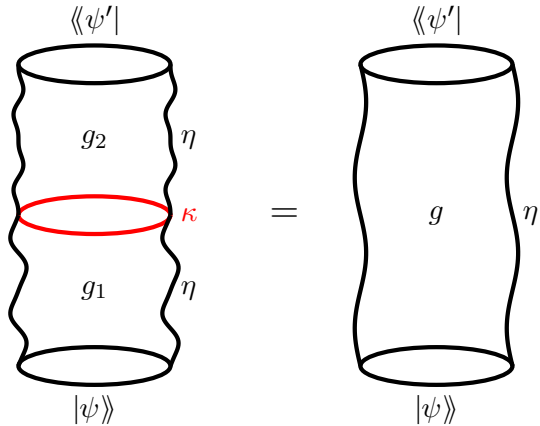

The remedy is obvious from this perspective: we must implement a gauge-fixing which identifies $\Sigma$ uniquely in some diffeomorphism-invariant way. As we described in section \ref{ssec:introKG} above, this is precisely what the maps $\kappa$ (such as the Klein-Gordon inner product) should do.   The equation $\eta\kappa\eta=\eta$ given in \eqref{eq:etakappaetaintro} is a manifestation of this process of cutting and gluing path integrals with a good gauge choice, shown in figure \ref{fig:nkn}. Each $\eta$ on the left represents the path integral between $\Sigma_1$ and $\Sigma$ or between $\Sigma$ and $\Sigma_2$, while $\kappa$ is an insertion of a gauge-fixing $\delta(\chi)$ which chooses a physical embedding of $\Sigma$ uniquely, along with an appropriate  (positive) measure factor $\Delta_\mathrm{FP}$. Then the product $\eta\kappa\eta $ (sewing along $\Sigma$) recovers $\eta$ on the right hand side, the integral over spacetimes between $\Sigma_1$ and  $\Sigma_2$. Thus, we can think of $\kappa$ as giving a `completeness relation' dual to the group average inner product $\eta$, taking gauge redundancy into account. But as emphasised already, this relation can be violated by spacetimes where $\kappa$ does not select a unique $\Sigma$.

\subsection{Inner products in the BRST formalism}

The BRST formalism provides a very general framework for systems with constraints. It naturally incorporates both the invariant and co-invariant states described above (and arbitrary mixtures of the two). It also provides a mechanism to describe the inner products introduced above (including the particular cases of group averaging and the Klein-Gordon form) along with a physical interpretation in terms of gauge-fixing. In doing so, it provides a certain rigidity absent from our considerations so far, removing much of the ambiguity in choosing an inner product, and providing further justification for  the group-averaging prescription. Finally, it is often a powerful technical tool for identifying the correct gauge-invariant form of inner products and observables.

In this formalism, for each constraint $\hdens$ we introduce corresponding fermionic ghost variables $c,b$, which act as raising and lowering operators respectively on a two-level Hilbert space spanned by $|\uparrow\rangle$, $|\downarrow\rangle$. The central object is a BRST operator $Q = \int c\hdens +\cdots$ built from the ghosts and constraints, with $\cdots$ determined by the requirements of nilpotency $Q^2=0$ and Hermiticity $Q^\dag=Q$. The physical Hilbert space is then given by the cohomology of $Q$, $\ker Q/\im Q$ (at ghost number zero). This gives different ways to represent physical states depending on the state of the ghosts. At the `lower level' $|\downarrow\rangle$ the constraints are important for determining $\ker Q$, giving rise to the invariant Hilbert space. At the  `upper level' $|\uparrow\rangle$ the constraints instead determine the denominator $\im Q$, giving us cosets corresponding to co-invariant states. So, the myriad of possible ways of imposing constraints correspond to choices of ghost state in a BRST cohomology class.

When we calculate inner products in this formalism, we immediately recover the natural pairing $\li\psi|\psi'\rc$ between invariants and co-invariants. However, inner products between two states of the same type are more subtle: a straightforward approach gives us an indeterminate form $\infty \times 0$, where $\infty$ comes from the original phase space variables and $0$ from the ghosts. We can interpret this as arising from a failure to gauge-fix, with $\infty$ coming from the volume of the unfixed gauge group. Following \cite{Marnelius:1990eq,Batalin:1994rd,Shvedov:2001ai}, the key idea to fix this is that the degeneracy can be lifted by inserting a trivial operator $\exp[Q,\Psi]_+$ (the exponential of a BRST-exact operator), physically equivalent to the identity. Here $\Psi$ is a fermionic operator of ghost-number $-1$ which we are free to choose, called the `gauge-fixing fermion'. An appropriate choice of $\Psi$ recovers each of the inner products discussed above. The ghost-free part of $[Q,\Psi]_+$ gives the generator of a gauge transformation, while the other terms give us insertions of ghosts $b$ (for co-invariant states) or $c$ (for invariant states) along with gauge-fixing Jacobians. The result is invariant under infinitesimal changes of $\Psi$ (the Fradkin-Vilkovisky theorem \cite{Fradkin:1975cq}).

However, the result is not totally independent of $\Psi$: there is a loophole to the Fradkin-Vilkovisky theorem, as the result can jump for finite changes when passing through a singular value (most notably $\Psi=0$, where we get the original indeterminate result). This is the cause (from the BRST perspective) of the signs discussed in section \ref{ssec:introKG}, including indefiniteness of the Klein-Gordon inner product.

 More specifically, we use an extended Hamiltonian form of the BRST formalism known as the Batalin-Fradkin-Vilkovisky method \cite{Batalin:1977pb,Fradkin:1975cq,Henneaux:1985kr,Henneaux:1994lbw}. In this BFV-BRST formalism, we work in an extended phase space which retains the Lagrange multipliers which impose constraints --- the lapse $N$ and shift $N_\perp^i$ in gravity (or the timelike component of the gauge field $A_0$ in gauge theories) ---  along with their conjugate momenta $\Pi$. These extra variables are ultimately removed by also including an extra constraint that $\Pi=0$, but nonetheless prove helpful to include both technically and conceptually. In particular, the freedom in the gauge-fixing fermion $\Psi$ can be used to choose a gauge, with the gauge-fixing condition coming from the equation of motion for $\Pi$ (so that integrating out $\Pi$ and other momenta reproduces the Faddeev-Popov path integral). For example, we can choose $\Psi$ to compute group average inner products using various familiar coordinate conditions (e.g., synchronous or de Donder gauges) including Gaussian averaged versions. The further technical virtue of this formalism is that it naturally gives us a group averaging procedure for rather general algebras of constraints, including those with field-dependent structure constants as arise in gravity (discussed further in section \ref{sec:disc}).

\subsection{Perturbative and non-perturbative physics in the semi-classical limit}

With the exception of simple toy models, we must face the fact that exact calculations in gravity are essentially impossible. This is not only a matter of technical difficulty: the non-renormalisablity of GR means that new physics is inevitably necessary in the UV, which surely replaces spacetime with some more fundamental description.\footnote{Holographic duality achieves this, but we still face the challenge of extracting local bulk physics from the available dual observables, which presumably requires understanding gravitational variables of some sort.} So, our tools are only useful if we can practically apply them in a semi-classical approximation. In particular, we  would like to sketch how the ideas we have discussed can give a semi-classical gravitational Hilbert space satisfying the criteria listed at the start of this introduction.

At leading order in $G_N$ we would like states to be described by local data on a specific Cauchy surface in a classical background spacetime. The Wheeler-DeWitt prescription of invariant states does not achieve this (at least by itself), because such wavefunctions are superpositions over all possible Cauchy surfaces, obscuring locality. However, we do get such a local description from co-invariant wavefunctions with a gauge-fixing ansatz as sketched in section \ref{ssec:GFintro} (perhaps obtained as the image of some $\kappa$ map as mentioned at the end of section \ref{ssec:introKG}).  We can think of this as a quantum version of initial data on a Cauchy surface specified by a gauge condition $\chi=0$. This correspondence to classical initial data is sharpest if we choose coherent wavefunctions (which means we do not impose $\chi=0$ as a sharp delta function), for example using Gaussian averaged versions of the $\kappa$ map, so that our states are well-localised in phase space.\footnote{A similar possibility is to specify a state by an invariant wavefunction, but only giving the data of the wavefunction (and perhaps a finite number of its derivatives) evaluated on the gauge-fixed surface $\chi=0$ in superspace, in the hope that this picks out a unique solution to the Wheeler-DeWitt equation. This is closely related to our proposal, differing by insertions of the measure factor $\Delta_\mathrm{FP}$ (which may depend on data away from $\chi=0$, in particular requiring derivatives of the invariant wavefunction normal to $\chi$). So, this alternative should be satisfactory in perturbation theory (and locally in superspace), but the co-invariant perspective is more robust, and useful in particular for understanding non-perturbative effects (in the group average inner product, for example).} Working with such gauge-fixed states around a classical background achieves the first two aims in our wish list: a description which reduces to local QFT when $G_N\to 0$, with a systematic perturbation theory to compute corrections order-by-order in $G_N$. In particular, group averaging supplies the inner product, and the BRST formalism provides the tools for computing this perturbatively in any desired gauge.

For a pair of states which are perturbations around the same background and defined on the same gauge-fixed Cauchy surface, the main  semi-classical contribution to the inner product comes from short time evolution (expanding around zero lapse and shift). It is this piece that we expect to give back the local QFT inner product (and perturbative corrections will still be local, albeit containing higher order spatial derivatives). But  there can also be semi-classical saddle-points which involve a finite time evolution, and  these are responsible for several virtues of this approach.

 Most simply, if we have a pair of states defined for two different gauge slices on the same classical background, there will be a real Lorentzian saddle point contributing to the inner product (corresponding to the piece of spacetime between the respective Cauchy surfaces). This inner product essentially becomes the matrix elements of a time-evolution operator: the evolution is purely a gauge transformation in gravity, but in this way we can nonetheless recover a notion of (approximately) local dynamics.
 
Additionally, the group average path integral can have saddle-points with complex metrics. The defining contour runs over real Lorentzian metrics, but this may be deformed to a steepest-descent contour which passes through a saddle-point with complex-valued metric. Such a saddle must be exponentially suppressed (its action will have positive imaginary part), but can nonetheless describe interesting physical effects. In particular, they can give the leading contribution to processes (such as decay by quantum tunneling) which are classically completely forbidden. If we allow our integral to include certain mild relaxations of Lorentzian geometries, these processes could potentially include topology change (as discussed in section \ref{sec:disc}). More generally, non-trivial saddle-points provide contributions to the inner product which are qualitatively different from the perturbative piece described above because they are completely non-local in space. We speculate that such corrections could help to explain the  null state phenomenon mentioned in the introduction.


\subsection{Paper outline}

The remainder of this paper explains the details of the ideas presented above in the simplest case of one-dimensional (or minisuperspace) theories of gravity with a single constraint, which we introduce in section \ref{sec:1D}. In section \ref{sec:states} we describe different approaches for imposing constraints to give a space of physical states, and examples of interesting states in each approach. We then turn to the inner product in section \ref{sec:IP}, including explicit calculations in a particularly simple class of examples with one-dimensional target space. A brief discussion of physical operators (in both Dirac and BRST formalisms) is given in section \ref{sec:ops}, which is mostly a preliminary for understanding the BFV-BRST formalism in section \ref{sec:BFVIP}. We then discuss applications in the semi-classical limit in section \ref{sec:SC}. We conclude with discussion in section \ref{sec:disc}.

Appendices \ref{app:string} and \ref{app:Maxwell} connect the ideas in this paper to analogous concepts in the string worldsheet  and abelian gauge theory respectively. Other appendices contain some additional technical details.

\section{One-dimensional (or mini-superspace) gravity}\label{sec:1D}

For simplicity and clarity we will spend most of our time illustrating the ideas in one-dimensional gravity (or mini-superspace) models. This is sufficient to discuss the gauging of reparameterisations of time, or the Hamiltonian constraint. While a full higher dimensional theory of gravity adds significant technical complication, the conceptual issues associated with defining Hilbert spaces and inner products are well illustrated by these simpler one-dimensional models.

\subsection{Coupling quantum mechanics to one-dimensional gravity}

To motivate the following, consider a `matter' theory in one dimension with Lagrangian $L(q,\dot{q})$ (where $q$ denotes a collection of $n$ configuration space variables), and minimally couple it to one-dimensional gravity. The only component of a one-dimensional metric $g =-N(t)^2 dt^2$ is the `lapse' $N(t)$, or equivalently the derivative $N=\frac{d\tau}{dt}$ of proper time $\tau$ with respect to coordinate time $t$. (Another possible name and notation for $N$ is an `einbein' $e$, defined so that $e(t) dt$ is a unit one-form.) We construct a diffeomorphism-invariant Lagrangian $\mathbf{L}$ by `minimal coupling' of the matter to the dynamical metric,
\begin{equation}\label{eq:1Dlag}
    \mathbf{L} = N L(q,N^{-1}\dot{q}).
\end{equation}
There is no pure gravitational `Einstein-Hilbert' term since there is no curvature in one dimension; the only possible term involving only $N$ is a `cosmological constant' $\Lambda N$, but this can be absorbed into a constant shift of $L$. The usual Legendre transform gives us the Hamiltonian
\begin{equation}\label{eq:properHam}
    \mathbf{H} = N H(q,p),
\end{equation}
where $H(q,p)$ is the `matter' Hamiltonian constructed from $L(q,\dot{q})$ in the usual way. Since the Lagrangian $\mathbf{L}$ does not depend on derivatives of $N$, we have the primary constraint that its conjugate momentum vanishes, $\Pi = p_N = 0$, and the Poisson bracket of $\mathbf{H}$ with  $\Pi$ generates the secondary constraint $H=0$.

The upshot is that the gravitational theory is constructed from the `matter' theory by imposing the Hamiltonian constraint $H=0$. Classically, this means that states are given by the `symplectic quotient' of phase space, meaning the space of zero energy trajectories (points on the constraint surface $H(q,p)=0$, with any two points along the same flow generated by $H$ identified). Quantum mechanically, we can take essentially any ordinary `matter' quantum mechanics theory with Hilbert space $\hilb_0$, inner product $\langle\cdot|\cdot\rangle$, and Hermitian Hamiltonian $H$ (the only assumption we need to make is that $H$ has continuous spectrum in a neighborhood of zero energy) and couple it to 1D gravity by imposing $H=0$ at the quantum level. The main point of this paper is to review and explore the details of how we do this.

For much of the paper we do not include the lapse as a separate degree of freedom, regarding $N$ as simply a Lagrange multiplier imposing the $H=0$ constraint. Alternatively, we can interpret this as eliminating the $\Pi=0$ primary constraint by fixing the gauge $N=1$. However, for some purposes it is useful to use a `non-minimal' formulation which explicitly retains $N$, $\Pi$ and the constraint $\Pi=0$; we return to this in section \ref{sec:BFVIP}.

The constraint $H$ generates a gauge symmetry, but note that in the Hamiltonian formalism this is not quite the full group of diffeomorphisms of time (or spacetime in higher dimensions) that we might expect from the Lagrangian formulation. Rather, it generates only the part that acts on the $t=0$ slice where the wavefunction is defined. More precisely, it gives the group of diffeomorphisms modulo the diffs that act as the identity at $t=0$. In one dimension, this becomes simply the one-dimensional group of rigid time translations generated by $H$.

Finally, note that we `quantise first', constructing an unconstrained quantum theory and then imposing $H=0$ quantum mechanically, rather than `constraining first', which means quantising the constrained symplectic quotient phase space. These two operations do not commute in general: in particular, `constrain first' can eliminate interesting and important non-perturbative effects (we discuss an example in section \ref{ssec:nonpert}).

\subsection{Examples: sigma models}\label{ssec:sigma}

For a very concrete set of examples we can consider a sigma-model:
\begin{equation}
    \mathbf{L} = \frac{1}{2N} G_{ab}(q) \dot{q}^a\dot{q}^b - N V(q).
\end{equation}
Here $G_{ab}$ is the metric on some `target space' manifold $\mathcal{M}$, which is non-degenerate but can have any signature; in particular, it is not required that $H$ is bounded from below to get a physically sensible theory. $V$ is a `potential' on this manifold. This is a general bosonic model with a Lagrangian  quadratic in velocities and time-reversal symmetry (relaxing this constraint allows an additional a `background magnetic field' term $A_a(q)\dot{q}^a$). Classically, the Hamiltonian is
\begin{equation}\label{eq:sigmaH}
    H = \frac{1}{2} G^{ab}(q) p_a p_b + V(q).
\end{equation}
The quantum Hamiltonian has an operator ordering ambiguity; for definiteness we make the choice $H = -\frac{1}{2}\nabla^2 + V$ where $\nabla^2$ is the Laplacian on target space $\mathcal{M}$ with metric $G$. Then, the matter Hilbert space is the square integrable (scalar) functions $L^2(\mathcal{M})$ of $\mathcal{M}$ with the measure $d^n q \sqrt{|\det G|}$.\footnote{A subtlety with this choice is that $p_a$ is not equal to $-i \frac{\partial}{\partial q^a}$, since this is not Hermitian. Instead we should take $p_a =-i \frac{\partial}{\partial q^a} - \frac{i}{4}(\partial_a
\log G)$ (with $G=|\det G|$). More generally, for any vector field $v$ on target space, the operator $-i v^a \partial_a - \frac{i}{2}(\nabla_a v^a)$ (obtained by acting with $v^a$ and covariant derivative $\nabla_a$ and averaging over the two orders) is Hermitian; we get the above by taking $v= \partial_a$. An alternative takes wavefunctions to be `half-densities' so that $|\psi|^2$ transforms as a volume form under changes of target space coordinates. This gives $p_a=-i \frac{\partial}{\partial q^a}$. In either case, the operator ordering for the kinetic term is $G^{-\frac{1}{4}} p_a G^{ab}G^{\frac{1}{2}} p_b G^{-\frac{1}{4}}$.} It is natural to also consider a quantum correction $\xi R$ to the potential, proportional to the curvature of target space; in particular, the `conformal coupling' value $\xi=-\frac{1}{8}\frac{n-2}{n-1}$ may be preferred \cite{Halliwell:1988wc,Moss:1988wk}. In any case, we can always absorb this in the definition of $V$.

This set of models give reasonable analogues to higher dimensional theories of gravity since the Hamiltonian constraint is quadratic in momenta. For this, we interpret the target space $\mathcal{M}$ as Wheeler's `superspace', which is the collection of all possible spatial geometries (Riemannian manifolds modulo diffeomorphisms). In fact, we obtain a theory of this kind by a mini-superspace truncation of a higher-dimensional theory. This means restricting the integral over metrics to a finite-dimensional family of homogeneous spatial metrics $\gamma_{ij}(q)dx^i dx^j$ preserving some symmetries (parameterised by the coordinates $q$). Making a metric ansatz  $ds^2 = -N(t)^2 dt^2 + \gamma_{ij}(q(t))dx^i dx^j$ (along with other fields respecting the same symmetry, depending only on time), we arrive at a one-dimensional theory of gravity precisely of the form \eqref{eq:sigmaH}.  Typically the target space metric $G$ has Lorentzian signature, with the timelike direction corresponding roughly to the volume of space. An example is an FLRW cosmology, for which the Hamiltonian constraint $H=0$ is the Friedmann equation and the scale factor $a$ is a timelike coordinate in superspace.

\subsection{One dimensional gravity vs worldline QFT}\label{ssec:worldlineQFT}

In cases where the target space metric $G$ is Euclidean or Lorentzian, the model \eqref{eq:sigmaH} is also closely related to free bosonic QFT on $\mathcal{M}$ where the potential becomes a spacetime dependent mass squared $V=\frac{1}{2}m^2$. We then think of our quantum mechanics as living on the worldline of particles, and interpret its wavefunction as a field on target space. The QFT or target space action is given by the expectation value of $H$,
\begin{equation}
    S_\mathrm{target}[\Phi] = \int d^n q \sqrt{|\det G|} \left[ \frac{1}{2}G^{ab}(q) \partial_a \Phi^*(q) \partial_b \Phi(q) + V(q) |\Phi(q)|^2\right].
\end{equation}
The Hamiltonian $H$ of our one-dimensional theory becomes the equation of motion $H\Phi=0$ for the quantum field $\Phi$. We have written a theory of a complex scalar; for a real scalar we can restrict to states invariant under time-reversal of the worldline theory (which is unrelated to target-space time reversal).  We will discuss this worldline or `first quantised' formulation of QFT later, in particular emphasising that the perspective of a one-dimensional theory of gravity naturally leads to a different Hilbert space than QFT (see also \cite{Casali:2021ewu}).

\section{States: invariants and co-invariants}\label{sec:states}

\subsection{States and the Hamiltonian constraint}

Starting from the unconstrained Hilbert Space $\hilb_0$, there are in fact two natural ways to construct a physical gauge-fixed Hilbert space by imposing the constraint $H=0$. 

The first is the `Dirac' approach \cite{Dirac:1950pj}, which is perhaps more familiar and standard in the literature. For this we restrict to `gauge-invariant states' which are left unchanged by time translation $e^{-i H t}$. These are  annihilated by the constraint $H$, so they are solutions to the Wheeler-DeWitt equation $H|\psi\rangle=0$. We denote these \emph{invariants} by $|\psi\ri$, giving a space of states
\begin{equation}\label{eq:Vinv}
    \Vinv = \ker(H) = \{ |\psi\ri: H |\psi\ri=0\}.
\end{equation}
Since the spectrum of $H$ is continuous at zero, strictly speaking these invariants are not elements of $\hilb_0$ since they are not normalisable (we comment on the precise mathematical definition of $\Vinv$ in the next subsection). Energy eigenstates $|\psi_E\rangle$ are $\delta$-function normalised, $\langle\psi_{E'}|\psi_E\rangle \propto \delta(E-E')$, so the $\hilb_0$ inner product of two invariants contains an infinite factor $\delta(E=0)$. The non-normalisability means that we do not get a sensible inner product on $\Vinv$ simply by restriction of the $\hilb_0$ inner product to the invariants,  so $\Vinv$ is (at this point) only a vector space of states, not a Hilbert space. We will discuss several ways to put an inner product on $\Vinv$ later.

The second natural construction of gauge-invariant states is the space of \emph{co-invariants}, where the name (following \cite{Chandrasekaran:2022cip}) reflects that these are dual to the invariants (in the sense of linear algebra). For this, instead of imposing a restriction on the states we impose an equivalence relation under action of the constraints:
\begin{equation}
    \Vco  = \hilb_0/\Im(H),
\end{equation}
so states in $\Vco$ (which we denote by $|\psi\rc$) are cosets
\begin{equation}\label{eq:coinv}
    |\psi\rrangle = \{|\psi\rangle + H|\chi\rangle : |\chi\rangle \in \hilb_0 \}.
\end{equation}
In other words, $|\psi\rrangle$ is an equivalence class of states under gauge transformations generated by $H$. We have written this in terms of infinitesimal gauge transformations, though we get the same by imposing equivalence under finite transformations $|\psi\rangle \sim e^{-iHt}|\psi\rangle$ for any $t$. Like the invariants, $\Vco$ is for now only a vector space of states, and we will later discuss methods for equipping it with an inner product.

To see that the space of invariant states $\Vinv$ is dual to the co-invariants $\Vco$, note that there is a natural sesquilinear pairing $\li \phi | \psi\rc$ between an invariant $|\phi\ri$ and a coset $|\psi\rc$ (or equivalently the complex conjugate $\lc \psi | \phi\ri$). This pairing is simply the inner product in $\hilb_0$, $\li \phi | \psi\rc := \langle \phi | \psi\rangle$ where we may take $|\psi\rangle$ to be any representative of the coset; this is well-defined since $H$ is Hermitian on $\hilb_0$ and $|\psi\rangle$ is invariant. This gives an anti-linear map $|\phi\ri\mapsto \li \phi|\ \cdot\  \rrangle$ from $\Vinv$ to $\Vco^*$, the space of linear functionals $\Vco\to \CC$, which is bijective (every functional is given by pairing with some invariant, and every non-zero invariant gives a non-zero functional). Hence we have an anti-linear isomorphism $\Vinv \simeq \Vco^*$ between invariants and the dual of co-invariants. In fact, in a more mathematically precise treatment this isomorphism can be taken to be the \emph{definition} of $\Vinv$, as we sketch now.


\subsection{Technical comments}\label{ssec:technical}

Since there are no normalisable states annihilated by $H$ (under the assumption of continuous spectrum at zero energy), a precise definition of $\Vinv$ requires an enlarged space of distributional states. For this, we first choose a dense subspace $\Phi \subset \hilb_0$ of `test functions' (equipped with an appropriate topology). For example, in a sigma model as in section \ref{ssec:sigma} we might take $\Phi$ to be the smooth compactly supported functions on target space $\target$ (though this is more restrictive than necessary). Then distributional states are defined as elements in the dual space $\Phi^*$ of continuous linear functionals on $\Phi$, though in  fact it will be more convenient to use the complex-conjugate space $\bar{\Phi}^*$ of antilinear functionals. So, $|\phi\rangle\in \bar{\Phi}^*$ is defined via the map $|\psi\rangle\mapsto \langle\phi|\psi\rangle^* = \langle\psi|\phi\rangle$ taking $|\psi\rangle\in \Phi$ to $\mathbb{C}$.

If $H$ maps test functions to test functions (and is continuous with respect to the chosen topology on $\Phi$), then using its self-adjointness we can define $H$ on $\bar{\Phi}^*$ via duality: $H|\phi\rangle$ is defined by the antilinear map taking $|\psi\rangle\in\Phi$ to $\langle \phi|H|\psi\rangle^*$. Invariants are then defined as distributions annihilated by $H$, $\Vinv = \ker H \subset \bar{\Phi}^*$; these are the antilinear functionals which vanish on the image of $H$. To get a nice pairing between invariants and co-invariants, we must construct $\Vco$ not from all of $\hilb_0$ but only from test functions, so $\Vco = \Phi/(\im H)$. Thus, in the equation $\li \phi | \psi\rc := \langle \phi | \psi\rangle$ we should read the right hand side as a pairing between $|\phi\rangle \in \Vinv \subset \bar{\Phi}^*$ and $|\psi\rangle \in \Phi$. From these definitions, an alternative characterisation of $\Vinv$ is the complex conjugate  of the dual of co-invariants $\overline{V}_\mathrm{co}^*$ (continuous antilinear functionals on $\Vco$, with continuity defined by the quotient topology on $\Vco$). In other words, we can use the isomorphism at the end of the last subsection to \emph{define} $\Vinv$.

There is a lot of freedom in choosing $\Phi$ and the choice certainly has an impact on various of the constructions below. But once we have chosen an inner product and passed to the completion to get the physical Hilbert space, many of the precise details will not matter. So, we will not distract ourselves too much with these details, only making a few comments in footnotes.

See \cite{Marolf:1995cn} for a much more detailed discussion of technical issues in systems with a single constraint.

\subsection{Invariants and co-invariants in BRST quantisation}\label{ssec:BRSTstates}

These two spaces of states arise naturally in the framework of BRST quantisation. Associated with the constraint $H=0$ there is a single pair of (Hermitian) ghosts $b,c$ with anti-commutator $[b,c]_+=1$, represented on a two-dimensional ghost Hilbert space $\hilb_g$:
\begin{equation}
    \hilb_g \text{ spanned by }|\downarrow\rangle,|\uparrow\rangle, \text{ with } c|\downarrow\rangle = |\uparrow\rangle, \: b|\uparrow\rangle = |\downarrow\rangle, \: b|\downarrow\rangle = c|\uparrow\rangle=0.
\end{equation}
Physical states are then described by the cohomology of the BRST operator $Q = cH$ acting on $\hilb_0\otimes\hilb_g$:
\begin{equation}\label{eq:HBRSTco}
    \hilb_\mathrm{BRST} = \frac{\ker Q}{\im Q}, \qquad Q=c H.
\end{equation}
This Hilbert space further decomposes into sectors of fixed `ghost number' $N_g = cb-\frac{1}{2}$ (with the constant chosen so that $N_g^\dag=-N_g$), so states $|\psi\rangle\otimes|\downarrow\rangle$ and $|\psi\rangle\otimes|\uparrow\rangle$ have ghost number $-\frac{1}{2}$ and $+\frac{1}{2}$ respectively for $|\psi\rangle \in \hilb_0$. The BRST operator has ghost number one, with $Q|\psi\rangle\otimes|\downarrow\rangle =H|\psi\rangle\otimes|\uparrow\rangle $ and $Q|\psi\rangle\otimes|\uparrow\rangle=0$. At the `lower level' of cohomology the image of $Q$ is trivial (only $0$) so physical states are simply given by the kernel of $Q$, which is the space of invariants $\Vinv$. At the `upper level' $Q$ annihilates everything, so the cohomology consists of cosets with respect to the image of $Q$, which is the space of co-invariants $\Vco$:
\begin{equation}\label{eq:HBRST}
    \hilb_\mathrm{BRST} \simeq \Vinv \otimes |\downarrow\rangle \oplus \Vco \otimes |\uparrow\rangle.
\end{equation}
A physical Hilbert space should consist only of states at some fixed ghost number, so we should take either invariants or co-invariants, not both. 

Note that for more complicated theories with multiple constraints, we have many choices in how to implement them: all invariant (\`a la Dirac), all co-invariant, or a mixture (the typical approach in gauge theories). Correspondingly, in BRST we have a single ghost for every constraint. We comment on these choices for gravity in the discussion \ref{sec:disc}, along with the additional technicalities (non-abelian constraint algebras and field dependent structure constants) that arise in BRST with several constraints. We will focus throughout on ideas which readily generalise to this more interesting context.

This BRST Hilbert space has a natural inner product induced from inner products on $\hilb_0$ and on $\hilb_g$. The latter is determined (up to normalisation) by Hermiticity of the ghost operators $b,c$ to be $\langle \uparrow | \downarrow \rangle = \langle \downarrow | \uparrow \rangle = 1$, with $\langle \uparrow | \uparrow \rangle = \langle \downarrow | \downarrow \rangle = 0$. This means that states in the lower level $|\downarrow\rangle$ pair with states in the upper level $|\uparrow\rangle$ with the opposite ghost number. The result is simply constructed from the above sesquilinear pairing $\li \cdot|\cdot \rc$, $\lc \cdot|\cdot \ri$ between invariants and co-invariants. It does not induce an inner product at any fixed ghost number and it is not positive-definite, so is not a candidate for the physical inner product of the constrained theory. Any pairing between co-invariants should naturally be accompanied by a $b$ ghost insertion, and a pairing between invariants should come with a $c$ ghost; we will see later how these arise from the residual diffeomorphism symmetry.

\subsubsection{Non-minimal BRST}\label{sssec:nonmin}

We will later make use of the non-minimal formalism \cite{Henneaux:1994lbw} where we retain the lapse $N$ as a dynamical variable, along with the additional constraint that its canonical momentum vanishes, $\Pi=0$. Before imposing constraints, states are wavefunctions $\psi(q,N)$ depending on $N$ (which we allow to take any real value, allowing for proper time to run in the opposite direction to coordinate time).\footnote{Apart from being conceptually important later for constructing inner products, this also has the technical benefit that the constraint $\Pi= -i \partial_N $ is essentially self-adjoint for real-valued $N$. If we were to impose the restriction $N>0$, then $\Pi$ would have no self-adjoint extension, which later leads to problems with constructing inner products.} In the BRST formalism, an additional pair of ghosts $\hat{b},\hat{c}$ are associated with the extra constraint. The ghost number is $N_g=cb-\hat{b}\hat{c}$, and the BRST charge becomes $Q= cH + \hat{c}\Pi$. The ghost Hilbert space is then four-dimensional, with a state $|\downarrow\downarrow\rangle$ at ghost number $N_g=-1$ annihilated by both $b,\hat{b}$, two states $|\uparrow\downarrow\rangle= c|\downarrow\downarrow\rangle$ and $|\downarrow\uparrow\rangle= \hat{c}|\downarrow\downarrow\rangle$ at $N_g=0$, and $|\uparrow\uparrow\rangle= c\hat{c}|\downarrow\downarrow\rangle$ at $N_g=+1$.

In this formalism, we take the physical states to be the BRST cohomology at ghost number $N_g=0$. For now we consider states proportional to either $|\downarrow\uparrow\rangle$ or $|\uparrow\downarrow\rangle$, and not superpositions of the two.\footnote{We could more generally look for closed and exact states with a superposition of $|\downarrow\uparrow\rangle$ and $|\uparrow\downarrow\rangle$, which might lead to equivalences between the two sectors. This is a little subtle, depending on technical choices regarding which wavefunctions are considered.  We will revisit such mixing of the sectors when considering the construction of inner products in section \ref{sec:BFVIP}.}

In the first case where we take the ghost state to be $|\downarrow\uparrow\rangle$, we find matter/lapse states which are invariant with respect to $H$ and co-invariant with respect to $\Pi$. Demanding that $|\psi\rangle\otimes|\downarrow\uparrow\rangle$ is BRST closed requires $H|\psi\rangle=0$, while $\Pi|\phi\rangle\otimes|\downarrow\uparrow\rangle$ is a BRST exact state $Q(|\phi\rangle\otimes|\downarrow\downarrow\rangle)$ for any matter/lapse state $|\phi\rangle$ satisfying $H|\phi\rangle=0$. This means that a physical wavefunction $\psi(q,N)$ must obey the Wheeler-DeWitt equation with respect to $q$ (for any fixed $N$), and among such states any derivative with respect to $N$ is regarded as trivial. This means that the physical state is determined only by the integral $\int_{-\infty}^\infty \psi(q,N) dN$, a function depending on $q$ only which solves the Wheeler-DeWitt equation. In particular, we can always choose a representative of the cohomology class which is an eigenstate of $N$ (with definite lapse $N=N_0$ for any fixed $N_0$), with wavefunction $\psi(q,N)\propto \delta(N-N_0)$. We can write such a state as $|\psi\ri\otimes|N=N_0\rangle\otimes |\downarrow\uparrow\rangle$, where $|\psi\ri$ is an invariant state as discussed above.

Similarly, states proportional to $|\uparrow\downarrow\rangle$ are co-invariant with respect to $H$, and invariant with respect to $\Pi$ so the   wavefunction $\psi(q,N)$ must be independent of $N$. The physical states of this form can be written as $|\psi\rc\otimes|\Pi=0\rangle\otimes |\uparrow\downarrow\rangle$, where $|\Pi=0\rangle$ is the state annihilated by $\Pi$ in the lapse sector (an equal superposition over all $N$).

 Combining these, the cohomology at $N_g=0$ is a direct sum of invariants and co-invariants (equivalent to \eqref{eq:HBRST}):
\begin{equation}\label{eq:hilbBFV}
	\hilb^{N_g=0}_\mathrm{BRST} \simeq  \Vinv \otimes |N=N_0\rangle \otimes |\downarrow\uparrow\rangle \oplus \Vco \otimes |\Pi=0\rangle \otimes |\uparrow\downarrow\rangle.
\end{equation}
In addition, there is a natural inner product on this  $N_g=0$ cohomology. But once again, this does not give us a norm on invariants or on co-invariants, because the nonzero inner products on the $N_g=0$ ghost Hilbert space are $\langle \uparrow\downarrow|\downarrow\uparrow\rangle$ and $\langle\downarrow\uparrow| \uparrow\downarrow\rangle$. Instead, we get an anti-symmetric  `off-diagonal' pairing between $\Vco$ and $\Vinv$, reproducing the natural pairing we saw above.

\subsection{Natural invariant and co-invariant states}\label{ssec:naturalstates}

We have two different methods of describing physical states. In the next section, we will see how they become equivalent to one another. Nonetheless, some states are simpler or more natural to describe in one language or the other. The summary is that states that involve arbitrary time-evolution in their definition often give rise to invariants (with the evolution projecting onto solutions to the Wheeler-DeWitt equation) \cite{Halliwell:1990qr}, while states defined without any evolution (e.g., from `initial data' at a single moment in time) can be interpreted as co-invariants.

\subsubsection{Scattering states as invariants}

The language of invariants is well-suited to describing states defined asymptotically, in the distant past or future. 
In the sigma models of section \ref{ssec:sigma}, the main examples are scattering states in $\hilb_0$: in-states defined at $t\to -\infty$, or out-states at $t\to +\infty$.

 Concretely, if target space is asymptotically flat in some direction (so $G_{ab}(q)$ and  $V(q)$ approach constant values $\eta_{ab}$ and $\frac{m^2}{2}$ as $q\to\infty$ in that direction), we can define states
\begin{equation}\label{eq:scatstate}
    |p_{\mathrm{in},\mathrm{out}}\ri := \lim_{t\to \pm \infty} e^{- iHt}|p\rangle,
\end{equation}
where $|p\rangle$ is the plane wave wavefunction $e^{i p_a q^a}$ (up to some convenient normalisation, perhaps). The relevant direction taking $q\to\infty$ is $q^a \propto \pm \eta^{ab}p_b$, with the $\pm$ for out and in states respectively. The momentum should satisfy $\eta^{ab} p_a p_b+m^2=0$ to obey the constraint in the `free' region of target space (a more familiar definition of a scattering state would involve an opposite evolution $e^{+i H_0 t}$ with a free Hamiltonian, but we do not need this as $H_0$ annihilates $|p\rangle$). With this condition, we generally expect \eqref{eq:scatstate} to approach a nonzero limiting wavefunction which satisfies the constraints.

For an alternative definition of scattering states (which generalises to target spaces with different asymptotics), we can work more directly in a position basis. Intuitively, we define a state that was at some definite point in target space $q$ in the past (for an in-state), and take $q$ to infinity in some direction. To make this precise, define
\begin{equation}\label{eq:inStates}
    |p_{\mathrm{in},\mathrm{out}}\ri := \lim_{q\to \infty} \mathcal{N}(q) \int _{t_0}^{\pm \infty} dt\, e^{- iHt}|q\rangle,
\end{equation}
where $\mathcal{N}(q)$ (usually a phase) is chosen so that the limit exists and is nonzero and $t_0$ is arbitrary since the result comes from large $|t|$ in the limit. Choosing $t_0=0$, we can write the integral as an operator $\frac{1}{iH \pm \epsilon}$ acting on $|q\rangle$; this gives a Green's function which does not solve the constraints for finite $q$, but instead gives a delta-function source on the right hand side. But when we take $q\to\infty$, this source gets pushed off to infinity and so the resulting state solves the Wheeler-DeWitt equation with no source.

In the most physically important examples of such scattering states, the relevant $q\to\infty$ region of superspace corresponds to a large spatial volume, with the direction specifying the conformal geometry and the behaviour of matter fields. An example is defining states in asymptotically de Sitter universes by data at future/past infinity $\scri_\pm$ (see \cite{Held:2024rmg} for an  example in de Sitter JT gravity).

Note that there may be some direction $q^0$ in target space which is natural to call `target space time' (particularly if $G_{ab}$ is Lorentzian). This need not be correlated with our `worldline' time. In particular, for an asymptotically Minkowski target space (with $V(q)\sim \frac{m^2}{2}>0$ as $q^0\to\pm\infty$) we expect four classes of scattering states: in- and out-states, with $p_0$ positive or negative. From the perspective of worldine QFT, these four choices correspond to solutions to the equations of motion which are purely positive or negative frequency in the asymptotic past or future of target space. These will not be independent states; the `S-matrix' of the worldline quantum mechanics relating in- and out-states describes the Bogoliubov coefficients of the QFT. We discuss these states explicitly in section \ref{ssec:1Dtarget}.

\subsubsection{Finite time states as co-invariants}

While invariants are a useful description of asymptotic states, co-invariants give us a natural way to define states at some finite time. For this, we simply choose some arbitrary wavefunction $|\psi\rangle\in \hilb_0$ in the unconstrained Hilbert space as a representative of the coset $|\psi\rc$. In particular, we can choose states corresponding to a definite $q=q_0$ (e.g., fixed spatial metric in a minisuperspace model), with wavefunction $\psi(q)=\delta(q-q_0)$. Other useful choices  include coherent states which are well-localised in both position and momentum space (e.g., both metric and extrinsic curvature) in the classical limit.

This is somewhat analogous to defining a classical spacetime metric by initial data at one time. The advantage is that we can describe physics in a way which is local in time. The price we pay is that the same state may be described in many different ways: for any $t$, $e^{-i H t}|\psi\rangle$ describes the same coset $|\psi\rangle$ defined at a different initial time (and we can take any superposition over such time translations). We comment more on the connection between this and gauge fixing in the next subsection.

 \subsection{States in the path integral formalism}\label{ssec:statesPI}

In the path integral formalism, states are associated with choices of initial or final boundary condition.  The different classes of states we have discussed above both appear naturally in the path integral formalism, but in slightly different ways. Understanding this is very useful for giving a proper Hilbert space interpretation to various gravitational path integrals.

First, the co-invariant states (particularly those with delta-function representative wavefunctions $\psi(q)=\delta(q-q_0)$ as above) appear very naturally in the path integral as boundary conditions at a fixed finite time. For example, we might be studying a path integral in pure gravity with a Dirichlet boundary condition of fixed spatial metric $g$ at initial time $t=0$, and (for a properly gauge-invariant path integral) we can interpret this boundary condition as a co-invariant state $|g\rc$. Given the redundancy inherent in such states, it is natural to `gauge fix' by restricting to some class of boundary conditions, which we discuss in a moment.

In contrast, the invariant scattering states introduced above can be defined in the path integral by boundary conditions in the infinite asymptotic past or future. This is manifest from \eqref{eq:scatstate}, for example, which corresponds to Neumann boundary conditions fixing $\dot{q}$ (in a configuration space path integral) or $p$ (in a first-order phase space path integral) at $t\to\pm \infty$.

Given this, it is straightforward to define a path integral which computes the pairing $\lc \cdot |\cdot \ri$ between co-invariant and invariant (scattering) states: we simply integrate over all fields on a semi-infinite half-line (say, $t<0$). We can remove the diffeomorphism invariance by applying a local gauge-fixing condition at each point in the bulk, most simply fixing lapse $N=1$  so that coordinate time coincides with proper time. This does not leave any residual gauge symmetries: rigid time translation is allowed locally, but disallowed by fixing the special $t=0$ boundary. Nor does it leave any moduli to integrate over (since there is a unique metric on the half-line up to diffeomorphisms). These facts mean that we do not have to make any further choices to define the path integral computing something like $\lc q|\psi_\mathrm{in}\ri$. This straightforwardness is a reflection of the fact that this pairing between invariants and  co-invariants is canonical. Finally, note that $\lc q|\psi_\mathrm{in}\ri$ can also be interpreted as the wavefunction $\psi_\mathrm{in}(q)$ of the invariant state (in the same way that $\langle q|\psi\rangle$ is the position-space wavefunction in ordinary quantum mechanics), which will satisfy the Wheeler-DeWitt equation as a concrete differential equation (rather than a more abstract operator version).

These boundary conditions also naturally give the correct states on the ghosts in the BRST formalism. The Faddeev-Popov ghost field $c(t)$ (associated with the Hamiltonian constraints) is a fermionic counterpart to the time component of vector fields $\xi(t)\partial_t$ which generate diffeomorphisms. For boundary conditions defined at a fixed finite time $t=0$, these diffeomorphisms must vanish on the boundary, $\xi(t=0)=0$. This Dirichlet boundary condition is inherited by the ghosts giving $c(t=0)=0$, which in the BRST Hilbert space results in states satisfying $c|\psi\rangle=0$. This defines the `top' level of cohomology, where the co-invariants live. There is no such constraint for the scattering states, leading to the opposite `Neumann' boundary condition $b=0$.

\subsection{Co-invariants and gauge-fixing}\label{ssec:coinvGF}

 While co-invariant states can be useful for making locality in time manifest, we must deal with the enormous ambiguity from the equivalence under acting with constraints. But such ambiguity is not unfamiliar: it is the usual redundancy from gauge-invariance, which we typically can deal with by choosing a gauge.

Quantum mechanically, gauge-fixing a state  means we would like to specify some way to choose a unique representative from each co-invariant coset. For example, in a sigma model we can divide our configuration space coordinates into a `clock' $\chi$ and transverse coordinates $q_\perp$, and specify the state at a time when $\chi=0$. Classically, this means specifying an $H=0$ trajectory in phase space by a point at which it intersects the surface $\Sigma_\chi$ defined by $\chi=0$: that is, by initial data satisfying the gauge condition $\chi=0$. A quantum version of this is to make an ansatz (with two terms because $H$ is quadratic in momenta, as explained below)
\begin{equation}\label{eq:coinvGF}
    \psi(q) = \Vert d\chi\Vert \delta(\chi) \psi_1(q_\perp) + \delta'(\chi) \psi_2(q_\perp).
\end{equation}
The factor $\Vert d\chi\Vert$ of the norm of the derivative of $\chi$ (under the target space metric) is chosen to make this invariant under reparameterisations of the clock variable. An analogue in a higher-dimensional theory of gravity is to choose a function $\chi$ of spatial metric and matter fields at each position in space, so that (classically) $\chi=0$ defines a Cauchy surface for the spacetimes of interest.

Ideally, each coset in $\Vco$  will have a unique wavefunction representative of this form: that is, the map from a pair of functions $(\psi_1,\psi_2)$ of $q_\perp$ to $\Vco$ defined by \eqref{eq:coinvGF} is bijective. This could fail in two ways. Either some cosets do not have a representative (the map is not surjective), which means that this ansatz will not describe the whole physical Hilbert space; the classical analogue is a trajectory that never encounters the surface $\Sigma_\chi$ in phase space. Alternatively, there may be several states that give rise to the same coset (so that there is some nonzero `pure gauge' $(\psi_1,\psi_2)$ in the image of $H$). In that case we have not completely eliminated the redundancy in the co-invariant description; the classical analogue is a trajectory which crosses $\Sigma_\chi$ several times.

Our ansatz is chosen to satisfy this existence and uniqueness locally at $\chi=0$ and perturbatively in diffeomorphisms, by which we mean that there is a unique representative among wavefunctions supported on the $\chi=0$ surface, imposing triviality of the image of $H$ among such states (which is weaker than using finite transformations $e^{-i H t}$). We included terms proportional to $\delta(\chi)$ and $\delta'(\chi)$ since $H$ is quadratic in momenta (a quantum analogue of initial data requiring both position and velocity), allowing us to remove second and higher derivatives of  $\delta(\chi)$. It is reasonable to expect that a sufficiently smooth wavepacket supported in a neighborhood of $\chi=0$ will also have a representative of the form \eqref{eq:coinvGF}, at least approximately (for example, one strategy is to Fourier transform from $\chi$ to the conjugate momentum space variable $p_\chi$, split the resulting wavefunction $\psi$ into even and odd parts in $p_\chi$, and eliminate $p_\chi^2$ using the constraint).\footnote{We can investigate whether $\chi=0$ provides a  `good gauge' non-perturbatively via the pairing with invariants, checking that the ansatz \eqref{eq:coinvGF} preserves non-degeneracy of the form $\li \phi |\psi\rc = \int \phi^*\psi$. That is, the pairing should vanish for all invariants  $|\phi\ri$  only if $\psi_1=\psi_2=0$ (otherwise there are residual gauge equivalences), and it should vanish for all $\psi_{1,2}$ only if $|\phi\ri=0$ (otherwise, there are co-invariants without a representative of this form). The pairing results in an integral over the surface $\Sigma_c$ (here meaning a surface in configuration space $\target$ rather than phase space $T^*\target$):
\begin{equation}\label{eq:invcoform}
    \li \phi |\psi\rc = \int_{\Sigma_\chi} dq_\perp \left[\phi^* \psi_1 -(\nabla_N \phi^*)\psi_2\right].
\end{equation}
The measure is the volume form on $\Sigma_\chi$ induced from the target space metric $G$, and $\nabla_N$ the normal derivative. Resemblance to the Klein-Gordon form (which we will meet below) is not a coincidence. The non-degeneracy of \eqref{eq:invcoform} is equivalent to existence and uniqueness of solutions to $H|\phi\ri=0$ given initial data $(\phi,\nabla_N \phi)$ on $\Sigma_\chi$. This holds if $\target$ is Lorentzian and globally hyperbolic and $\Sigma_\chi$ is a Cauchy surface \cite{bar2010linear}, so in such cases \eqref{eq:coinvGF} is a good gauge (but is likely to fail otherwise).}

For applications to more realistic theories of gravity, we propose this as a practical way to describe the physical Hilbert space by local data, working in perturbation theory around some classical background. This means making a gauge-fixed ansatz analogous to \eqref{eq:coinvGF}, where the choice corresponds to a gauge which is good classically for spacetimes sufficiently close to the background. For this, it is probably most convenient to replace the delta functions with narrow Gaussians, so that the states correspond to classical field configurations (coherent states, well-localised in position and momentum). We will later give a systematic procedure to construct such an ansatz from a classical gauge choice (including Gaussian averaging). We then expect the gauge-fixed ansatz to take care of all redundancy in perturbation theory. Nonetheless, there can still be non-perturbative redundancies; we propose these as a possible gauge symmetry explanation for the existence of null states in effective gravity. We revisit this point in the discussion section \ref{sec:disc}.

\subsubsection{Maximal volume gauge}\label{sssec:maxvol}

Of course, there are many possible choices of gauge condition $\chi=0$, where $\chi(q,p)$ can depend on momenta as well as positions. Each gauge condition should lead to a different possible ansatz like \eqref{eq:coinvGF}. An important example for higher-dimensional theories of gravity is the `maximal volume' (or zero mean curvature) gauge choice \cite{Witten:2022xxp}. For this, write the spatial metric $g=\Omega^2 \hat{g}$ in terms of a conformal metric $\hat{g}$ (fixed to have unit determinant, for example) and conformal factor $\Omega$, and choose a wavefunction to be independent of $\Omega$. This means that the state is a delta function in the canonical conjugate to $\Omega$, which is proportional to the mean curvature (trace of extrinsic curvature) $\Tr K$. An analogue for a one-dimensional sigma model \eqref{eq:sigmaH} is to choose a gauge which is linear in momentum, such as $\chi= p_0$ where $p_0$ is the momentum conjugate to some target space coordinate $q^0$. In a mini-superspace model, $q^0$ might be interpreted as parameterising the volume of space  (in which case it is often the timelike direction in a Minkowski-signature target space). This is a good gauge when classical trajectories involve $q_0$ increasing to a unique maximum before decreasing. This could model a big bang/big crunch cosmology, or alternatively a family of Cauchy surfaces in an asymptotically AdS universe all sharing the same asymptotics at spatial infinity. In two-dimensional JT gravity (either for closed universes or a two-sided Hilbert space with asymptotically AdS boundary conditions),  a similar interpretation applies to the geodesic states studied in \cite{Harlow:2018tqv,Iliesiu:2024cnh}, with $\chi$ being the momentum conjugate to the dilaton (which is proportional to extrinsic curvature). In our language these can be interpreted as co-invariants with wavefunctions chosen to be independent of the dilaton (or a delta function in its canonical conjugate, the extrinsic curvature).

\section{Gravitational inner products}\label{sec:IP}

From the above discussion of spaces of states, two issues remain to be resolved in order to define a physical Hilbert space of the constrained theory. The first is that we apparently have two different ways to define a space of states, so we must decide whether to use invariants or co-invariants. The second is that we do not immediately have an inner product on either of these spaces of states.

In this section we point out that both of these are resolved together: an inner product on either $\Vco$ or $\Vinv$ gives us an equivalence between these two spaces of states. We then discuss specific proposals for the inner product. First we describe the `group average' inner product and related ideas, which we believe to be the most natural construction for the gravitational Hilbert space. We then discuss the Klein-Gordon form for sigma models, which was the originally proposed inner product of DeWitt \cite{DeWitt:1967yk}. We explain how each of these arise in the path integral formulation, in particular how the Klein-Gordon form is obtained from gauge fixing \cite{Woodard:1989ac} and its relation to the gauge-fixed co-invariants in section \ref{ssec:coinvGF}. We then illustrate these ideas explicitly in models with one-dimensional target spaces. Finally, we contrast these constructions with the Hilbert space obtained in  the worldline formulation of QFT.

\subsection{Inner products and equivalences between invariants and co-invariants}

The spaces of co-invariants and invariants are not unrelated: as noted above they are dual spaces, with the natural nondegenerate sesquilinear pairing $\lc\cdot |\cdot\ri$ between $\Vinv$ and $\Vco$ (and hence the anti-linear bijection $\Vinv\simeq\Vco^*$). With this, a Hermitian form on $\Vco$  is equivalent to a linear Hermitian map $\eta:\Vco\to \Vinv$ (usually called a \emph{rigging map}) by defining
\begin{equation}\label{eq:etaIP}
    \lc \psi_2|\psi_1\rc_\eta  := \lc\psi_2| \eta\psi_1\ri=\li \eta\psi_2  |\psi_1\rc, \qquad |\psi_{1,2}\rc\in \Vco\,.
\end{equation}
The second equality is a requirement on $\eta$ to ensure Hermiticity of $\lc \cdot|\cdot\rc_\eta$. It is equivalent to $\eta^\dag = \eta$, with $\dag$ defined with respect to the pairing $\li\cdot|\cdot\rc$ (so $\eta^\dag:\Vco\to \Vinv$ is defined by $\llangle\psi_2 |\eta^\dag \psi_1\ri = \li \eta \psi_2|\psi_1\rrangle$).

If this form is positive definite we have an inner product on co-invariants, and passing to the completion with respect to the resulting norm on $\Vco$ gives us a Hilbert space $\hco$ of co-invariants.\footnote{To get a good Hilbert space it is sufficient for $\lc \cdot|\cdot\rc_\eta$ to be positive \emph{semi}-definite, in which case the resulting Hilbert space can informally be thought of as the quotient $\hco\simeq \Vco/\ker \eta$ of all co-invariants by the `null states' $\ker \eta$ (in addition to the null states $\im H \subseteq \hilb_0$ already implicit in the definition of $\Vco$). But we see no reason to expect additional null states, unless there is some additional (perhaps discrete) gauge symmetry which could be accounted for in the original definition of $\Vco$.} We will see interesting examples below which do not obey this, but only satisfy a weaker condition that the form is non-degenerate (i.e., if $\lc\psi_2|\psi_1\rc_\eta=0$ for all $|\psi_2\rc$, then $|\psi_1\rc=0$), which is equivalent to $\eta$ being injective ($|\eta \psi\ri=0\implies |\psi\rc=0$). This means that $\eta$ is a linear isomorphism between co-invariants $\Vco$ and a subspace $\im \eta\subseteq \Vinv$ of invariants. It is reasonable that this subspace might not be the whole of $\Vinv$, since we would like to exclude invariant wavefunctions corresponding to non-normalisable states, for example (comments on examples are given in footnote \ref{foot:domains}). The upshot is that putting an inner product on co-invariants gives us an equivalence between our two ways of describing states, and vice-versa.

It is slightly more concrete to think of $\eta$ as an operator on the original Hilbert space $\hilb_0$,\footnote{More precisely, it's a distributional operator which maps test function states $\Phi$ to distributional states in  $\bar{\Phi}^*$, in the notation of section \ref{ssec:technical}.} and this is the perspective we will take for the most part. To get this from the above map $\Vco\to\Vinv$ we simply map any state $|\psi\rangle\in \hilb_0$ to its co-invariant coset $|\psi\rc$ before applying $\eta$, and regard the resulting invariant $|\eta\psi\ri$ as a (distributional) state in $\hilb_0$. In this way of thinking we require $\eta H = 0$ (because $\eta$ is well defined on co-invariant cosets) and $H\eta=0$ (because $\eta$ maps to invariants). It will also be Hermitian $\eta^\dag=\eta$ with respect to the usual $\hilb_0$ inner product (which makes the two properties above equivalent), and positive semi-definite to get a good Hilbert space (with $\im H$ giving null states).

The map $\eta$ also induces an equivalent Hilbert space structure on invariants. Assuming $\eta$ is nondegenerate as above, it has an inverse $\eta^{-1}:\im\eta \subseteq\Vinv\to \Vco$.  we get a Hermitian form on invariants by inverting $\eta$:
\begin{equation}
	\li\psi_2|\psi_1\ri_\eta = \li  \psi_2| \eta^{-1}\psi_1\rc = \lc \eta^{-1}  \psi_2| \psi_1\ri = \lc \eta^{-1}\psi_2 |\eta^{-1}\psi_1\rc_\eta, \qquad |\psi_{1,2}\ri\in \im\eta .
\end{equation}
As for co-invariants, if $\eta$ is positive-definite this is an inner product, and by taking the completion we turn $\im\eta\subseteq\Vinv$ into a Hilbert space $\hinv$, which is naturally isomorphic to $\hco$ using $\eta$.

Once again, we would like to make this more concrete and practical by thinking about $\eta^{-1}$ as a map on the original Hilbert space $\hilb_0$. This means we want an operator which takes a solution $|\psi\rangle$ to the Wheeler-DeWitt equation $H|\psi\rangle$, and gives us some state $|\phi\rangle$ such that $\eta|\phi\rangle=|\psi\rangle$. But this leaves us with a lot of freedom: we can choose any gauge-equivalent representative $|\phi\rangle$ from its co-invariant coset. As a result, $\eta^{-1}$ is represented concretely by a whole family of maps, any one of which we call $\kappa$. The condition for $\kappa$ to be compatible with $\eta$ is that it is a generalised inverse, meaning
\begin{equation}\label{eq:etakappaeta}
    \eta \kappa \eta = \eta\,.
\end{equation}
One way to read this equation is that $\kappa$ is a right inverse to $\eta$ (i.e., $\eta\kappa$ is the identity) when acting on the image of $\eta$ (in particular, on invariant wavefunctions). Another way to read it is that $\kappa$ is a left inverse to $\eta$, up to addition of a null state in $\ker\eta$.

Since $\kappa$ selects some particular representative from a co-invariant coset, we can think of it as a \emph{gauge-fixing map}. For example, following the discussion of section \ref{ssec:coinvGF} we might like to choose $\kappa$ to fix to the $\chi=0$ gauge, which means that it always maps invariants to wavefunctions of the form \eqref{eq:coinvGF}. Such a map is nice for recovering a  description which is local in time (on a particular gauge-fixed Cauchy surface) from a Wheeler-DeWitt wavefunction which is highly non-local in time, involving a superposition over all possible Cauchy surfaces. We will see later how to construct such a $\kappa$ systematically from any gauge condition. The challenge  (arising from Gribov ambiguities, and the impossibility of finding a universally applicable gauge condition) is to perform such a gauge-fixing while simultaneously satisfying the compatibility condition \eqref{eq:etakappaeta}.

The various maps and inner products introduced in this section and the relations between them were sketched in figure \ref{fig:introcoinv}.


This discussion is rather abstract, so we now turn to examples which illustrate these ideas. At this point there is a great deal of freedom in the selection of an inner products ($\eta$ or $\kappa$) with the above properties, so these examples may seem like arbitrary choices. We will give a more principled justification of these choices later when we discuss how these inner products arise in the BRST formalism, which provides a much more rigid framework.

\subsection{Group averaging inner product}\label{ssec:GA}

There is one choice of a rigging map which does not depend on any details of the unconstrained theory besides the abstract Hamiltonian constraint $H$, and automatically gives us a positive inner product. It also has a simple, natural and general path integral description (as we will describe in section \ref{ssec:IPPIs}), and also emerges naturally in the BRST formalism (section \ref{sec:BFVIP}). For these reasons, we advocate that this is the leading candidate for the physical inner product.

 As a map on $\hilb_0$, we can write this formally as a delta function of the constraint:
\begin{equation}
    \eta = 2\pi\delta(H).
\end{equation}
This is manifestly Hermitian, positive semi-definite (since $\delta$ is a non-negative distribution), and satisfies $\eta H=H \eta=0$ as required for a rigging map.

Another suggestive way to write this is as a `group average', integrating over the action of the gauge group of time translations:
\begin{equation}\label{eq:GA}
    2\pi\,\delta(H) = \int_{-\infty}^{\infty} dt\, e^{-iH t}.
\end{equation}
This integral will converge when we take the matrix elements of states with sufficiently smooth energy resolution. This construction generalises rather nicely to many other groups $G$ of gauge transformations. For that, take an unconstrained Hilbert space $\hilb_0$ which furnishes a unitary representation $U(g)$ of $G$; then $\eta$ is constructed by integrating $U(g)$ over $G$ with an appropriate measure \cite{Giulini:1998kf}. In the case that $G$ is compact (so that it has a discrete set of irreducible unitary representations), $\eta$ is simply the orthogonal projection onto the trivial representation, the subspace $\Vinv\subseteq \hilb_0$ of gauge-invariant states. But the construction applies also for many non-compact groups, such as the case $\RR$ of interest here, as well as more complicated situations encountered in gravity including constraint algebras with field-dependent structure constants \cite{Shvedov:2001ai}.

Note that this generalisation also naturally incorporates global information about the gauge group; this is in contrast to BRST constructions which depend only on the Lie algebra of infinitesimal gauge transformations.

This $\eta$ can be thought of either as a map  $\Vco \to \Vinv$, which `projects' a general state onto a solution of the Wheeler-DeWitt equation,  or as an inner product on $\Vco$.

As explained above, this $\eta$ also induces an equivalent dual inner product $\kappa$ on invariants. To understand this informally, recall that energy eigenstates in $\hilb_0$ are delta-function normalised, with $\langle \psi_{E'}|\psi_E\rangle \propto \delta(E-E')$. In particular, invariants  (zero energy eigenstates) have an infinite norm due to a factor of $\delta(E=0)$. The $\kappa$ dual to the group-average $\eta$ gives a finite inner product by dividing out by this infinite factor (times $2\pi$).

The `spectral analysis' inner product makes this mathematically precise \cite{Marolf:1994ae}. We  use the spectral theorem for a self-adjoint constraint $H$, which tells us that the Hilbert space decomposes as a direct integral of energy eigenspaces $\hilb_E$:\footnote{Here $dE$ is the usual flat measure on $\RR$ (up to an additional factor of $2\pi$ appearing in the inner product), and we have assumed that $H$ has a purely continuous spectrum. The construction goes through unaltered if $H$ has additional discrete spectrum, as long as there is a neighborhood of $E=0$ with no discrete eigenvalues.}
\begin{equation}\label{eq:Hdirectint}
    \hilb_0 = \int^\oplus dE \,\hilb_E \,.
\end{equation}
This means that a state $|\psi\rangle\in \hilb_0$ is given by specifying a family of states $|\psi_E\rangle\in \hilb_E$ for each energy $E$, and the inner product is computed by the integral $\langle\psi'|\psi\rangle = \int \frac{dE}{2\pi} \langle \psi'(E)|\psi(E)\rangle$. We can then simply define the invariants as the zero energy eigenspace, $\Vinv := \hilb_{E=0}$. This space already comes equipped with an inner product, and this is precisely the $\kappa$ described informally above. In this language, $\eta=2\pi\delta(H)$ is the map taking the family $\{|\psi(E)\rangle\}_E$ onto its zero energy component $|\psi(E=0)\rangle$. A generalised inverse $\kappa$ takes a zero-energy state $|\psi(E=0)\rangle$, and extends it in an arbitrary way to a smooth family $\{|\psi(E)\rangle\}_E$.

However, this spectral analysis construction of the inverse only directly generalises for abelian constraints. One of the challenges for practically applying the group averaging procedure is that there is not more generally an obvious recipe to write down or systematically compute the dual inner product $\kappa$ on invariant states. This is one motivation for exploring other possibilities and understanding how they relate to group averaging.

\subsection{The Klein-Gordon form}\label{ssec:KG}

While there are many reasons to advocate for the group averaging inner product, it was not the first proposal for gravity. This was the `Klein-Gordon' form on the space of invariants suggested by DeWitt in analogy to QFT \cite{DeWitt:1967yk}. This has the drawback that it is not positive-definite, so it is not truly an inner product (we will discuss the relation to the inner product in free QFT in section \ref{ssec:QFTIP}).  But while it has indefinite signature, it is non-degenerate so most of the above formalism continues to apply. Some useful references in the context of QFT are \cite{Hollands:2014eia,Wald:1995yp}.

For this construction, we need more structure than the abstract Hamiltonian $H$. We specialise to the sigma-model examples of section \ref{ssec:sigma}, for which the Hamiltonian is given by $H = -\frac{1}{2}\nabla^2 +V(q)$, where $\nabla^2$ is the Laplacian on a pseudo-Riemannian manifold $\target$ and $V$ a potential function on that target space. In fact, we will assume that the target space metric $G_{ab}$ is Lorentzian and such that $\target$ is  globally hyperbolic. This means that the constraint equation $H|\psi\rangle=0$ has a good Cauchy problem. Specifically, these conditions guarantee that solutions $\psi(q)$ are in one-to-one correspondence with initial data on a Cauchy surface $\Sigma$ (the value $\Phi=\psi|_\Sigma$ and normal derivative $\Pi=\nabla_N\psi|_\Sigma$ of the field $\psi$). For a short review of relevant technical results, see \cite{bar2010linear}. We emphasise that this is an initial value problem with respect to `target space time', not to be confused with the `worldline' time $t$ which is the physical time from the perspective of our one-dimensional theory of gravity.

In this specific context, the invariant states are precisely the solutions to the Wheeler-DeWitt equation $H|\psi\rangle=0$. More precisely, we may take $\Vinv$ to be the space of complex solutions $\psi(q)$ with smooth compactly supported initial data $(\Phi,\Pi)$ on some  Cauchy surface $\Sigma$ (a condition which is independent of $\Sigma$). On this space of invariants define the `symplectic bilinear form'
\begin{equation}\label{eq:Omega}
    \Omega(\psi_2,\psi_1) = \frac{1}{2}\int_\Sigma (\Pi_2\Phi_1-\Phi_2 \Pi_1),
\end{equation}
where $\Sigma$ is any Cauchy surface (we always take the measure of integrals on $\Sigma$ as the volume form induced from the target space metric $G$).  This is independent of the choice of $\Sigma$. Note that the definition requires us to choose a time-orientation for $\target$, since the initial velocity $\Pi=\nabla_N \phi$ is the normal derivative of the field with future-pointing unit normal vector $N$. By combining  with the `complex structure' $\psi \mapsto \bar{\psi}$ we define the Klein-Gordon form
\begin{equation}\label{eq:KGIP}
    \li\psi_2|\psi_1\ri_\mathrm{KG} := i \, \Omega(\bar{\psi}_2,\psi_1),
\end{equation}
which is a non-degenerate (but not positive) Hermitian form on $\Vinv$. The constant of proportionality must be imaginary for Hermiticity, but is otherwise a matter of convention.

Following the earlier general discussion, we can also characterise the Klein-Gordon form by relating it to structures involving the dual space $\Vco$. If $\Vinv$ is the space of solutions to $H\psi=0$, we can think of $\Vco$ as the space of \emph{sources} $f$ for the inhomogeneous equation $H\psi = f$. Take co-invariants $|f\rc$ to be represented by functions (or more generally distributions) $f$ on target space $\target$ with the natural pairing $\lc f|\psi\ri = \int_\target \bar{f}\psi$.  For this to be well-defined, we take $f$ to have compact support in time,\footnote{This means that the support of $f$ must be bounded by a pair of Cauchy surfaces. We could be less strict and allow non-compact support with sufficiently fast decay in the future, but this wouldn't add anything since every coset in fact has a representative with compact support in time.} and to define $\Vco$ we quotient by the image of $H$ in this space (since $\lc Hg|\psi\ri=0$ for any solution $\psi$). The compactness of $f$ in time is a very important condition, since without it any function would be in the image of $H$ and the space of co-invariants would become trivial. Now, the abstract results above tells us that the Klein-Gordon form is equivalent to an isomorphism between co-invariants and invariants (maps $\eta:\Vco\to\Vinv$ and $\eta^{-1}=\kappa:\Vinv\to\Vco$), and to a Hermitian form $\lc\cdot|\cdot\rc_\mathrm{KG}$ on co-invariants.

To make this concrete, consider the `causal propagator' (or Pauli-Jordan distribution) $\Delta$ which constructs solutions $\psi = \Delta f$ from a source $f$. This solution is defined as the difference between retarded $\mathcal{G}_+ f$ and advanced $\mathcal{G}_- f$ solutions to the constraint equation with source $f$. In  other words,
\begin{equation}
    \begin{gathered}
        \Delta f = \mathcal{G}_+ f- \mathcal{G}_-f, \\
        H \mathcal{G}_\pm f = f, \quad \mathcal{G}_\pm f=0 \text{ to the past($+$)/future($-$) of the support of } f.
    \end{gathered}
\end{equation}
This $\Delta$ gives us a linear map $\Vco\to\Vinv$ from co-invariants to invariants, since $\Delta f$ solves the constraint equation, and if $f$ is in the image of $H$ then  $\Delta f=0$. For the latter, note that if $f = Hg$ for some compactly supported $g$ then $g$ serves as both the advanced and retarded solution with source $f$, so $\mathcal{G}_+f = \mathcal{G}_-f = g$. It is also injective: suppose $\Delta f=0$ and let $g = \mathcal{G}_+f = \mathcal{G}_-f$; then $g$ is compactly supported and $f=Hg$, so $f\in \im H$. Surjectivity will be proved by explicitly constructing an inverse in a moment, so we have an isomorphism.\footnote{Follow the construction as outlined in section \ref{ssec:technical}, we might choose to start with test functions $\Phi$ being smooth and compactly supported (or rapidly decaying) wavefunctions, with a dual space $\Phi^*$ of distributions (or tempered distributions) on $\target$. Then $\im \eta_\mathrm{KG}$ will consist of solutions with smooth and compactly supported (or rapidly decaying) initial data. This would be a strict subspace of $\Vinv$, which would contain all (tempered) distributional solutions with no requirements on smoothness or decay of initial data. For many of the states in $\Vinv$ but not in $\im \eta_\mathrm{KG}$, the Klein-Gordon form will not converge so it is necessary to exclude them. Similar comments apply to group averaging, except that this does not respect target space lightcones so $\im\eta$ is not restricted to compactly supported initial data.\label{foot:domains}}

The causal propagator is related to the Klein-Gordon form by the identity
\begin{equation}\label{eq:fpsiOmega}
    \int_\target \psi f = \Omega( \psi,\Delta f),
\end{equation}
which we prove in appendix \ref{app:KGid}.
 In terms of our pairing between invariants and co-invariants we can rewrite \eqref{eq:fpsiOmega} as 
\begin{equation}
    \li \psi|f\rc = -i \li \psi|  \Delta f \ri_\mathrm{KG} \implies \eta_\mathrm{KG}|f\rc = -i|   \Delta f\ri,
\end{equation}
where in the second line we have identified the map $\eta_\mathrm{KG}:\Vco\to\Vinv$ associated with the Klein-Gordon form as $ -i \Delta$ . This immediately gives us a Hermitian form on co-invariants, which we can write very concretely in terms of the kernel $\Delta(q_2,q_1)$ of the causal propagator,
\begin{equation}\label{eq:KGcoinv}
    \lc f_2|f_1\rc_\mathrm{KG} = -i\int_\target \bar{f}_2(q_2)\Delta(q_2,q_1)f_1(q_1) dq_1dq_2.
\end{equation}

Note that the group average inner product can also be written as a difference between two propagators \cite{Casali:2021ewu}:
\begin{align}\label{eq:FminusAF}
    \eta_{GA} = 2\pi\delta(H) = i\left(\frac{1}{H+i\epsilon}-\frac{1}{H-i\epsilon}\right).
\end{align}
In term of the group average integral \eqref{eq:GA}, the terms come from positive and negative $t$ respectively.\footnote{This split is rather special to one dimension, where evolution is either forward or backward in proper time. In higher dimensions you also get contributions from a mixture, going forward in some regions of space and backwards in others.} This expression is `Feynman minus anti-Feynman' rather than `advanced minus retarded'. We can think of this as coming from a fixed-energy causal (retarded minus advanced) propagator for $H$ with respect to \emph{worldline time} (while the Klein-Gordon norm comes from the causal propagator with respect to target space). This hints at a relationship between the Klein-Gordon form and group averaging, which we will see in various guises later: they agree when the proper (worldine) time and target space time flow in the same direction, and disagree by a sign when they flow in opposite directions.

Finally, we can also  construct the corresponding map $\kappa$ from invariants to co-invariants. Concretely, this means finding a family of maps taking a solution $\psi$ to a source $f$ satisfying $\psi = 
-i\Delta f$ (with any two such sources differing by $Hg$ for some compactly supported $g$). Since the Klein-Gordon form can be expressed as an integral on a Cauchy surface $\Sigma$, it is natural to guess that we should be able to find a source $f$ for any $\psi$ with compact support contained in $\Sigma$. 
Assuming for now that such a source exists, restricting the solution $\psi = -i\Delta f$ to the future of $\Sigma$ would give us just the retarded solution: defining a step function $\Theta_\Sigma(q)$ equal to 1 to the future of $\Sigma$ and zero to the past, we have $\Theta_\Sigma(q) \psi(q) = -i \mathcal{G}_+f(q)$ (plus a compactly supported distribution $g$ on $\Sigma$, perhaps). Then we can recover $f$ (plus $Hg \in \im H$ which leads to the same coset in $\Vco$) by acting with the Klein-Gordon operator $H$:
\begin{equation}\label{eq:fSigma}
    f_\Sigma = i H (\Theta_\Sigma \psi).
\end{equation}
This explicitly constructs the desired compactly supported distribution inverting $\eta|f\rc = |\psi\ri$, so $\kappa |\psi\ri = |f_\Sigma\rc$. Despite appearances, $f_\Sigma$ is not in the image of $H$ because $\Theta_\Sigma \psi$ is not compactly supported in the future. More generally, we can replace $\Theta_\Sigma$ with any function equal to one in the future and zero in the past (or any other constant values differing by unity). We can also use this result to write the Klein-Gordon form in an alternative way,
\begin{equation}\label{eq:kappaSigma}
    \kappa_{\Sigma} = i [H,\Theta_\Sigma],
\end{equation}
where we write a commutator so that $\kappa_{\Sigma}$ is a more manifestly  Hermitian operator (though this still needs care, since $H$ is not Hermitian on invariants due to boundary terms when integrating by parts).
To recover \eqref{eq:Omega}, note first that integrals on $\Sigma$ can be rewritten in term of integrals on $\target$ (with appropriate measures) using the step function $\Theta_\Sigma$:
\begin{equation}
    \int_\Sigma N_a V^a = \int_\target V^a\nabla_a \Theta_\Sigma.
\end{equation}
We have $[H,\Theta_\Sigma] = -\frac{1}{2} (\nabla^a \nabla_a \Theta_\Sigma + \nabla_a \Theta_\Sigma \nabla^a)$, so the matrix elements of this commutator (after integrating by parts so the first term acts to the left) indeed give us \eqref{eq:KGIP}.

If we choose a `clock function' $\chi$ as one of our coordinates as in section \ref{ssec:coinvGF} so that the Cauchy surface $\Sigma$ corresponds to $\chi=0$ and $\Theta_\Sigma(q)=\Theta(\chi)$, then the wavefunction \eqref{eq:fSigma} resulting from the $\kappa$ map will precisely take the form of the ansatz \eqref{eq:coinvGF} we gave for our gauge-fixed co-invariant representative. Concretely, this arises from the product rule for the Laplacian in $H$ where either one derivative or two hit the factor of $\Theta(\chi)$, leading to $\delta(\chi)$ or $\delta'(\chi)$ terms (and terms where $\Theta$ is not differentiated do not contribute, using $H\psi=0$). This suggests a general way to construct such a gauge-fixing ansatz: we require that the wavefunctions are in the image of some particular $\kappa$ map. Later we will see how to construct much more general classes of $\kappa$ map systematically from any gauge choice, allowing us to write ansatze analogous to \eqref{eq:coinvGF} for any gauge.

\subsection{Path integrals, inner products and gauge fixing}\label{ssec:IPPIs}

Several of the constructions we have mentioned here arise naturally in the path integral formulation. To understand the various possibilities, we first recall from section \ref{ssec:statesPI} how co-invariant and invariant states often arise as boundary conditions in the path integral: co-invariants often are associated with boundaries at a finite time, while invariants  often arise from boundary conditions in the infinite past or future. When we try to pair these with one another, for one-dimensional gravity we arrive at spacetimes which look like various types of intervals. From a pairing of two co-invariants (where we expect an $\eta$ map) we get a finite interval $[T_i,T_f]$; from two invariants (expecting a $\kappa$ map) we get the whole real line $\RR$; if we have one of each we get a semi-infinite interval $[T_i,\infty)$ or $(-\infty,T_f]$. We already saw in \ref{ssec:statesPI} that the last option naturally and immediately gives us the canonical pairing between the two types of state $\li\cdot|\cdot\rc$, but for the other two possibilities different issues arise. Before we look at each of these, string theorists may like to note an analogy with various possible worldsheet topologies: for the propagation of closed strings these intervals are analogous to a cylinder, a sphere two-point amplitude and a disc one-point amplitude respectively. More details of this analogy are described in appendix \ref{app:string}.

\begin{figure}
    \centering
    \input{Figures/pairings}
    \caption{Pairings between various types of states in one-dimensional gravity arise naturally from path integrals over different sorts of intervals. A path integral over a finite interval (left) gives a pairing between co-invariants; fixing to constant lapse gauge leaves a single modulus, and integrating this over $\RR$ gives the group-average $\eta$. A path integral over an interval which is infinite at both ends (middle) naturally describes a pairing between invariants (described by asymptotic boundary conditions); there is a residual time-translation symmetry which must be gauge-fixed by insertion of a $\kappa$ map. Finally, a semi-infinite integral has neither a modulus  nor a residual symmetry, and it leads to the simple canonical pairing between one invariant and one co-invariant state.}
    \label{fig:Pairings}
\end{figure}
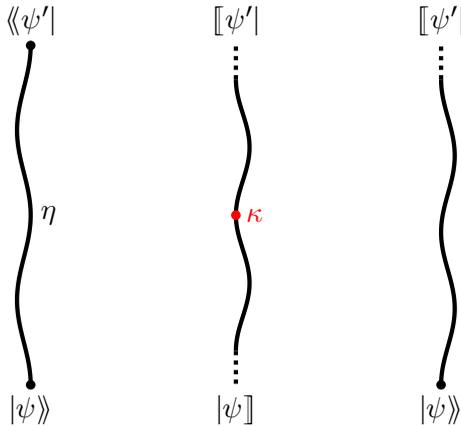

In the case of a finite interval, different spacetime metrics are classified up to diffs (which are required to fix the endpoints) simply by the total proper time $T$, which is the integral of the lapse $T= \int N$. The path integral over matter fields on this fixed metric computes matrix elements $\langle \psi_2|e^{-iH T}|\psi_1\rangle$ of time evolution between initial and final states. But this leaves us with the gravitational path integral over all metrics, which here is simply the integral over the `modulus' $T$. The natural measure for this integral is the obvious flat measure $dT$ \cite{Cohen:1985sm}, but we also have to decide on a range of integration. A full exploration of the possibilities is in \cite{Casali:2021ewu}. The only one which leads to a good pairing between physical states (i.e., which only depends on the co-invariant cosets $|\psi_{1,2}\rc$ of $|\psi_{1,2}\rangle$) is to integrate over the whole real line, including positive and negative $T$. The upshot is that the path integral over all spacetimes between initial and final boundaries at finite time lands us on the group averaging integral \eqref{eq:GA} (up to an arbitrary normalisation which can be absorbed into the measure):
\begin{equation}
	\lc \psi_2|\psi_1\rc_\mathrm{GA} = \int\frac{\mathcal{D}g \mathcal{D}q}{\operatorname{Diffs}} e^{iS}.
\end{equation}
Such a relation formally holds also in higher dimensions, and the path integral in fact provides a reasonable definition for the group average in gravity. We can interpret this path integral as an integral over the gauge group, which consists of diffeomorphisms of spacetime modulo diffs which vanish on a specific fixed $t$ Cauchy surface where a wavefunction is defined (in this case, the initial or final Cauchy surface corresponding to the end of the interval). See further discussion of this path integral in \cite{Araujo-Regado:2022gvw}. Finally we note that in the BRST formalism, the integral over an unfixed modulus is associated with insertion of a $b$-ghost (see \cite{Polchinski:1998rq}, for example), which is precisely what we require to get a nonzero pairing $\langle \uparrow|b|\uparrow\rangle$ between co-invariant states in the ghost sector.

For the last possibility we consider pairings between invariants, thought of as a path integral where spacetime is the whole real line (infinite in both directions). In this case, there is a single diffeomorphism class of metrics, so we have no moduli. However, after fixing to the gauge $N=1$ where coordinate and proper time agree (for example), we are still left with a one-parameter family of residual diffeomorphisms: the rigid time translations. The infinite norm $\propto\delta(E=0)$ of invariants using the inner product on the unconstrained Hilbert space $\hilb_0$ arises from the volume of this residual gauge group. In higher dimensions, the analogue is that we choose a local gauge condition which determines classical evolution completely (e.g., synchronous gauge  or harmonic gauge), but we are still left with the residual diffeomorphisms corresponding to choice of $t=0$ slice (and coordinates on that slice). Note that this residual gauge group is precisely the group described above (diffs modulo those which fix $t=0$), which is also the space of possible Cauchy surfaces (including choice of spatial coordinates on the surface) in a given spacetime.

 We see two possible ways to deal with the residual gauge group and get a finite result for the path integral. The first is to directly divide out by the infinite volume of the gauge group: this is what the spectral analysis inner product does, but we do not know how to generalise this beyond our simple 1D gravity models (it seems likely that the infinite volume of the unfixed gauge group would be field-dependent in higher dimensional models, if one could make sense of it). The more obvious alternative is to implement a further gauge condition which selects the $t=0$ slice (and in higher dimensions, coordinates on that slice to fix momentum constraints).

For one possible class of gauge conditions, we can select a clock field $\chi(q)$ (depending only on target space coordinates $q$ for now) and  fix the $t=0$ slice as a moment of time when $\chi$ vanishes, setting $\chi(q(t=0))=0$. This means inserting $\delta(\chi(q(0)))$ in the path integral. But to make the result independent of gauge choice, we should also insert a Faddeev-Popov measure factor, which is the Jacobian determinant of the derivatives of the gauge-fixing functions with respect to gauge transformations. For our one-dimensional models, this `determinant' is just the proper time derivative $\dot{\chi}$, so with this gauge choice we insert $\delta(\chi)\dot{\chi}$ at $t=0$ into the path integral. Using the classical equations of motion, the measure factor becomes $\dot{\chi}=p_N$, the component of momentum normal to the surface $\chi=0$. Quantum mechanically, $\delta(\chi)$ and $p_N$ do not commute so we must choose an ordering prescription to define the quantum operator corresponding to our insertion. The average (or Weyl) ordering $\frac{1}{2}(\delta(\chi)p_N+p_N\delta(\chi))$ is a reasonable prescription (which we will later justify in the BRST formalism). This operator is precisely our Klein-Gordon form $\kappa_\mathrm{KG}$ \eqref{eq:KGIP}. We can see this most directly by noting that our insertion $\delta(\chi)\dot{\chi}$ can also be written as the time derivative of $ \Theta(\chi)$, and comparing  to our expression \eqref{eq:kappaSigma}. This connection between the Klein-Gordon form and gauge fixing was noted in \cite{Woodard:1989ac} and  discussed more recently in \cite{Witten:2022xxp}. In string theory, essentially the same calculation makes sense of sphere/disk amplitudes with two closed/open string vertex operators respectively \cite{Erbin:2019uiz} (see appendix \ref{app:string} for more on the analogy with the string worldsheet).

The path integral interpretations of group averaging and Klein-Gordon inner products give us a nice interpretation of our compatibility condition $\eta\kappa\eta=\eta$ for inner product on invariants and co-invariants, shown in figure \ref{fig:nkn}. The right hand side $\eta$ represents a path integral over all spacetimes computing a group average inner product. As usual in the path integral formalism, Hilbert spaces arise when we cut a path integral, dividing it into a `ket' piece computing one wavefunction at a given time and a `bra' piece computing the other wavefunction, while the inner product sews these pieces back together. This gives us an interpretation of $\eta\kappa\eta$ on the left hand side, where each piece of the path integral gives a factor of $\eta$ thought of as preparing an invariant wavefunction, and $\kappa$ represents the sewing procedure.\footnote{We have given an interpretation in terms of invariant states created by group averaging some initial and final co-invariant wavefunctions. An alternative interpretation of the cutting in terms of co-invariants is that $\kappa$ is a completeness relation, inverse to the matrix of inner products $\eta$ on co-invariants.} This ability to cut and sew back together embodies a notion of locality, and positivity of the inner product embodies unitarity (allowing a sensible probabilistic interpretation), and these are the reasons why Hilbert spaces are important. For QFT in a fixed background we can simply cut along any Cauchy surface $\Sigma$ that we choose, and sewing back together simply requires identifying fields on $\Sigma$ and integrating over them: this is a simple `$L^2$ inner product' of fields on $\Sigma$. But naively following this in gravity leads to a huge overcounting (responsible for the divergent nature of $\eta^2$), because each history will be counted once for every possible cutting surface. We have already seen the resolution: we must choose a gauge condition to select a unique slice $\Sigma$ for every spacetime appearing in the path integral, leading to the insertion of $\kappa$ as our gluing inner product.


However, there is one significant problem with using the Klein-Gordon form for such a gauge fixing  (already noted in \cite{Woodard:1989ac}), which is responsible for a disagreement between Klein-Gordon and group averaging inner products (so $\eta_\mathrm{GA}\kappa_\mathrm{KG}\eta_\mathrm{GA} \neq \eta_\mathrm{GA}$). The problem is that the Faddeev-Popov determinant $\dot{\chi}$ does not always give the correct measure factor: it is correct only up to a sign, and really we should take the absolute value. The upshot is that we expect compatibility between group averaging and Klein-Gordon inner products in cases where $\dot{\chi}>0$ when crossing the gauge surface $\chi=0$, and discrepancy by a sign when  $\dot{\chi}<0$ there.  We could also have situations where $\chi=0$ is not a complete gauge fixing, with a particular spacetime configuration crossing the gauge surface several times (or not at all). In such cases, $\kappa_\mathrm{KG}$ will count the number of crossings with signs (necessary for invariance under deformations of $\chi$, which will create or destroy $\chi=0$ crossings in cancelling pairs). Semi-classically, we can simply deal with this deficiency by hand: no problems will arise in perturbation theory around a classical solution for which $\chi=0$ is a good gauge, and this is sufficient for most applications. But we do not expect that inserting the Jacobian with an absolute value $|\dot{\chi}|$ by hand (as suggested in \cite{Marolf:1996gb}, for example) will make sense in an exact quantum theory (since in the UV, we expect the typical path to become very jagged like Brownian motion, crossing the $\chi=0$ slice infinitely many times).

Once we have this gauge-fixing interpretation there is nothing particularly special about Klein-Gordon inner products: we can choose $\chi$ to be any real function of $q$ and $p$ (or any Hermitian operator). An example is a zero-momentum gauge $\chi=p_0$ introduced in section \ref{sssec:maxvol} in analogy to  the maximal volume gauge $\chi=\Tr K$, for which \cite{Witten:2022xxp} gives an extensive discussion of the gauge fixing and Faddeev-Popov determinant.  Our choice will be guided by our desire that $\chi=0$  gives a good gauge, preferably with all relevant trajectories in the path integral crossing the gauge slice in the correct direction from $\chi<0$ to $\chi>0$ so no wrong sign appears in the Faddeev-Popov measure. But we expect that no choice will satisfy this universally (gravity has an insoluble Gribov problem); the  best we can do is make a case-by-case choice which works semi-classically for a particular problem.


\subsection{Example: a one-dimensional target space}\label{ssec:1Dtarget}

To make the above constructions more explicit and compare the resulting inner products, we can look at the simple example of one-dimensional target spaces. In such cases, our sigma-model Hamiltonian \eqref{eq:sigmaH} can be written as
\begin{equation}\label{eq:H1D}
    H= -\tfrac{1}{2}p^2+V(q).
\end{equation}
This is the familiar non-relativistic particle in a potential, apart from the `wrong sign' kinetic term (from $q$ being a `timelike' coordinate with a one-dimensional Lorentzian metric $G_{00}=-1$). Take care to note that because of this, momenta and velocities are related with a sign, $\dot{q}=-p$. 

Note that examples with higher-dimensional target space can also reduce to this form if the `timelike' and `spacelike' dependence factorise.
A familiar example takes $\target$ to be Minkowski spacetime with constant potential $\frac{1}{2}m^2$; this is the worldline theory of free massive QFT. Working in a basis of fixed spatial momentum $k$  reduces this to a family of decoupled effectively one-dimensional Hamiltonians \eqref{eq:H1D} with constant $V(q) = \frac{1}{2}\omega^2$ with $\omega^2 =|k|^2+m^2$.

For connections to gravity, a possible interpretation to have in mind (borne out in mini-superspace examples) is to interpret $e^q$ as a volume of space, so movement to the left/right corresponds to contraction/expansion of the universe.

First consider the simple case of a constant potential $V(q) = \frac{1}{2}\omega^2$ with $\omega>0$ (which includes the Minkowski example as described above). Translation-invariance of $H$ makes the states very simple to describe in momentum space. Co-invariant states are given by (smooth, say) momentum-space wavefunctions $\tilde{\psi}(p)$ modulo functions of the form $(p^2-\omega^2)f(p)$, which are simply the wavefunctions that vanish at $p=\mp \omega$. So the coset of $\tilde{\psi}(p)$ is determined by its values $\tilde{\psi}(\mp \omega)$ at those points, and $\hco$ is two-dimensional. Invariant states are linear combinations of $\tilde{\psi}_\pm(p) \propto \delta(p\pm \omega)$, or of $\psi_\pm(q) \propto e^{\mp i \omega q}$ in position space: positive- and negative-frequency solutions to the Wheeler-DeWitt equation $H\psi=0$ (noting that positive frequency means negative $p$ and vice-versa, due to the negative kinetic term).

Using the rigging map $\eta=2\pi\delta(H)$ directly gives
\begin{equation}
\begin{gathered}
     \eta = 2\pi\delta\left(\tfrac{1}{2}(p^2-\omega^2)\right) = \frac{2\pi}{\omega}(\delta(p-\omega)+\delta(p+\omega)),\\
     \langle q|\eta |q'\rangle = \frac{2}{ \omega}\cos(\omega(q-q')).
\end{gathered}
\end{equation}
The position space result in the second line is a Fourier transform of the momentum space expression, or can be obtained by the (conditionally convergent) group average integral of the free non-relativistic propagator $\langle q'|e^{-i H t}|q\rangle$ over $t\in\RR$. With this inner product, an obvious orthonormal basis $\{|+\rc,|-\rc\}$ of co-invariants is given by the `positive frequency' coset $|+\rc$ of momentum-space wavefunctions with $\psi(p=-\omega)=\sqrt{\frac{\omega}{2\pi}}$ and $\psi(p=+\omega)=0$, and the `negative frequency' coset $|-\rc$ with $\psi(p=-\omega)=0$ and $\psi(p=+\omega)=\sqrt{\frac{\omega}{2\pi}}$. Under $\eta$, these cosets map to invariants $|\pm \ri = 2\pi\delta(H)|\pm \rc$ with (position space) wavefunctions $\psi_\pm(q)=\frac{1}{\sqrt{\omega}} e^{\mp i \omega q}$, which (by definition) is an orthonormal basis of the dual inner product on invariants. This agrees with the spectral analysis inner product as expected: $\psi_\pm(q)$ are the $E= 0$ value of $\psi_{E,\pm}(q) = \frac{1}{\sqrt{ |p|}} e^{i p q}$, where $p = \mp \sqrt{\omega^2-2E}$. These are eigenstates of $H$ with eigenvalues $E$ (for $E<\frac{1}{2}\omega^2$), normalised by $\langle\psi_{E',\epsilon'}|\psi_{E,\epsilon}\rangle=2\pi\delta(E-E')\delta_{\epsilon\epsilon'}$ (with $\epsilon,\epsilon'=\pm$ specifying the sign of $p$). Note that this differs from the usual momentum basis by $\sqrt{|p|}$ in the denominator, because we normalise by delta-functions in $E$ (not $p$).

The Klein-Gordon inner product on invariants for one-dimensional target spaces is 
\begin{equation}
    \li \psi_2|\psi_1\ri_\mathrm{KG} =-\frac{i}{2} (\psi^*_2(q)\psi_1'(q)-\psi'^*_2(q)\psi_1(q)),
\end{equation}
noting that the negative kinetic term means that the normal derivative defining $\Pi$ in \eqref{eq:Omega} is $\partial_N = - \partial_q$. This is the Wronskian of a pair of solutions $\psi_1$, $\psi^*_2$ to $H\psi=0$, so is independent of $q$ (for any potential $V$). For our simple constant $V$ example,  we find $\li \pm|\pm \ri = \pm 1$ (and $\li \pm|\mp \ri=0$), so negative frequency solutions have negative Klein-Gordon norm. The corresponding form on co-invariants is
\begin{equation}
\begin{gathered}
     \eta_\mathrm{KG} = 2\pi\frac{\delta(p+\omega)-\delta(p-\omega)}{\omega} = -2\pi\sgn(p)\delta(H),\\
     \langle q|\eta_\mathrm{KG} |q'\rangle = \frac{2\sin(\omega(q-q'))}{ i\omega}.
\end{gathered}
\end{equation}
We have written $\eta_\mathrm{KG}$  as $-\sgn(p)$ times the group averaging rigging map to emphasise that they differ only by a sign: this is precisely the sign in the measure we commented on in the previous subsection.
The position space result may be familiar (up to field normalisation) as the commutator $[\Phi(q),\Phi(q')]$ of a harmonic oscillator `position' variable $\Phi$ at different values of target space `time' $q$ (since this is the target space one-dimensional `QFT' corresponding to $H=\frac{1}{2}(\omega^2-p^2)$, see sections \ref{ssec:worldlineQFT} and \ref{ssec:QFTIP}).

We can generalise this analysis to a potential which is constant only asymptotically in the distant past and future, $V(q) \to \frac{1}{2}\omega^2$ for $|q|\to\infty$ \cite{Hartle:1997dc}. We will summarise the upshot in terms of invariants. There is still a two-dimensional space of invariants, but now there are two different natural bases of solutions to $H|\psi\ri=0$. These are in states $|\mathrm{in},\pm\ri$ defined as a solution coming in from $ q\to \pm\infty$ in the far past on the worldline, and out states $|\mathrm{out},\pm\ri$ similarly going to $q\to\pm\infty$ in the future: see section \ref{ssec:naturalstates}. The asymptotics of the in state wavefunctions are as follows:
\begin{equation}\label{eq:psiIn}
    \begin{aligned}
        \psi_{\mathrm{in},+}(q) &\sim \begin{cases}
            \frac{1}{\sqrt{\omega}}\left(e^{i \omega q} + S_{++}e^{-i \omega q}\right) & q\to +\infty \\
            \frac{1}{\sqrt{\omega}}S_{-+}e^{ i \omega q} & q\to -\infty
        \end{cases}\\
        \psi_{\mathrm{in},-}(x) &\sim \begin{cases}
            \frac{1}{\sqrt{\omega}}S_{+-}e^{-i \omega q} & q\to +\infty \\
            \frac{1}{\sqrt{\omega}}\left(e^{-i \omega q} + S_{--}e^{i \omega q}\right)  & q\to -\infty
        \end{cases}
    \end{aligned}
\end{equation}
The out-states are given by complex conjugation (time-reversal) of out-states, $\psi_{\mathrm{out},\pm} = \psi_{\mathrm{in},\pm}^*$. Either of these bases are orthonormal for the group average (spectral analysis) inner product, and the two bases are related by the $2\times 2$ unitary S-matrix $S_{\pm\pm}$ of our 1D quantum mechanics (which is also symmetric, $S_{+-}=S_{-+}$, due to time-reversal symmetry):
\begin{equation}
    \li \mathrm{out},\epsilon'|\mathrm{out},\epsilon\ri_\mathrm{GA} = \delta_{\epsilon\epsilon'}, \quad \li\mathrm{in},\epsilon'|\mathrm{in},\epsilon\ri_\mathrm{GA} = \delta_{\epsilon\epsilon'}, \quad \li\mathrm{out},\epsilon'|\mathrm{in},\epsilon\ri_\mathrm{GA} = S_{\epsilon'\epsilon}.
\end{equation}
The Klein-Gordon inner product is simplest to express in a mixture of the two bases, as overlaps between in-kets and out-bras:
\begin{equation}\label{eq:KGoutin}
\li\mathrm{out},\pm|\mathrm{in},\pm\ri_\mathrm{KG} = 0, \qquad \li\mathrm{out},\pm|\mathrm{in},\mp\ri_\mathrm{KG} = \pm \li\mathrm{out},\pm|\mathrm{in},\mp\ri_\mathrm{GA}.
\end{equation}
We can interpret this as saying that the Klein-Gordon form gives a vanishing contribution for trajectories which `reflect', coming from $q=+\infty$ in the distant past and leaving to $q\to +\infty$ in the future or the same for $q\to -\infty$. Trajectories which start at $q\to -\infty$ and `transmit' to $q\to+\infty$ have positive contribution (equal to the result from group averaging), and those in the opposite direction contribute negatively. One way to express this is a relation (equivalent to that given in \cite{Hartle:1997dc})
\begin{equation}\label{eq:KGSAsigns}
    \li \psi'|\psi\ri_\mathrm{KG}  =\li \psi'|\tfrac{1}{2}(\epsilon_\mathrm{in} + \epsilon_\mathrm{out})|\psi\ri_\mathrm{SA},
\end{equation}
where $\epsilon_{\mathrm{in,out}}$ are operators which give $\sgn(p)$ in the asymptotic future or past (with in and out states as eigenstates respectively). These don't commute (because a incoming wave from one direction scatters into a superposition of both out states).

These results could have been anticipated from the gauge-fixing interpretation given for the Klein-Gordon form in section \ref{ssec:IPPIs}. The two inner products agree when a constant $q$ gauge-fixing surface is crossed once in a positive direction, they disagree with a sign when it is crossed once in a negative direction so the Jacobian is negative, and the Klein-Gordon form gives zero when a gauge-fixing surface is not crossed at all (or crossed an equal number of times in each direction). This interpretation is clear only when we compute overlaps between in- and out-states as above, since the contributing trajectories all begin and end in a definite direction of target space: the Klein-Gordon overlap between two in-states, for example, is more murky (until we re-express one in a basis of out states).

 Another interesting case is when $V=\frac{1}{2}\omega^2$ for $q\to-\infty$ as before, but $V<0$ for $q\to\infty$. This might correspond to `bang-crunch' cosmologies that expand and recollapse. In this case, the allowed space of invariant states (defined by the image of $\eta$ starting from  smooth compactly supported wavefunctions, or by requiring that the form $\kappa$ is well-defined) is quite different depending on the choice of $\eta$ or $\kappa$. Group averaging (or spectral analysis) now gives a single state, corresponding to the solution of the Wheeler-DeWitt equation that decays at $q\to\infty$. This leads to physics in accord with semi-classical expectations \cite{Marolf:1994wh}, exponentially suppressing classically-forbidden configurations with large $q$. However, in the context of the Klein-Gordon norm it is natural to allow any initial data and hence any solution to $H|\psi\ri=0$ (the usual requirement is compactly supported initial data, which is vacuous in 1D), leading to a two-dimensional space of states. Alternatively, it has been suggested to keep only the solution which is purely positive frequency at $q\to-\infty$ \cite{Wald:1993kj,Higuchi:1994vc}. In either case, we include solutions which grow at $q\to\infty$ (in the associated target-space QFT described in the next section, this would corresponds to uncontrolled tachyonic particle production).\footnote{The target space interpretation (where $q$ is interpreted as time in target space) is a`free QFT' in one spacetime dimension, with a scalar field $\Phi$ and time-dependent Hamiltonian
\begin{equation}
    H_\mathrm{target}(q) = \frac{1}{2}\Pi(q)^2 + V(q) \Phi(q)^2,
\end{equation}
where $\Pi$ is the momentum conjugate to $\Phi$. This is a Harmonic oscillator with time-dependent frequency $V = \frac{1}{2} \omega^2$.} This would appear (depending on the interpretation of the wavefunction) to favour large $q$, where there are no classical solutions, which seems challenging for a sensible semi-classical limit. We view this as further support for group averaging.

\subsection{The QFT inner product}\label{ssec:QFTIP}

The motivation of DeWitt \cite{DeWitt:1967yk} and \cite{Wald:1993kj} for proposing the Klein-Gordon inner product was by analogy to quantum field theory. Specifically, as explained in section \ref{ssec:sigma}, the Hamiltonian we are considering arises as the equation of motion of a free field $\Phi(q)$ (which we take to be real here) on the manifold $\target$, with Lagrangian $\frac{1}{2}(\nabla\Phi)^2 + V \Phi^2$. Here, we explain the connection between this QFT and a quantum mechanical `worldline theory' with the constraint $H=0$ in our language of invariants and co-invariants. We will not use this in the remainder of the paper, so the reader can safely skip this part.

This connection depends on first choosing a `vacuum state' $|\Omega\rangle_{QFT}$ of the QFT Hilbert space. In free QFT, it is natural to take $|\Omega\rangle_{QFT}$ to be a Gaussian state, meaning that the one-point function of $\Phi$ vanishes and the two-point (Wightman) function
\begin{equation}
    W_\Omega(q_2,q_1) := \langle \Omega|\Phi(q_2)\Phi(q_1)|\Omega\rangle_{QFT}
\end{equation}
determines all higher-point functions by a sum over Wick contractions. We can then define a `single particle' Hilbert space $\hilb_{SP}\subset \hilb_{QFT}$ by acting on $|\Omega\rangle_{QFT}$ with an operator linear in the field $\Phi$, 
\begin{equation}\label{eq:fSP}
    |f\rangle_{QFT} := \int dq f(q) \Phi(q) |\Omega\rangle_{QFT} \in \hilb_{SP}.
\end{equation} The $n$ particle Hilbert space created by acting with a product of $n$ such operators (minus Wick contractions) is then the symmetric product $\operatorname{Sym}^n \hilb_{SP}$, and the full QFT Hilbert space is the Fock space given by the direct sum of all these sectors, $\hilb_{QFT} \simeq \bigoplus_{n=0}^\infty \operatorname{Sym}^n \hilb_{SP}$. In this construction there is a great deal of freedom in the choice of vacuum state $|\Omega\rangle$, and hence in the subspace $\hilb_{SP}$ of single particle states. Generically (in the absence of target space time-translation symmetry for example) there is no preferred vacuum, due to the familiar phenomenon of particle production. Different vacua are related by Bogoliubov transformations, which are the QFT interpretaton of the S-matrix of the worldline theory (for example, the coefficients in \eqref{eq:psiIn} above).

Once the choice of $|\Omega\rangle_{QFT}$ has been made, we can relate the single particle Hilbert space $\hilb_{SP}$ to the constrained state spaces $\Vco$, $\Vinv$. First, note that the single-particle states constructed in \eqref{eq:fSP} are naturally associated with co-invariants $|f\rc$, since the field obeys the equation of motion $H \Phi(q)=0$. Taking $f=Hg \in \Im H$, integrating by parts and using the equation of motion tells us that $\int f \Phi = \int g H\Phi=0$, so all states $|f\rc$ in the same co-invariant coset lead to the same QFT state. Second, note that we can identify a single particle state $|\Psi\rangle_{QFT}$ with an invariant state $|\psi\ri$ by taking the wavefunctions to be matrix elements of the field operator, $\psi(q) = \langle\Omega|\Phi(q)|\Psi\rangle$. This gives a nonzero wavefunction for any single particle state of $\hilb_{QFT}$, and zero for other multi-particle sectors. We can summarise this by the following linear maps:
\begin{gather}
    \Vco \to \hilb_{SP}:|f\rc \mapsto |f\rangle_{QFT} := \int_\target dq f(q) \Phi(q) |\Omega\rangle_{QFT}  \qquad (\text{surjective}),\\
     \hilb_{SP}\to\Vinv:|\Psi\rangle_{QFT} \mapsto |\psi\ri ,\quad \psi(q) = \langle \Omega |\Phi(q)|\Psi\rangle_{QFT} \qquad (\text{injective}).
\end{gather}
These maps are `dual', meaning that they relate the canonical pairing between $\Vco$ and $\Vinv$ and the QFT inner product:
\begin{equation}
    \langle f|\Psi\rangle_{QFT} = \int dq f^*(q) \langle\Omega|\Phi(q)|\Psi\rangle = \lc f|\psi\ri.
\end{equation}

From this, the QFT inner product on $\hilb_{SP}$ induces an inner product on $\Vco$ by pullback under the above linear map,
\begin{equation}
\begin{gathered}
    \lc f_2|f_1\rc_\Omega =\langle f_2|f_1\rangle_\mathrm{QFT} =  \int dq_1dq_2 f_2^*(q_2)f_1(q_1) W_\Omega(q_2,q_1) \\
    \implies \eta_\Omega = W_\Omega.
\end{gathered}
\end{equation}
This identifies the map $\eta_\Omega$ from co-invariants to invariants as the Wightman function $W_\Omega$. This inner product is positive semi-definite, but unlike the examples of $\eta$ discussed above it is degenerate. The means that the image of $\eta_\Omega$ (the space of invariant wavefunctions $|\psi\ri$ that arise from a state $|\Psi\rangle_{QFT}$ in the QFT) does not cover all invariants. Instead, it will cover only half of  $\Vinv$. When $|\Omega\rangle_{QFT}$ is the vacuum state in a static target space these are the positive frequency wavefunctions. The resulting co-invariant Hilbert space contains additional null states besides $\Im H$.

For the dual Hilbert space of invariants, for the inner product we can in fact use the Klein-Gordon form \eqref{eq:KGIP}. Instead of modifying the inner product, we simply restrict the space of allowed states to a `positive frequency' subspace $V_+\subset \Vinv$ on which $\kappa_{KG}$ is positive-definite. This selects `half the states' since we must have $\Vinv = V_+ \oplus V_-$ where the negative frequency space $V_-=V_+^*$ consists of complex conjugates of positive frequency solutions. The choice of $V_+$ is an alternative characterisation of the state $|\Omega\rangle_{QFT}$.

\section{Operators on the physical Hilbert space}\label{sec:ops}

In quantum mechanics, Hermitian operators play two (closely related) roles: they are observables, and they generate unitary flows. The most prominent example of the latter is the Hamiltonian, which generates time translation. Gravity is particularly unusual in this respect since time translation is locally a gauge symmetry (though we can still consider physical time translation operators in the presence of timelike boundaries such as the conformal boundary in asymptotically AdS spacetimes). Nonetheless, in any gauge theory  we must consider the constraints when we discuss operators and their flows on the physical Hilbert space. There are two complementary considerations, both classically and quantumly. First, we are constrained because we want the flow generated by a physical operator to map physical states (satisfying constraints) to physical states. On the other hand, we have additional freedom because we can always add a gauge transformation to a flow, leading to a different but physically equivalent point on a gauge orbit.

Commonly, the criterion given for physical operators is that they are gauge-invariant, commuting (or Poisson commuting in the  classical theory) with the constraints. This is convenient (and perhaps sufficient) for observables, and it certainly suffices to satisfy the above constraint of mapping physical states to physical states (whether using the invariant or co-invariant Hilbert space). However, it is unnecessarily and overly restrictive, particularly for operators generating flows like Hamiltonians, since it does not allow us the freedom of a gauge choice. A simple example is $U(1)$ gauge theory, where the usual Hamiltonian contains terms of the form $E^i \partial_i A_0 $ which do not commute with the primary constraint $\Pi^0 $ (the momentum conjugate to $A_0$), along with possible additional gauge-fixing terms: see appendix \ref{app:Maxwell} for details of this example. This is even more true in gravity, where the non-abelian algebra of Hamiltonian constraints makes it impossible to write an operator which generates time-translations and also commutes with constraints.

Given this motivation, in this section we review the more general conditions on physical operators, and comment on the action of the resulting physical operators on the Hilbert spaces of invariants and co-invariants discussed above. We do this first classically in the language of Dirac, and then explain how this relates to the BRST formalism (which provides a powerful and convenient way to ensure gauge invariance in the quantum theory). Of particular importance for us will be the trivial operators which act non-trivially on the unconstrained Hilbert space $\hilb_0$ but are physically equivalent to the zero operator. These can be interpreted as infinitesimal quantum gauge transformations. In the BRST formalism  we can exponentiate these trivial (BRST-exact) operators to finite gauge transformations, which we will use in the next section to construct inner products in a more systematic manner.

\subsection{Dirac observables}\label{ssec:GOops}

We first review observables  in theories with constraints in Dirac's formalism \cite{Dirac:1950pj}, starting with the classical theory in which observables are functions of phase space.

First, we have a constraint for a function $O(q,p)$ of phase space to be a physical observable: it should be constant on gauge orbits of physical states. This means that Poisson brackets of $O$ with the constraints (i.e., the gauge variation of $O$) should vanish \emph{weakly}, meaning they are zero on the constraint surface. In our case with a single constraint $H$, this means that $\{H,O\} = H \tilde{O}$ for some function $\tilde{O}$ (so the time derivative of $O$ vanishes if the energy is zero). This also means that the flow generated by $O$ preserves the constraint surface, and that gauge-equivalence is retained by the flow (i.e., two points starting on the same gauge orbit end on the same gauge orbit).

In addition to this constraint, we have some freedom. Two different functions of phase space correspond to the same physical observable if they are weakly equal (equal on the constraint surface), meaning we have the equivalence $O\sim O+H \xi$ for any $\xi$.  Note that if $O$ is a good observable then so is $ O+H \xi$, since $\{H,H\xi\} = H\{H,\xi\}$ is weakly zero.\footnote{More generally, this is true for any set of first class constraints.} 

Quantum observables are similar, though we need to take operator ordering into account. Depending on whether we take our observables to act on invariant  or co-invariant states, different choices of ordering are needed.  The conditions for an operator $O$ on $\hilb$ to give rise to a well-defined operator on the constrained Hilbert space, and the equivalences between such $O$, are the following:
\begin{equation}\label{eq:Ocoinv}
	\begin{aligned}
   O \text{ operator on } \hinv:&\qquad [H,O] = i \tilde{O}H, \qquad O\sim O+\xi H, \\
    O \text{ operator on }  \hco:&\qquad [H,O] = i H \tilde{O} , \qquad O\sim O+ H \xi .
\end{aligned}
\end{equation}
The condition in the first line ensures that $HO|\psi\rangle=0$  whenever $H|\psi\rangle=0$, and the equivalence arises because $\xi H$ annihilates all invariants. Similarly, the condition in the second line is required so that $O$ maps null states to null states, and the equivalence comes from identifying states in the same co-invariant coset. Note that these observables are in one-to-one correspondence: if $O_\mathrm{inv}$ gives a well-defined operator on invariants (with  $[H,O_\mathrm{inv}] = i\tilde{O} H$ as above), then $O_\mathrm{co} = O_\mathrm{inv}+i\tilde{O}$ is well-defined on co-invariants.

If we have several constraints, we can continue to define observables as equivalence classes of operators which weakly commute with constraints modulo weakly zero operators. For the quantum theory the ordering in these definitions is appropriate to invariance or co-invariance as in \eqref{eq:Ocoinv}, but may use different prescriptions for different constraints. For example, in the non-minimal formalism introduced at the end of section \ref{ssec:BRSTstates}, invariant states with respect to $H$ are treated as co-invariant with respect to $\Pi$ and vice-versa.

In such cases with several constraints and states with a mixture of invariance and co-invariance, this definition of observables has a significant drawback: it does not give us a well-defined algebra. For example, the product of $O$ with a weakly zero operator may not be weakly zero, even if $O$ weakly commutes with constraints. Taking the specific example of states invariant with respect to $H$ and co-invariant with respect to $\Pi$, both $H$ and $\Pi A$ (for any matter operator $A$) are weakly zero, but their product $H\Pi A $ may not be if $[H,A]$ is non-zero. In particular, we cannot guarantee the unitary flow constructed by exponentiating these operators will be gauge-invariant. This is resolved in the BRST formalism.

Before moving onto BRST observables, we make one comment on the narrower class of observables which exactly commute with constraints, for which none of these complications from operator ordering arise.\footnote{One may have to take more care when there are field-dependent structure constants as occurs in gravity, since then an operator might commute with the constraints but not with the BRST charge.} One class of  examples is constructed by averaging, such as $O = \int dt e^{+i H t}O_0 e^{-iH t}$ for any $O_0$ (such that the integral converges), which in the path integral might be written as the integral of a local quantity over all of spacetime \cite{DeWitt:1962cg,Giddings:2005id}. Such operators can act both on invariants and co-invariants. It is reasonable to demand that these two actions should be physically equivalent, meaning that $O$ commutes with the rigging map $\eta$. Imposing this requirement essentially uniquely selects $\eta$ to be the group average \cite{Giulini:1998kf}, lending further support to this proposal.

\subsection{Observables in BRST}
 
 The description of physical observables in section \ref{ssec:GOops} is reminiscent of cohomology: a requirement to weakly commute with constraints is reminiscent of closedness, while the equivalence up to weakly zero operators is reminiscent of exactness. This is realised precisely in the BRST formalism, where the physical operators are given by the operator cohomology (at ghost number zero).
 
Classically, consider functions $\mathbf{O}$ on the BRST extended phase space (depending on Grassmann ghost variables $b,c$ as well as $q,p$) with $N_g=0$. The ghost-free part of $\mathbf{O}$ (obtained by setting ghosts to zero) is an ordinary function $O$ of bosonic variables $q,p$ which will correspond to the `Dirac' observable discussed above. The Dirac condition that Poisson brackets of $O$ with the constraints are weakly zero is replaced by the condition that the Poisson bracket of $\mathbf{O}$ with the BRST charge is \emph{exactly} zero, $\{Q,\mathbf{O}\}=0$ (or $\mathbf{O}$ is BRST-closed). The equivalence between weakly equal operators is replaced with triviality of BRST-exact operators, defined as the Poisson bracket $\{Q,\Psi\}$ for any function $\Psi$ with $N_g=-1$. Hence the physical operators are elements of the BRST cohomology of closed $\mathbf{O}$ modulo exact $\mathbf{O}$ at $N_g=0$.

To see how this works, let's explicitly work through our case of a single constraint with $Q=c H$. We can write the most general $N_g=0$ function as $\mathbf{O} = O - i b c \tilde{O}$ with $O$, $\tilde{O}$ independent of ghosts, and we have $\{Q,\mathbf{O}\} = c (\{H,O\} -\tilde{O} H)$.\footnote{The Poisson bracket on the extended phase space is given by \begin{equation}
    \{f,g\} = \frac{\partial f}{\partial q^i}\frac{\partial g}{\partial p_i}-\frac{\partial f}{\partial p_i}\frac{\partial g}{\partial q^i} - i \left(\frac{\partial^R f}{\partial b}\frac{\partial^L g}{\partial c}+\frac{\partial^R f}{\partial c}\frac{\partial^L g}{\partial b}\right),
\end{equation}
where the $L,R$ in the ghost terms indicate left and right derivatives. This gives the usual correspondence between quantum (anti-)commutators and Poisson brackets, $[\cdot,\cdot]_\pm \sim i\{\cdot,\cdot\}$.\label{foot:bcPoiss}} From this we see that $O$ weakly commutes with the constraint if and only if it has a BRST-closed extension $\mathbf{O}$. For the BRST-exact operators we consider the Poisson bracket of $Q$ with the most general $N_g=-1$ operator $\Psi = b\xi$, giving $\{Q,\Psi\} = -i(H\xi - i bc \{H,\xi\})$; this means we can identify $O=H\xi$ (with $\tilde{O}=\{H,\xi\}$) as a trivial observable.

For more complicated cases with many constraints (particularly for non-abelian algebras and field-dependent structure constants), the Dirac condition of weakly commuting with constraints is not quite enough to get a physical observable. The requirement that $O$ can be completed to a BRST exact $\mathbf{O}$ by the addition of ghost terms gives the Dirac condition, but also imposes further conditions at higher order in ghosts.

Quantum mechanically, we simply replace Poisson brackets in the above discussion with (anti-)commutators. The BRST formalism then automatically takes care of the different ordering prescriptions discussed above. A BRST observable $\mathbf{O} = O - i b c \tilde{O}$ is closed ($[Q,\mathbf{O}]=0$) if $[H,O] = i H \tilde{O}$ as appropriate for co-invariants in \eqref{eq:Ocoinv}; and indeed co-invariants are accompanied by ghost states annihilated by $c$, so $\mathbf{O}$ acts on these states simply as the ghost-free part $O$. Similarly, a BRST-exact operator $[Q,b\xi]_+ = H\xi  +bc [\xi, H]$ generates the equivalences appropriate for co-invariants. On the other hand, on invariant states (with ghosts annihilated by $b$) the same operator $\mathbf{O}$ will instead act as $O_\mathrm{inv} = O-i\tilde{O}$, giving the correspondence between operators acting on invariants and co-invariants discussed after \eqref{eq:Ocoinv}. The BRST observable $\mathbf{O}$ packages both possibilities (and all possible mixtures if there are more constraints) into a single object.

Furthermore, in this formalism the physical observables in fact form an algebra: the product of two BRST-closed operators is closed, and the product of a BRST-closed (physical) operator with a BRST exact (trivial) operator is BRST exact (trivial). This follows simply from the super-Jacobi identity: for example, $[Q,\mathbf{O}_1\mathbf{O}_2] = [Q,\mathbf{O}_1]\mathbf{O}_2+\mathbf{O}_1[Q,\mathbf{O}_2]=0$ if $\mathbf{O}_{1,2}$ are closed. An important consequence is that this allows us to discuss the unitary flow generated by a Hermitian operator $O$: the exponential $e^{is\mathbf{O}}$ is a physical (BRST closed) operator, with the extra ghost terms in the operator providing the Hamiltonian version of the Faddeev-Popov determinants familiar from the path integral. An important special case which we will use in the next section is a finite quantum gauge transformations, meaning the exponential of a BRST exact operator $\exp[Q,\Psi]_+$, which is BRST-equivalent to the identity.

%
%
%
%
%
%

\section{Inner products from the BFV-BRST formalism}\label{sec:BFVIP}

In section \ref{sec:IP} we discussed general properties of inner products on $\hinv$ and $\hco$, along with some examples. From that discussion alone there is a great deal of freedom in defining an inner product, and the choice seems rather arbitrary. In this section we describe a way to obtain inner products in a more systematic way from the BFV-BRST formalism \cite{Batalin:1977pb,Fradkin:1975cq}, which removes much of that freedom.

Useful references for general tools on the BFV-BRST formalism include \cite{Henneaux:1985kr} and the book \cite{Henneaux:1994lbw}. The main ideas for the application to inner group averaging follow \cite{Marnelius:1990eq,Batalin:1994rd,Shvedov:2001ai}.

\subsection{Sketch of the main ideas}

For this, we will use the non-minimal BFV version of the BRST formalism introduced around equation \eqref{eq:hilbBFV}, where we include the lapse $N$ as a dynamical degree of freedom compensated by the constraint $\Pi=0$, requiring that the momentum conjugate to $N$ vanishes. One way to motivate this choice is that now we can choose the physical states to reside at ghost number $N_g=0$, which is necessary to get a non-zero inner product (since $N_g$ is anti-Hermitian). But this doesn't work immediately; let's see what happens when we na\"ively try to take the inner product of two Wheeler-DeWitt (invariant) states in the BFV Hilbert space \eqref{eq:hilbBFV}:
\begin{equation}
    |\psi\rangle \otimes|\psi_N\rangle \otimes |\uparrow\downarrow\rangle\in \hilb_0\otimes\hilb_N\otimes\hilb_\mathrm{ghost} \text{ with }H|\psi\rangle=0 \longrightarrow \underbrace{\langle \psi'|\psi\rangle }_{=\infty} \langle\psi_N'|\psi_N\rangle \underbrace{\langle \uparrow\downarrow|\uparrow\downarrow\rangle}_{=0} \,.
\end{equation}\label{eq:IPindeterminate}
We find an indefinite form: the first factor $\langle \psi'|\psi\rangle$ from the unconstrained Hilbert space gives infinity since invariant wavefunctions are non-normalisable, as remarked below equation \eqref{eq:Vinv}. But this is multiplied by $\langle \downarrow\uparrow|\downarrow\uparrow\rangle=0$ from the ghost sector.\footnote{The lapse sector inner product $\langle\psi_N'|\psi_N\rangle$ could be finite, zero, or infinity depending on what representatives of the BRST cohomology class we choose, but in any case the result is indefinite.} The idea is to resolve this degeneracy by inserting a finite `gauge transformation'  $e^{[Q,\Psi]_+}$, the exponential of a BRST exact (and hence physically trivial) operator, in such a way that we get a finite result.  We then hope to interpret this finite result as a physical inner product between invariants. Similar remarks apply to co-invariant states (with the infinity coming from the norm of the lapse wavefunction $|\Pi=0\rangle$ annihilated by $\Pi$). We will see precisely how this works in examples below.

This construction has several important properties associated with BRST-invariance. The first is that it depends only on the BRST cohomology class of the states, since acting on a BRST exact state with $e^{[Q,\Psi]_+}$  gives another BRST exact state. This means that the result for invariant states should be independent of the lapse-dependence of wavefunctions (except for the integral over $N$), and for co-invariant states will depend only on the co-invariant coset as desired. The second property is that infinitesimal changes in $\Psi$ do not affect the matrix elements of $e^{[Q,\Psi]_+}$ between BRST-closed states (the Fradkin-Vilkovisky theorem \cite{Fradkin:1975cq}). At least na\"ively, this means that the result should be independent of the choice of $\Psi$. This means that there is not in fact much freedom in choosing an inner product; there is an essentially unique choice.\footnote{\label{foot:rescaling}This is true once we have chosen a BRST charge $Q$. For our simple one-dimensional gravity model this still leaves some freedom, for example by rescaling the Hamiltonian $H$ by a function of fields to get a new classically-equivalent constraint $\tilde{H}$, and defining a different BRST operator $\tilde{Q}=c\tilde{H} + \cdots$: a specific example is discussed in section \ref{sssec:GA}. We regard this choice of $Q$ as part of the ambiguity inherent in quantising a classical theory. Nonetheless, in richer models with non-abelian constraint algebras and spatial dependence we expect the requirement of a local nilpotent Hermitian BRST charge to be much more restrictive, so there will be less remaining ambiguity (in perturbation theory, hopefully only the usual ambiguity of renormalisation of parameters, including generating higher-derivative terms).}

However, we will also show below how to obtain both group-average and Klein-Gordon inner products in this way, and these are different! More precisely (since they are inner products on different Hilbert spaces $\hco$ and $\hinv$), they are not compatible by the $\eta\kappa\eta=\eta$ generalised inverse criterion. The resolution is a loophole to the Fradkin-Vilkovisky theorem that allows for discrete jumps as we vary $\Psi$,  since the theorem does not tell us about finite changes in $\Psi$ if the amplitude passes through an indeterminate form (most notably the original $\Psi=0$ in  \eqref{eq:IPindeterminate}).

There is a more physical perspective on these algebraic manipulations related to gauge fixing. The original infinity we encountered in the inner product of invariants can be thought of as the infinite volume of the gauge group, as remarked in section \ref{ssec:IPPIs}. To get a finite result we need to fix this overcounting of equivalent configurations, so that we integrate over every gauge-orbit exactly once. In the construction we are discussing here, the bosonic part of the generator $[Q,\Psi]_+$ typically has precisely this effect, with a gauge condition appearing after integrating out $\Pi$ and/or $N$ (though with this BRST approach we never need to deal directly with the resulting second-class constraints, Dirac brackets and so forth). For this reason, $\Psi$ is usually called the \emph{gauge-fixing fermion}. The ghost terms in  $[Q,\Psi]_+$ have the job of fixing an invariant measure on the resulting gauge-fixed slice, by contributing the Faddeev-Popov determinant $\Delta_\mathrm{FP}$.

However, as we already saw in section \ref{ssec:IPPIs} for the example of the Klein-Gordon form, $\Delta_\mathrm{FP}$ is not quite the correct measure, since it is not automatically positive and we should take its absolute value. Furthermore, we cannot simply fix this by choosing a sign in the final result, since  $\Delta_\mathrm{FP}$ can vanish and its sign can change as we move between different gauge orbits or change the gauge condition. This indicates the our choice of $\Psi$ corresponds to a gauge condition which is not globally valid: a Gribov problem \cite{Gribov:1977wm}. A given gauge-orbit may not have any representative satisfying the gauge condition, or it may have several (perhaps with different signs for $\Delta_\mathrm{FP}$). These phenomena give an intuitive explanation for the possibility of jumps as we vary the gauge-fixing fermion $\Psi$. It also suggests a criterion for selecting a unique physical inner product: it should arise from some choice of $\Psi$, give a positive measure to every gauge-orbit (leading to a positive semi-definite inner product), and count each orbit exactly once. These criteria are satisfied by the group-averaging proposal for our one-dimensional gravity models, which is perhaps the strongest evidence that this is the physically correct inner product. It is plausible that this remains true in more complicated theories, though there are significant technical obstacles to making this completely precise in realistic theories of gravity, some of which are discussed in section \ref{sec:disc}.

\subsection{The non-minimal formalism and gauge-fixing}\label{ssec:nonminGF}

Before introducing the full BRST machinery, we first give a brief explanation of the reasons for using the non-minimal BFV formalism, and the connection to gauge-fixing. This will be helpful for interpreting the results below.

We first compare the classical variational principles coming from first-order actions, using the example of the Hamiltonian $\mathbf{H}=N H$  generating proper time evolution as given in \eqref{eq:properHam} (obtained by a na\"ive canonical recipe from the action \eqref{eq:1Dlag}), with and without including a momentum $\Pi$ conjugate to $N$:
\begin{align}
    S_\mathrm{minimal}[q,p,N] &= \int dt \left[p \dot{q}-NH(q,p) \right], \\
     S_\mathrm{non-minimal}[q,p,N,\Pi] &= \int dt \left[p \dot{q} + \Pi \dot{N}-NH(q,p)  \right].\label{eq:Snonmin}
\end{align}
Varying the `minimal' action (in which $\Pi$ is absent, and $N$ acts only as a Lagrange multiplier to enforce $H=0$) gives equations of motion which do not have a unique solution: $N$ is completely undetermined. This is nothing but the gauge symmetry under diffeomorphisms of time. But for the non-minimal action, the additional equation of motion from varying $\Pi$ means that initial data for $q,p,N,\Pi$ determine a unique time evolution. The extra variable $\Pi$ acts as a Lagrange multiplier fixing a gauge condition, in this case the `synchronous' gauge $\dot{N}=0$.

This simple observation raises the question: can we modify the action to describe evolution in other gauges? The discussion in section \eqref{ssec:GOops} makes it clear that we can achieve  this by adding terms to the Hamiltonian which are weakly zero, i.e., which vanish when constraints are satisfied. Specifically, by adding a term $F(q,p,N)\Pi$ linear in $\Pi$ to get the full Hamiltonian $\mathbf{H}=N H +\Pi F $, we get a Hamiltonian that continues to generate time evolution but now in a much more general gauge (with $\dot{N} = F$). We can also add higher order terms in $\Pi$ (e.g., a multiple of $\Pi^2$): this will not change the classical dynamics when on the constraint surface $\Pi=H=0$, but  can be useful in the quantum theory  where this gives us the Gaussian-averaged gauges familiar from gauge theory.

Gravity is special in this context, because time evolution (except at spatial boundaries) is a gauge transformation: the Hamiltonian $\mathbf{H}=N H$ itself is weakly zero, so these flows simply move us between different points on the same gauge-orbit. In non-gravitational theories (or for gravity with asymptotic boundaries where time evolution is physical), the Hamiltonian will be a non-trivial Dirac observable as discussed in section \eqref{ssec:GOops}, i.e.~a function which weakly commutes with the constraints, ambiguous up to addition of a weakly zero function. The example of Maxwell theory is discussed in appendix \ref{app:Maxwell}.

In gravity (or any other theory) we can consider gauge transformations generated by any weakly zero function $G(q,p,N,\Pi)$, which need not include a term $NH$ allowing us to interpret the flow in terms of time translation. For example, for any $\chi(q,p)$ we can consider the flow generated simply by $G = \Pi\chi(q,p)$. On the constraint surface (with initial condition satisfying $\Pi=0$), the classical evolution leaves $q,p$ unchanged and relates the change in the lapse $N$ to $\chi(q,p)$: in particular, classical solutions with equal initial and final values of lapse $N$ exist only if $\chi=0$. So under these conditions on initial and final $N$ states, the transformation generated by $\Pi\chi(q,p)$ can be interpreted as imposing the gauge condition $\chi=0$ at a single moment of time. This motivates using such a flow to impose gauge fixing on invariant states as discussed in section \ref{ssec:IPPIs}.

When we pass to the quantum theory, guaranteeing gauge-invariance is more subtle due to issues of operator ordering (or the measure in the path integral approach). In section \ref{sec:ops}, this showed up by the failure of Dirac observables to form an algebra.  The BRST formalism resolves this: in place of a weakly-zero generator $G(q,p,N,\Pi)$ as in the examples discussed above we use a BRST exact operator $[Q,\Psi]_+$ for some gauge-fixing fermion $\Psi$. We get back the weakly zero function $G$ from the ghost-free part of $[Q,\Psi]_+$ (in the classical limit when ordering becomes unimportant).

\subsection{Non-minimal BRST}

With these preliminaries, we are now ready to study the quantum theory in the non-minimal BRST formalism. First, we briefly recall the key features already introduced earlier. For the bosonic part of the Hilbert space, we have wavefunctions depending on the original unconstrained variables $q$  as well as the real-valued lapse $N$. Associated with the two constraints $H$ and $\Pi$ are two sets of ghosts  $c$ and $\hat{c}$ and their conjugate momenta $b$ and $\hat{b}$, with non-trivial anti-commutation relations $[c,b]_+=[\hat{c},\hat{b}]_+=1$ represented on a four-dimensional ghost Hilbert space with basis $\{|\downarrow\downarrow\rangle,|\uparrow\downarrow\rangle=c|\downarrow\downarrow\rangle,|\downarrow\uparrow\rangle=\hat{c}|\downarrow\downarrow\rangle,|\uparrow\uparrow\rangle=c\hat{c}|\downarrow\downarrow\rangle\}$. The extended BRST operator and ghost number are
\begin{equation}\label{eq:QBFV}
    Q = c H + \hat{c} \Pi,\qquad    N_g = cb + \hat{c}\hat{b}-1.
\end{equation}
The latter says that $c,\hat{c}$ raise $N_g$ by $1$ while $b,\hat{b}$ lower it by one, and the constant is chosen to make $N_g$ anti-Hermitian. We take physical states to all lie at $N_g=0$.

For help comparing to literature, we note that often only $c,b$ are referred to as the ghosts and ghost momentum, while $\hat{b},\hat{c}$ respectively are the anti-ghosts and anti-ghost momentum. Additionally, an alternative convention is common in which (anti-)ghost momenta are defined as anti-Hermitian operators such as $\rho=i b$  (with anti-commutation relations $[c,\rho]_+=i$). This has the advantage that no factors of $i$ are required in  the classical theory (e.g., for Poisson brackets as in footnote \ref{foot:bcPoiss}), but we choose our current notation as more likely to be familiar to most readers.

We also need to specify Hermiticity relations for the operators, which determine an (indefinite signature) inner product on the extended BRST state space. For this we will make choices that may seem rather strange at first sight \cite{Arisue:1981qs,Shvedov:2001ai}. These  will be necessary to end up with sensible Hermitian physical inner products, and in fact matches with standard prescriptions in gauge theory, albeit in a guise that may be unfamiliar. We will take the lapse $N$ and its corresponding conjugate momentum $\Pi$ to be \emph{anti-Hermitian} operators.\footnote{An equivalent prescription is to redefine $N,\Pi$ by a factor of $i$, which leads to Hermitian operators with wavefunctions defined on the imaginary axis. It is not entirely clear to us which is more natural.} Similarly we will take the ghosts $\hat{c}$ and $\hat{b}$ (corresponding to the constraint $\Pi=0$) to be anti-Hermitian, so that the BRST charge $Q$ is Hermitian. We retain the standard representation of $N,\Pi=-i\partial_N$ acting on wavefunctions $\psi(N)$ depending on $N\in\RR$, but the anti-Hermiticity leads to a modified inner product with a sign flip:
\begin{equation}\label{eq:imagNorm}
\langle \psi_2|\psi_1 \rangle  =\int_{-\infty}^{\infty} \psi^*_2(-N)\psi_1(N) dN.
\end{equation}
 Notably, this inner product is not positive-definite: even wavefunctions have positive norm, while odd wavefunctions have negative norm. Nonetheless, just like for the ghost sector this is not a problem on the extended BRST state space, as long as we end up with a positive-definite physical inner product on BRST cohomology. Similarly, the inner product on the ghosts are fixed by their (anti-)Hermiticity. Of particular interest are the inner products between states with $N_g =0$,
\begin{equation}
   \langle\downarrow\uparrow|\downarrow\uparrow\rangle= \langle\uparrow\downarrow|\uparrow\downarrow\rangle=0,\quad \langle\uparrow\downarrow|\downarrow\uparrow\rangle=\langle\downarrow\uparrow|\uparrow\downarrow\rangle=-1.
\end{equation}
The Hermiticity relations imply that the value of $\langle\uparrow\downarrow|\downarrow\uparrow\rangle$ must be real, but the precise value is a normalisation convention (the sign is chosen for later convenience).

To connect this apparently strange choice of lapse quantisation to something more familiar, we recall some ideas from a covariant (e.g., Gupta-Bleuler) quantisation of gauge fields. The (Hermitian) gauge fields $A_\mu$ are decomposed into annihilation/creation operators $a_\mu,a^\dag_\mu$ for each momentum, with covariant commutation relations $[a_\mu,a_\nu^\dag]\propto \eta_{\mu\nu}$, and states are built on a vacuum annihilated by $a_\mu$. This means that the timelike component (corresponding to $A_0$, the Lagrange multiplier which imposes the Gauss law constraint) obeys a harmonic oscillator algebra with a wrong sign. If we try to write states in terms of wavefunctionals of $A_0$, this leads to a vacuum state wavefunction with a wrong-sign Gaussian, $\psi \propto \exp(+\# A_0^2)$. To make sense of this (and match the inner products obtained from the oscillator representation), we can take this wavefunction to be defined for \emph{imaginary} values of $A_0$, so $A_0$ is a Hermitian operator with imaginary eigenvalues. In our conventions, $N$ is analogous to $i A_0$: an \emph{anti}-Hermitian operator with real-valued wavefunctions. So, our anti-Hermitian lapse with its unusual inner product is essentially the same as the  wrong sign appearing in timelike oscillators of gauge fields. For more details, see appendix \ref{app:Maxwell}. The usual uniqueness of the representation of position and momentum operators on Hilbert space (the Stone-von Neumann theorem) does not apply because the inner product is not positive-definite.


Now we introduce some notation for the states in this BRST formalism which correspond to co-invariants and invariants as discussed in section \ref{sec:states}:
\begin{align}
    |\psi_\mathrm{BRST} \rc &:= |\psi\rc \otimes |\Pi=0\rangle_N\otimes |\uparrow\downarrow\rangle_{g} \,, \\
    |\psi_\mathrm{BRST}\ri &:= |\psi\ri \otimes |N=0\rangle_N\otimes |\downarrow\uparrow\rangle_{g}\,.
\end{align}
The first line gives co-invariant states corresponding to the BRST cohomology with ghost state fixed to be $|\uparrow\downarrow\rangle_g$ (annihilated by $c$ but not by $\hat{c}$). To be BRST closed the state must be annihilated by $\Pi$, fixing the lapse sector state to be $|\Pi=0\rangle_N$, with wavefunction $\langle N|\Pi=0\rangle_N =1$ (independent of $N$). BRST-exact states of this form are given by $Q$ acting on $|\phi\rangle\otimes |\Pi=0\rangle_N\otimes |\downarrow\downarrow\rangle_g$, which identifies $|\psi\rangle\sim |\psi\rangle+ H|\phi\rangle$ recovering the co-invariant equivalence classes. The second line is a representation of invariant states by choosing ghost state $|\downarrow\uparrow\rangle_{g}$. To be BRST closed, such a state must be annihilated by $H$ so we get invariants. The equivalence under adding BRST-exact states (co-invariance under the $\Pi$ constraint) preserves only the integral $\int \psi(q,N) dN$ of the remaining wavefunction, and we have used this freedom to choose wavefunctions of the form $\psi(q)\delta(N)$, fixing the lapse state $|N=0\rangle_N$.


\subsection{Some examples}

We are now ready to choose at some specific examples of gauge-fixing fermions $\Psi$, calculate the matrix elements of the associated BRST-trivial operator between (co-)invariant BRST closed states, and interpret the results.

\subsubsection{Group averaging inner product}\label{sssec:GA}

For our first example, choose the gauge-fixing fermion $\Psi = -i b N$ (with a factor of $i$ required for Hermiticity since $N^\dag=-N$). This gives us a BRST-exact operator $[Q,\Psi]_+=-iNH + b\hat{c}$, which generates a finite gauge transformation $e^{[Q,\Psi]_+} = e^{-i N H}(1+b\hat{c})$. We calculate matrix elements of this operator between states $|\psi_\mathrm{BRST}\rc=|\psi\rangle\otimes |\Pi=0\rangle \otimes |\uparrow\downarrow\rangle$ in the co-invariant sector:
\begin{equation}
   \lc\psi'_\mathrm{BRST}|e^{[Q,\Psi]_+} |\psi_\mathrm{BRST}\rc = \langle\uparrow\downarrow|b\hat{c}|\uparrow\downarrow\rangle \int dN\langle\psi'|e^{-i N H}|\psi\rangle = \lc \psi'|\psi\rc_\mathrm{GA}.
\end{equation}
In the ghost sector, only the $b\hat{c}$ term contributes, giving us unity (interpreted as the trivial measure from fixing to synchronous gauge  $\dot{N}=0$). The integral over $N$ comes from the inner product between $|\Pi=0\rangle$ states. We end up with precisely the group average inner product \eqref{eq:GAIP} discussed in section \ref{ssec:GA}.

To interpret the action of $[Q,\Psi]_+$ note that the bosonic part is the canonical Hamiltonian $\mathbf{H}=NH$, so we can interpret the result as evolution for unit coordinate time in the gauge $\dot{N}=0$ as described in \ref{ssec:nonminGF}. The constant value of $N$  can then be interpreted as the total proper time, which is a modulus that we must integrate over. The simple ghost part gives us the flat measure $dN$ for this modulus \cite{Cohen:1985sm}.

To illustrate the invariance implied by the Fradkin-Vilkovisky theorem, we consider a few simple generalisations. First we can simply rescale the generator by inserting $e^{t[Q,\Psi]_+}$ for some real $t$, which has the interpretation of the coordinate time interval. The lapse integral from the bosonic piece (integrating over evolutions by proper time $Nt$) then results in an insertion of $2\pi \delta(t H)$, while the fermionic piece provides an extra factor of $t$: these almost precisely compensate for each other since $\delta(t H) = \frac{1}{|t|}\delta(H)$, together giving us a factor of $\sgn(t)$. This illustrates the invariance under small deformations (changes of $t$ when $t\neq 0$), but also the failure of invariance from discrete jumps that can occur when passing through a singular value: here, the change in sign as we pass through $t=0$.

Another simple generalisation is to add an extra term $ - \frac{\xi}{2} \hat{b}  \Pi$ to the gauge-fixing fermion $\Psi$ for some parameter $\xi$, which results in an extra Gaussian factor $e^{-\frac{\xi}{2}   \Pi^2}$ in the inner product (without affecting ghost terms). This gives us Gaussian averaged gauges (often used in quantisation of gauge theories). For the lapse variable, the bosonic piece $-\frac{\xi}{2}\Pi^2 -i HN$ of $[Q,\Psi]_+$ can  be written as a first-order differential operator acting on wavefunctions of $\Pi$. Solving the resulting time-dependent Schr\"odinger equation for the action of $e^{[Q,\Psi]_+}$ on initial lapse wavefunction $\delta(\Pi)$ gives us $e^{-\frac{\xi}{6}H^2}\delta(\Pi+H)$. So, ultimately we find that the insertion $e^{-i N H}$ is replaced by $\exp \left(- \frac{\xi}{6}  H^2  -i N H\right)$. The  result after integrating over $N$ is of course unchanged as expected, but the intermediate steps may be nicer, particularly in the semiclassical limit: the resulting integral over $N$ will typically be sharply peaked at the classical value, giving a rapidly convergent integral rather than a conditionally convergent oscillatory integral.

We can also choose a much more general gauge for time evolution, where we fix coordinate time in essentially any way we please (e.g., conformal time in an FLRW universe rather than proper time). For this, consider a gauge-fixing  fermion $\Psi= -i b N F$ where $F$ is a positive Hermitian operator on the original unconstrained Hilbert space (a function of $q$s and $p$s). This gives us $[Q,\Psi]_+ = -i b c NFH - i cb NHF +  b\hat{c} F$. Classically (i.e., using Poisson brackets), the bosonic part of this is $NFH$, so we can interpret this as generating evolution using a rescaled Hamiltonian $\tilde{H}=FH$, along with a redefinition of the lapse $N$ (given by $F^{-1}N$ in terms of the original lapse). The $\Pi$ equation of motion sets this new rescaled lapse to be constant.  The upshot is that we are choosing a gauge for coordinate time $t$ such that $\frac{d\tau}{d t} \propto F$ (where $\tau$ is the original proper time).\footnote{Alternatively, we could follow the approach sketched in section \ref{ssec:nonminGF} by choosing $\Psi = -ib N - i\hat{b}F$, which gives $[Q,\Psi]_+=-i(NH+\Pi F) +(\text{ghosts})$. Then, $N$ is the usual lapse (not rescaled), but $[Q,\Psi]_+$ generates time translations in a gauge $\dot{N}=F$. This seems more natural, and makes closer contact with standard gauge-fixing approaches (see appendix \ref{app:Maxwell} for $U(1)$ gauge theory in Lorenz gauge). Unfortunately, it is technically trickier to understand the resulting inner product.} This is rather common in mini-superspace models (for example choosing $F(q)$ to make the target space metric flat, or using conformal time), and we will comment on this below.

Exponentiating this operator is slightly trickier because we have three terms which may not commute, but as we show in appendix \ref{app:BRST} its matrix elements between $ |\uparrow\downarrow\rangle$ ghost states relevant for co-invariants can be computed, finding
\begin{equation}\label{eq:BRSTrescaled}
    \langle\uparrow\downarrow|e^{[Q,\Psi]_+} |\uparrow\downarrow\rangle = F^{\frac{1}{2}}\exp\left(-i N F^{\frac{1}{2}} H F^{\frac{1}{2}} \right) F^{\frac{1}{2}}.
\end{equation}
Taking the matrix elements of this result between $|\Pi=0\rangle$ lapse states (i.e., integrating over $N$) corresponds to a group average with rescaled Hamiltonian $F^{\frac{1}{2}} H F^{\frac{1}{2}}$, with extra compensating `measure factors' of $F^{\frac{1}{2}}$ (which perhaps could be absorbed into a redefinition of the wavefunctions and unconstrained inner product). By invariance under choice of $\Psi$, the result $F^{\frac{1}{2}}\delta(F^{\frac{1}{2}} H F^{\frac{1}{2}})F^{\frac{1}{2}}=\delta(H)$ is simply a new representation of the original group average rigging map, though the propagator in the integrand before averaging is different.\footnote{\label{foot:ESA}A subtlety is that $F^{\frac{1}{2}} H F^{\frac{1}{2}}$ may not be essentially self-adjoint so the exponential may not be defined for all $N$, and hence $\delta(F^{\frac{1}{2}} H F^{\frac{1}{2}})$ may not make sense, or may be ambiguous: see comments in the discussion section \ref{sec:disc}.}

For sigma-models, if we take $F$ to be a  function $F(q)=\Omega(q)^{-2}$ of $q$ only, then the rescaled Hamiltonian $\tilde{H}$ is classically obtained by a conformal transformation of the target space metric $\tilde{G}_{ab} = \Omega^{2}G_{ab}$ and a rescaled potential $\tilde{V} = \Omega^{-2}V$. Quantising including a `conformally coupled' curvature term to the potential (as mentioned in \ref{ssec:sigma}) \cite{Halliwell:1988wc,Moss:1988wk}, the action of the Hamiltonian is changed by a simple rescaling $\tilde{H} = \Omega^{-1-\frac{n}{2}}H\Omega^{\frac{n}{2}-1}$ (which is Hermitian under the inner product with rescaled measure $\sqrt{\tilde{G}}d^nq =\Omega^n\sqrt{G}d^nq $). This gives a simple correspondence between \mbox{(co-)invariant} wavefunctions under $H$ or $\tilde{H}$: $\tilde{\psi}_\mathrm{inv} =\Omega^{1-\frac{n}{2}}\psi_\mathrm{inv} $ is annihilated by $\tilde{H}$ if $\psi_\mathrm{inv}$ is annihilated by $H$, and similarly $\tilde{\psi}_\mathrm{co-inv} =\Omega^{-1-\frac{n}{2}}\psi_\mathrm{co-inv} $ induces a well-defined map between co-invariant cosets. Furthermore, taking a group averaging integral with $\tilde{H}$ gives us essentially the same calculation as \eqref{eq:BRSTrescaled}, up to conjugation by $\Omega^{\frac{n}{2}}$ (to account for the change in measure), and we arrive at\cref{foot:ESA} $\delta(\tilde{H}) =  \Omega^{1-\frac{n}{2}}\delta(H)\Omega^{1+\frac{n}{2}}$. The upshot is that group averaging with a rescaled Hamiltonian $\tilde{H}$ gives us the same physical Hilbert space, once we appropriately rescale wavefunctions.

\subsubsection{Klein-Gordon form (and generalisations)}

Next, consider choosing the gauge-fixing fermion $\Psi = -i\hat{b} \chi$ for some Hermitian operator $\chi$ on the unconstrained Hilbert space. For the Klein-Gordon form, we  take $\chi$ to be given by a target space `time function' $\chi(q)$ of the coordinates on $\target$, but as discussed in section \ref{sec:IP} we can generalise this to an operator depending also on momentum. We find $[Q,\Psi]_+ = -i\Pi \chi +i [H,\chi] \hat{b}c$, noting the appearance of the `Faddeev Popov' measure factor $i[H,\chi]$ in the ghost part. We now insert $e^{[Q,\Psi]_+}$ between two states to get an inner product (which is slightly trickier than for the group average because the two terms may not commute).  We can write this exponential as
\begin{equation}\label{eq:GFFKG}
    e^{ [Q,\Psi]_+} = e^{-i \Pi \chi} +  c\hat{b} \left[H,\Pi^{-1} e^{-i \Pi \chi}\right].
\end{equation}
One way to verify this is to replace $[Q,\Psi]_+$ by $s[Q,\Psi]_+$, and check that the result solves the `Schr\"odinger equation' that the differentiating with respect to $s$ gives the same as acting with $[Q,\Psi]_+$. We also explain a more explicit calculation in appendix \ref{app:BRST}, which illustrates some general useful ideas for these sorts of calculations. Note that while $\Pi^{-1}e^{-i \Pi \chi}$ is singular at $\Pi=0$, the singular part $\Pi^{-1}$ does not contribute since it commutes with $H$.

The $c\hat{b}$ term will give us a non-zero result for matrix elements between ghost states $|\downarrow\uparrow\rangle$ corresponding to invariants. The lapse sector is easiest to evaluate in the momentum representation (wavefunctions of $\Pi$). Taking the simplest $|N=0\rangle$ lapse wavefunctions ($\delta(N)$ in the $N$ basis and  a constant $\frac{1}{\sqrt{2\pi}}$ in the $\Pi$ basis), we need the integral
\begin{equation}
    \frac{1}{2\pi}\int d\Pi \frac{e^{-i \Pi \chi}}{\Pi+i\epsilon} = - i \Theta(\chi),
\end{equation}
where the $i\epsilon$ in the denominator is an arbitrary prescription to regulate the $\Pi=0$ singularity (other possible prescriptions differ by a constant which drops out when  we take the commutator with $H$, for example the principle value integral gives $-i\pi \sgn(\chi)$). From this, we find the result
\begin{equation}\label{eq:BRSTKG}
   \li \psi'_\mathrm{BRST}| e^{ [Q,i\hat{b}\chi]_+}|\psi_\mathrm{BRST}\ri = i \li \psi'|[H,\Theta(\chi)]_-|\psi\ri.
\end{equation}
This is precisely the expression $\kappa_\Sigma$ we found in \eqref{eq:kappaSigma} for the Klein-Gordon form on the surface $\Sigma$ defined by $\chi=0$.

We can also choose other lapse wavefunctions in the bra and ket states, which replaces  the step function $\Theta(\chi)$ with a more general function which becomes zero or one for large negative or positive $\chi$ respectively. Specifically, this function is obtained by convolving the lapse wavefunctions with a step function.  Thus we get the same result, but smoothed out over representations with shifted values of $\chi$. A specific example is to use different fixed-lapse wavefunctions $|N=N_i\rangle$ and $|N=N_f\rangle$, which has the effect of shifting $\chi$, replacing $\Theta(\chi)$ with $\Theta(\chi-N_f+N_i)$. The Gaussian average gauge (adding $-\frac{\xi}{2}\hat{b}\Pi$ to $\Psi$) similarly acts to smooth over a Gaussian window of width $\sqrt{\xi}$ in $\chi$.

To connect this with the gauge-fixing $\kappa$ map discussed in section \ref{sec:IP}, we simply take the inner product in the lapse and ghost sectors and treat the remaining object as an operator in the original unconstrained Hilbert space. A general Gaussian averaged gauge-fixing map looks like
\begin{equation}\label{eq:kappaxi}
    \kappa_{\chi,\xi} = \langle N=0|\otimes\langle\downarrow\uparrow|e^{[Q,-i\hat{b}\chi - \frac{\xi}{2}\hat{b}\Pi]_+} | N=0\rangle\otimes| \downarrow\uparrow\rangle = -\frac{1}{2\pi}\int_{-\infty}^\infty \frac{d\Pi}{\Pi} [H,e^{-i \Pi\chi}]e^{-\frac{\xi}{2}\Pi^2}.
\end{equation}

\subsubsection{Zero momentum gauge}\label{sssec:p0gauge}

For the above discussion, we took $\chi$ to be a general operator, with the specific example in mind of a function $\chi(q)$ of target space in a sigma-model to recover the Klein-Gordon form.  Here we discuss a different choice which is a  one-dimensional (or mini-superspace) analogue to gauge-fixing on the maximal-volume slice in higher dimensions \cite{Witten:2022xxp}, mentioned in section \ref{sssec:maxvol}.

For this, we choose $\chi = p_0$ where $p_0$ is the momentum in a timelike direction of target space. This is likely to  correspond to a good gauge in cases where the potential $V$ becomes large and negative for large $q^0$. In that case, a typical trajectory will come from negative $q^0$ with negative $p_0$ (which corresponds to positive $\dot{q}^0$ due to the negative kinetic term), reflect off the potential, and return to negative $q^0$ with positive $p_0$, crossing the $\chi=0$ slice once in the positive direction.

An analogous choice in higher dimensional gravity is to gauge-fix to  $K(x)=0$, where $K$ is the trace of the extrinsic curvature. This momentum corresponds to the timelike direction in superspace, and is canonically conjugate to the volume density $\sqrt{\det \gamma}$ where $\gamma$ is the spatial metric. In most cases, the potential ($\sqrt{\det \gamma}(2\Lambda-R)$ where $R$ is the spatial curvature, plus contributions from matter energy density) will become negative at large volume density for fixed conformal metric (the most notable exceptions occurring for positive cosmological constant). See \cite{Witten:2022xxp} for a detailed discussion.

For simplicity we restrict our calculations to the example of a one-dimensional target-space discussed in section \ref{ssec:1Dtarget} with $H= - \frac{p^2}{2}+V(q)$ (now returning to the standard sign convention for $p$ with $[q,p] = i$ so that $\dot{q}=-p$), and use the gauge-fixing function $\chi=p$. Our considerations straightforwardly extend to many more general cases: in particular, an immediate generalisation is a higher-dimensional target space whose metric $G_{ab}$ is independent of $q$ (so that $p$ corresponds to a Killing vector of target space), in which case the following goes through unchanged by suppressing dependence on other coordinates.

To set expectations, recall that ignoring issues of operator ordering would give us a combination of a gauge-fixing delta function $\delta(p)$  with a measure factor $\Delta_\mathrm{FP} = i[H,p] = -V'(q)$. But the two operators $\delta(p)$ and $\Delta_\mathrm{FP}$ certainly do not commute, so it is not very obvious what the precise result will be.

As we saw above, the inner product comes from the $c\hat{b}$ term in $e^{i[Q,\Psi]_+}$ given in \eqref{eq:GFFKG}. We calculate the matrix elements in the position basis, with the only contribution to the commutator coming from the potential term in $H$:
\begin{equation}
\begin{aligned}
    \langle q|[V,\Pi^{-1} e^{-i \Pi p}]|q'\rangle &= \Pi^{-1}\left(\langle q|V(q) |q'+\Pi \rangle- \langle q-\Pi|V(q') |q'\rangle\right) \\
    &= \delta(q'-q+\Pi) \frac{V(q)-V(q')}{\Pi},
\end{aligned}
\end{equation}
where we use the fact that $e^{-i\Pi p} $ is the operator that translates $q$ by $\Pi$. If we then integrate over $\Pi$ (taking the expectation value with lapse wavefunctions $|N=0\rangle$), we get matrix elements for the associated sesquilinear form
\begin{equation}\label{eq:kappap}
    \kappa_p(q,q') =  i \langle q|[H,\Theta(p)]|q'\rangle = -\frac{1}{2\pi}\frac{V(q)-V(q')}{q-q'}.
\end{equation}
We can easily generalise this to the Gaussian averaged version as in \eqref{eq:kappaxi},
\begin{equation}\label{eq:kappapxi}
    \langle q|\kappa_{p,\xi}|q'\rangle = - \frac{1}{2\pi}\frac{V(q)-V(q')}{q-q'}e^{-\frac{\xi}{2}(q-q')^2}.
\end{equation}
We can see that close to the diagonal $|q-q'|\ll 1$, these matrix elements become $\frac{1}{2\pi}$ times $-V'(q)$ which is the Faddeev-Popov determinant $\Delta_\mathrm{FP}$. This is a reasonable match with expectations because the position space matrix elements of $\delta(p)$ are simply constant, $\langle q|\delta(p)|q'\rangle = \frac{1}{2\pi}$.

Alternatively, we  can calculate matrix elements in momentum space (if the potential does not grow too fast) to get
\begin{equation}
    \langle p|\kappa_p|p'\rangle =  \frac{1}{4\pi i} (\sgn p - \sgn p')  \hat{V}(p-p'),
\end{equation}
where $\hat{V}(p) = \int dq\, e^{-i p q}V(q)$, and take a Fourier transform to recover the position space result.

Note that these expressions for $\kappa_p$ do not sharply project onto the $p=0$ component of the wavefunction in any obvious sense. For example, we do \emph{not} get the simplest possibility of an  `average' ordering of delta function and measure factor $\kappa = -\frac{1}{2}(\delta(p)V'(q)+V'(q)\delta(p))$. The way in which \eqref{eq:kappap} (or better, its Gaussian average version \eqref{eq:kappapxi}) gives a $p=0$ gauge-fixing becomes sharper in a semi-classical limit, as we'll see in section \ref{sec:SC}.


We can explicitly evaluate the overlap $\li \psi_2|\psi_1\ri_p$ of two invariant states using the sesquilinear form $\kappa_p$ in various cases, and check that the result accords precisely with expectations from the interpretation in terms of gauge-fixing function $p=0$. Here we just summarise results for various possible qualitative shapes of the potential.

First, in the motivating case where $V$ is a positive constant for $q\to-\infty$ and negative at $q\to +\infty$, we find that $\li \psi_2|\psi_1\ri_p = \li \psi_2|\psi_1\ri_\mathrm{GA}$, agreeing with the physical group average result. If instead $V$ is a positive constant at $q\to + \infty$ and negative at $q\to -\infty$, we find a relative minus sign. This is because trajectories come in from the right (with negative $\dot{q}$ and positive $p$) and leave from the left (with negative $p$), crossing the gauge-fixing slice in the decreasing direction. For a one-dimensional target space this is somewhat trivial (since the physical Hilbert space is one-dimensional), but this result also applies in more general cases where $\kappa_p$  gives a useful representation of the physical inner product.


We can also check the case where $V$ approaches a positive constant for both $q\to\pm \infty$, with scattering states as defined in \eqref{eq:psiIn}. Similarly to \eqref{eq:KGoutin} for the Klein-Gordon form, we can interpret the result most straightforwardly in terms of matrix elements between in and out states. We find
\begin{equation}
    \li \mathrm{out},\pm|\mathrm{in},\pm\ri_{p} = \mp  \li \mathrm{out},\pm|\mathrm{in},\pm\ri_\mathrm{GA}, \qquad \li \mathrm{out},\pm|\mathrm{in},\mp\ri_{p} = 0,
\end{equation}
precisely in line with expectations from the gauge-fixing interpretation.

\subsection{Mixing invariants and co-invariants}\label{ssec:imN}

It is instructive to look not only at the inner products constructed as above, but also at the states obtained from acting with our gauge transformations $e^{[Q,\Psi]_+}$. For the case of group averaging $\Psi = -ibN$ acting on co-invariants, we find the following (where we represent the matter and lapse sector as wavefunctions of $q,N$):
\begin{gather}
    e^{[Q,\Psi]_+} \psi_0(q) |\uparrow\downarrow\rangle = \psi(q;N) (|\uparrow\downarrow\rangle -|\downarrow\uparrow\rangle), \\
\text{where}\quad     i \frac{\partial}{\partial N}
\psi(q,N) = H \psi(q,N), \quad \psi(q,N=0)=\psi_0(q).\end{gather}
So, we arrive at a wavefunction $\psi(q,N)$  which is a solution to the time-dependent Schr\"odinger equation, with $N$ treated as the time!

The space of such wavefunctions gives us an alternative way to represent physical states. These are in a sense intermediate between invariants and co-invariants, because they are annihilated not by $H$ or $\Pi$, but by a linear combination $H+\Pi$. 
Our gauge transformation has converted our original co-invariant wavefunction $\psi_0(q)$ to a new but physically equivalent form, namely the corresponding solution of the time-dependent Schr\"odinger equation $\psi(q,N)$. Furthermore, this new representation is less singular, in the sense that we can \emph{directly} take inner products in the extended Hilbert space (remembering the anti-Hermicity of $N$ \eqref{eq:imagNorm}) without encountering an indeterminate form:
\begin{equation}
    \langle\psi|\psi\rangle =\int dq dN \psi^*(q,-N)\psi(q,N)  = \frac{1}{2}\int dqdt\, \psi_0^*(q) e^{-i H t} \psi_0(q) \propto \lc \psi_0|\psi_0\rc_\mathrm{GA}.
\end{equation}
This inner product is finite, and unsurprisingly gives us the group averaging result (up to a factor of 2, which is compensated for by the ghosts, arising because this norm involves two insertion of $e^{[Q,\Psi]_+}$ so that $t=2N$).

 Just like (co-)invariants, this new representation of physical states also comes with an identification under addition of a null state in the image of $H$. Because these states are annihilated by $H+\Pi$, this is equivalent to addition of something in the image of $\Pi$, namely any total derivative with respect to $N$. In particular, this tells us that wavefunctions $\psi(q,N)$ and $\psi(q,N+\Delta N)$ related by a constant time translation $\Delta N$ are physically equivalent.

In terms of ghosts, we are choosing a representation of the BRST cohomology class which is not annihilated by $c$ or $\hat{c}$, but by the non-hermitian combination $c-\hat{c}$. To make the resulting structure clearer, we can package the ghosts into a single complex (non-Hermitian) ghost $C\propto c-\hat{c}$  and its adjoint, and similarly package the two (anti-)Hermitian constraints into a single complex constraint $G$ as follows:
\begin{equation}
    G=\frac{H+\Pi}{\sqrt{2}},\quad C=\frac{c-\hat{c}}{\sqrt{2}} \implies  Q = C^\dag G + C G^\dag.
\end{equation}
Now, a BRST closed state which is annihilated by $C$ must be annihilated by $G$, and hence must be a solution of the time-dependent Schr\"odinger equation as above. Similarly, a BRST exact state which is annihilated by $C$ must be of the null form described in the above paragraph. The key difference in the ghost sector (compared to states annihilated by $c$ or $\hat{c}$) is that the state annihilated by $C$ has non-zero norm.

Note that this is similar to the situation more typically encountered in BRST quantisation of gauge theories. In that case, one usually decomposes ghosts and constraints into positive frequency parts (analogous to $C,G$) and their conjugate negative frequency parts, and builds states on a vacuum annihilated by positive frequency modes. See appendix \ref{app:Maxwell}  for details of the example of Maxwell theory. This means that we can immediately take inner products in the extended BRST Hilbert space. We expect a similar procedure to be useful for many problems in gravity (perturbation theory around Minkowski space being a simple example). However, it may not always be possible or convenient to pair up constraints in this manner. A simple example is in perturbation theory around an FLRW cosmology, where modes with non-zero momentum should be expected to pair up rather naturally in this way, but not spatially homogeneous zero modes, which may need to be treated separately perhaps using the ideas in this paper. Something similar occurs for the string worldsheet, where the analogues to $G$ are the non-Hermitian Virasoro modes $L_n$ for $n>0$, with $L_n^\dag = L_{-n}$, but the Hermitian zero modes $L_0$ are not paired in this way (see appendix \ref{app:string}). We expect such issues to only become more pronounced when working around an excited background with fewer symmetries (e.g., a black hole spacetime).

\section{Semiclassical limits}\label{sec:SC}

Our main examples of one-dimensional gravity are very useful to explore the main ideas without too much technical complication. But in some ways these examples are too simple: many calculations can be done exactly, which is rarely possible (perhaps even in principle) in realistic models for interesting applications. With this in mind we discuss the semi-classical limit in this section, aiming to illustrate ideas that generalise most readily to higher-dimensional theories of gravity.

We then illustrate these ideas concretely using the example of a one-dimensional target space (introduced in section \ref{ssec:1Dtarget}) with $H=-\frac{p^2}{2}+V(q)$, emphasising the analogy with the maximal-volume gauge $\Tr K=0$ as discussed in section \ref{sssec:p0gauge}.

In this section we restore factors of $\hbar$ to indicate how parameters scale in the semi-classical limit. Note that for gravity, the small parameter is $\hbar G_N$ (compared to a typical curvature scale), which we can scale separately from $\hbar$ if we would like to use a semi-classical limit from gravity while treating matter as fully quantum.

\subsection{General comments}

We start our discussion with some general comments about the semi-classical limit in general sigma model or mini-superspace examples \eqref{eq:sigmaH}. This outlines a general strategy for understanding the Hilbert space in semi-classical perturbation theory, which we expect to generalise nicely to higher-dimensional theories of gravity. This is only a sketch, and we do not attempt to work out all the details (an interesting exercise for the future). In the next subsection we specialise to the examples of one-dimensional target space introduced in section \ref{ssec:1Dtarget}, for which we implement these ideas explicitly.

\subsubsection{Group averaging}

We start by considering the group averaging rigging map and associated inner product. In the position basis (taking co-invariant states $|q\rc$ represented by $\delta$-function wavefunctions), this is an integral of the propagator $K$ (the position space matrix elements of $e^{-iHt/\hbar}$):
\begin{equation}
    \eta(q,q') = \lc q|q'\rc = \int_{-\infty}^\infty dt K(q,q';t), \qquad K(q,q';t) = \langle q|e^{-iHt/\hbar}|q'\rangle .
\end{equation}
Note that this is similar to the fixed-energy retarded propagator (or Green's function) $G(q,q';E) = i\int_0^\infty K(q,q';t)e^{iE t/\hbar}dt$ at $E=0$,
the difference being that we integrate over all real $t$ (not just $t>0$). (In fact, it is twice the imaginary part of $G$, or the difference of propagators as expressed in \eqref{eq:FminusAF}.) This means we can use some standard results from semi-classical  approximation in quantum mechanics (e.g., chapter 38 of \cite{ChaosBook}).

In the semi-classical limit, $K(q,q';t)$ for fixed $q,q',t$ is dominated by classical solutions beginning at $q'$ and ending at $q$, taking time $t$. Each such solution contributes
\begin{equation}
    K(q,q';t) \sim \frac{1}{(2\pi i \hbar)^\frac{n}{2}} \sqrt{D_K(q,q';t)} e^{\frac{i}{\hbar}S(q,q';t)},
\end{equation}
where $S = \int dt (p\dot{q}-H(p,q))$ is the action of the classical solution (or Hamilton's principal function). We will not make use of the details of the prefactor here, except that $D_K$ (the van Vleck determinant) is fixed in the $\hbar\to 0$ limit. For fixed $q,q',t$ we generically expect there to be a discrete set of solutions --- for given $q'$, there is an $n$ dimensional space of solutions from choosing initial momenta, the same as the dimension of endpoints $q$ --- and the contributions from these are summed. Now to get the rigging map $\eta(q,q')$ we perform the integral over $t$, which is dominated by points of stationary phase. The time derivative of the on-shell action (with fixed endpoints) is the energy $\frac{\partial S}{\partial t}=-E$, so the stationary points correspond to classical solutions satisfying the constraints $H=0$ (for any time $t$), each contributing
\begin{equation}\label{eq:etaSC}
    \eta(q,q') \sim \frac{1}{ (2\pi i \hbar)^\frac{n-1}{2}} \sqrt{D_\eta(q,q')} e^{\frac{i}{\hbar}S(q,q';E=0)}
\end{equation}
with a different determinant $D_\eta$.\footnote{More generally, taking the Fourier transform of $K$ with respect to $t$  by stationary phase gives the Legendre transform of the action $S(q,q';t)$ in the exponent, which is $S(q,q';E) = \int p dq$ regarded as a function of $E$ (Hamilton's characteristic function).}  For $q$ parametrically close to $q'$, there will also be a contribution from small $t$ which needs to be considered separately. 

\subsubsection{Invariant (Wheeler-DeWitt) wavefunctions}

Since $\eta$ maps general wavefunctions to invariant states, we expect that $\psi(q) = \eta(q,q')$ in \eqref{eq:etaSC} should be a semi-classical solution to the Wheeler-DeWitt equation for any fixed $q'$. To leading order, this means that the exponent $S(q) = S(q,q';E=0)$ solves the Hamilton-Jacobi equation
\begin{equation}\label{eq:HJ}
   H(q,p=\partial S(q)) =  \frac{1}{2}G^{ab}(q) \partial_a S(q)\partial_b S(q) + V(q)=0,
\end{equation}
since at leading order the momentum $p_a = -i \hbar \partial_a$ acting on $e^{\frac{i}{\hbar}S(q)}$ (times any one-loop prefactor) can be replaced by $\partial_a S$. And indeed, all (real) solutions of this equation correspond to the action of an $(n-1)$-dimensional family of classical trajectories solving the constraint $E=0$ (see \cite{Hartnoll:2022snh}, for example).

We can similarly get invariant states following the construction of scattering states in \eqref{eq:scatstate} or \eqref{eq:inStates}. In the case of \eqref{eq:inStates} the semi-classical calculation is precisely the same as \eqref{eq:etaSC}, except we only allow saddle-points at positive/negative $t$ for in/out-states respectively. We then push the initial point $q'$ to infinity (which means the saddle-points will be pushed to large $|t|$), and discard a divergent contribution to the action.

However, this does not make the closest possible contact with the classical theory, because $\psi(q) = \eta(q,q')$ does not represent a single classical solution: it corresponds to a superposition over all possible initial momenta compatible subject to the Hamiltonian constraint (or in the case of scattering states, a superposition over all impact parameters). To remedy this, we can consider the action of $\eta$ on coherent wavefunctions with well-defined position and momentum in the $\hbar\to 0$ limit. For example, we can choose an initial Gaussian wavefunction of the form $\psi_0(q) \propto e^{-\frac{A}{2\hbar}(q-q_0)^2+i p_0 \cdot q/\hbar}$ close to $q_0$, where $H(q_0,p_0)=0$.\footnote{Here $A(q-q_0)^2$ should be interpreted as $A_{ij}(q^i-q_0^i)(q^j-q_0^j)$ for some positive-definite Riemannian metric on target space at the point $q_0$ (not the mixed-signature target space metric $G$ appearing in $H$).} Then $\psi(q) = \int \eta(q,q') \psi_0(q')$  will be supported close to the trajectory of the classical solution specified by initial data $q_0,p_0$. The integral over $q'$ has a real saddle point (namely, $q'=q_0$) only if the classical solution with initial conditions $(q_0,p_0)$ passes through the final point $q$. Otherwise, the wavefunction is exponentially suppressed. In particular, this means that a pair of co-invariant states with Gaussian wavefunctions $\psi_1,\psi_2$ centered on phase space points $(q_1,p_1)$ and $(q_2,p_2)$ have an exponentially small inner product $\lc \psi_2|\psi_1\rc = \int dq_1 dq_2 \psi_2^*(q_2) \eta(q_2,q_1) \psi_1(q_1)$ unless $(q_1,p_1)$ and $(q_2,p_2)$ both lie on the same $E=0$ classical solution. The statement that generalises to a higher dimensional theory of gravity is that the inner product is exponentially suppressed unless  $(q_1,p_1)$ and $(q_2,p_2)$ correspond to initial data (metrics and extrinsic curvatures) for a pair of Cauchy surfaces $\Sigma_1$, $\Sigma_2$ in the same classical spacetime solving the Einstein equations. Such coherent states can also arise naturally from states prepared by  Euclidean path integrals (e.g., the Hartle-Hawking no boundary state): see further comments in section \ref{sec:disc}.

Note that the exponential tails of the wavefunction away from the classical trajectory can be interesting, perhaps giving an amplitude to tunnel to a classically forbidden region (an example will be given in a moment). Furthermore, these non-perturbative parts of the wavefunction (or of an inner product between co-invariants) may be computed by a saddle-point for which either the trajectory $q$ or the proper time/lapse $t$ (or both) are complex, corresponding in higher dimensions to complex (or perhaps Euclidean) spacetime metrics. But importantly, the original definitions of the quantities of interest are in terms of  integrals over real $t$ and $q$, enabling us to determine which saddle-points contribute by appropriate deformation of the contour of integration. In particular, there is no conformal factor problem \cite{Gibbons:1978ac} in such cases where there is a clear Hilbert space interpretation of the quantity of interest, with a corresponding Lorentzian path integral (even if the relevant saddle-points may correspond to Euclidean geometries). Similar ideas have been suggested many times before \cite{Louko:1988fi,Dasgupta:2001ue,Marolf:1996gb,Marolf:2022ybi}. Note also that in such cases, contributing complex saddle points can only give exponentially suppressed contributions, not enhanced \cite{Marolf:1996gb}. We study an example in detail in section \ref{ssec:nonpert}, and give further comments in section \ref{sec:disc}.

\subsubsection{Gauge-fixing maps or inner products on invariants}

It is relatively straightforward to apply the $\kappa$ gauge-fixing maps to the semi-classical invariant wavefunctions described above. Starting with an invariant wavefunction which is well-localised along a single classical trajectory $(q_c(t),p_c(t))$, we should choose a gauge condition $\chi=0$ which uniquely identifies a single point on the trajectory. For example, we might choose $\chi(q)$ to be a function of target space $\target$ so that the curve $q_c(t)$ on which the wavefunction is large crosses the $\chi=0$ surface exactly once. Then, applying the corresponding Klein-Gordon gauge-fixing operator $\kappa_\chi \sim \tfrac{1}{2}(p_\chi \delta(\chi)+\delta(\chi)p_\chi)$ gives us a wavefunction which is localised close to a single point on $\target$.

However, this is not convenient in the semiclassical limit because the resulting wavefunction $\kappa\psi$ is not well-localised in phase space: we have localised it \emph{too much} in $\chi$, so that it has large uncertainty in the conjugate momentum $p_\chi$. For example, a purely `positive frequency' state which corresponds to trajectories crossing the $\chi=0$ slice  with $\dot{\chi}>0$  requires an exponentially delicate balance between $\delta(\chi)$ and $\delta'(\chi)$ terms in such a wavefunction. This issue is worse still for gauge conditions involving momenta, when the wavefunction does not localise on target space (the example $\chi=p$ is discussed below). We would prefer a gauge-fixing prescription which gives rise to co-invariant wavefunctions with well-defined position and momentum in the semi-classical limit, which allows an interpretation as classical initial data.

A resolution to this uses the Gaussian averaged $\xi$ gauges  introduced in section \ref{sec:BFVIP}, where we smear the gauge-fixing delta function $\delta(\chi)$ to a Gaussian $\exp(-\frac{\chi^2}{2\hbar\xi})$. This localises $\chi$ to a width $\sqrt{\hbar\xi}$, and hence the conjugate momentum to  a width of order $\sqrt{\hbar\xi^{-1}}$, meaning that the resulting wavefunction can correspond to a well-defined point in phase space.

To show how this works in more detail, for some $\chi(q,p)$ consider the action of $\kappa_{\chi,\xi}$ as written in \eqref{eq:kappaxi} on a semiclassical invariant wavefunction. This is a wavefunction of the form $\psi(q)\sim e^{\frac{i}{\hbar}S(q)}$ (times a prefactor which is unaffected at one-loop order), where $S$ solves the Hamilton-Jacobi equation \eqref{eq:HJ}. Note that $S$ can be real (corresponding to a superposition of many classical solutions like in \eqref{eq:etaSC}), or have an imaginary part like our coherent wavefunctions introduced above in which case we can restrict attention close to the locus where $\Im S$ is minimised. 

The action of the operator $e^{-\frac{i}{\hbar}\Pi\chi}$ is computed by a path integral where we use $\chi$ as a `Hamiltonian' and flow by `time' $\Pi$. Classically, this means we evolve by $\frac{d q}{d \Pi} = \frac{\partial \chi}{\partial p}$, $\frac{d p}{\partial \Pi} = -\frac{\partial \chi}{\partial q}$, with initial condition $p(0)=S'(q(0))$ and final condition $q(\Pi)=q$ set by the point at which we evaluate the resulting wavefunction (in position basis). We will later integrate over $\Pi$ weighted by $e^{-\frac{\xi}{2\hbar}\Pi^2}$, which will be dominated by small $\Pi$ so we only need to expand the flow for short times, which is determined by local data in phase space. The new wavefunction is the exponential of the resulting on-shell action,\footnote{The one-loop determinant is $1+O(\Pi)$, and does not contribute at the order we are working.} which has small $\Pi$ expansion $S(q)-\Pi\chi(q,\partial S(q))+\frac{1}{2}\zeta(q)\Pi^2 +\cdots$ where $\zeta(q) = \frac{\partial\chi}{\partial p_a}\left(\frac{\partial\chi}{\partial q^a}+\frac{\partial\chi}{\partial p_b}\frac{\partial^2S}{\partial q^a\partial q^b}\right)$.  This gives
\begin{equation}
\begin{aligned}\relax 
    [H,e^{-\frac{i}{\hbar}\Pi\chi}] e^{\frac{i}{\hbar}S(q)} &\sim H\left(q,\partial S(q) - \Pi \tfrac{\partial}{\partial q}\chi(q,\partial S(q))+\cdots\right) e^{\frac{i}{\hbar}\left(S(q)-\Pi\chi(q,\partial S(q))+\frac{1}{2}\zeta(q)\Pi^2 +\cdots \right)} \\
    &\sim -\Pi \dot{\chi}(q) e^{\frac{i}{\hbar}\left(S(q)-\Pi\chi(q,\partial S(q))+\frac{1}{2}\zeta(q)\Pi^2\right)},
\end{aligned}
\end{equation}
where in the second line we use the Hamilton-Jacobi equation for $H$ to identify $\frac{\partial H}{\partial p_a} \frac{\partial}{\partial q^a}\chi(q,\partial S(q)) = \frac{\partial\chi}{\partial q^a}\dot{q}^a + \frac{\partial\chi}{\partial p_a}\dot{p}_a = \dot{\chi}$ as the time derivative of $\chi$ along the classical trajectory with $p = \partial S$. 

Now we perform the integral  $-\frac{1}{2\pi \hbar}\int \frac{d\Pi}{\Pi}  e^{-\frac{\xi}{2\hbar}\Pi^2}$ with the above result in the integrand as instructed in \eqref{eq:kappaxi} (restoring a factor of $\hbar$ from the time derivative). This is simply a Gaussian integral, as long as we can neglect higher-order terms in the expansion. This is self-consistent as long as the saddle-point value  $\Pi = \frac{\chi}{\zeta+i \xi}$ is small (in fact, $\Pi^3\ll \hbar$ to avoid contributions from higher-order terms at the one-loop order we are working). This means that $q$ is close to the surface $\chi(q,\partial S(q))=0$ where the gauge condition is satisfied (or alternatively, for generic $q$ if we choose $\xi\gg 1$). In that case, we find
\begin{equation}\label{eq:kappaxiSC}
    \kappa_{\chi,\xi} e^{\frac{i}{\hbar}S(q)} \sim \frac{\dot{\chi}(q)}{\sqrt{2\pi\hbar(\xi-i\zeta)}}  e^{\frac{i}{\hbar}S(q)-\frac{1}{2\hbar(\xi-i\zeta(q))}\chi(q,\partial S(q))^2}, \qquad (1\ll \xi \ll \hbar^{-1}).
\end{equation}
Elsewhere, the wavefunction is exponentially suppressed. Thus, $\kappa_{\chi,\xi}$ imposes the gauge condition  $\chi=0$ not as a delta-function but as a Gaussian with width roughly $\sqrt{\xi\hbar}\ll 1$, accompanied with the classical value  of the Faddeev-Popov factor $\dot{\chi}$. If we use $\kappa_{\chi,\xi}$ to compute the norm of the state, the integral $\int \psi^*\kappa_{\chi,\xi}\psi$ will be localised on the surface $\chi(q,\partial S(q))=0$.

The same calculation applies (using stationary phase approximation) for $\xi=0$ when $q$ is close to the $\chi=0$ surface, but the wavefunction is no longer suppressed when $\chi\neq 0$. At points where $\chi(q,\partial S(q))$ is not small there can still be a real saddle-point computing $\kappa_{\chi}\psi(q)$, but it is no longer at small $\Pi$ so it requires solving for the classical flow generated by $\chi$ for finite `time'. The Faddeev-Popov factor can then be more complicated and non-local in target space, which we have already seen for the case $\chi=p$ in \eqref{eq:kappap} (explored more in the semiclassical limit below). Even when $\chi(q,\partial S)$ is small, there can be additional saddle-points at large $\Pi$ which give further contributions to $\kappa_\chi\psi$ of comparable magnitude (which cannot happen for $\xi>0$).



The resulting wavefunctions \eqref{eq:kappaxiSC} can now be regarded as gauge-fixed representatives of co-invariant states, as outlined in section \ref{ssec:coinvGF}. But now we have a gauge fixing ansatz which applies for any gauge condition $\chi=0$, and which behaves nicely in the semiclassical limit.

We can then use group averaging to compute norms of these states. First, we can check what happens locally when we group average the co-invariant state in \eqref{eq:kappaxiSC}, by concentrating on the region where $\frac{\chi^2}{\xi-i\zeta}\ll 1$ and small $t$. Time-evolution simply translates the Gaussian packet along the classical flow defined by $S$ (with $p=\partial S$), so group averaging smears the wavefunction along each classical trajectory. We can change variable in this integral from $t$ to $\chi$ to get a Gaussian integral (with a sign if $\dot{\chi}<0$). The result is that we recover our original invariant wavefunction, up to a factor of $\sgn \dot{\chi}$. This verifies that $\kappa_{\chi,\xi}$ indeed acts as a generalised inverse to $\eta$, when $\chi=0$ is a good gauge. But $\eta$ will give the correct (positive) physical result in all cases. In particular, $\eta$ may have additional contributions from large times in the group average integral: for example, these may arise if  $\chi$ is a good gauge locally or perturbatively, but fails to be a good gauge globally or with non-perturbative effects (with trajectories crossing the $\chi=0$ surface several times).

\subsubsection{Comparison to symplectic quotient phase space}

At this level of approximation, we expect that the resulting Hilbert space to be equivalent to the quantisation of the symplectic quotient phase space (i.e., the phase space of classical zero-energy trajectories). We sketch one strategy for making this identification here.

It is simplest to quantise a phase space if it can be identified as the cotangent bundle $T^*\mathcal{Q}$ of some configuration space $\mathcal{Q}$, so we can take the Hilbert space $L^2(\mathcal{Q})$ to consist of wavefunctions defined on $\mathcal{Q}$. But there is no general way to do this for the symplectic quotient. An exception is for constraints which are linear functions of momenta. In that case, the gauge transformations generated by the flow act on configuration space (for us, the target space $\target$), since $\delta q$ is a function of $q^a$ only. The symplectic quotient can then be taken at the level of configuration space: it is canonically identified with $T^*(\target/\RR)$ (where $\RR$ is our one-dimensional gauge group generated by a single constraint). Of course, this does not apply to our gravitational constraints. However, we can reduce to such a situation if we limit ourselves to working perturbatively around a family of background solutions. For this, we first choose an $(n-1)$-dimensional family of trajectories. This sweeps out an $n$-dimensional surface in phase space, which we can (locally, and at generic points) write as $p=p_0(q)$. We then work with a constraint to linear order in $\delta p = p-p_0(q)$, namely $H \sim \frac{\partial H}{\partial p_a}(q,p_0(q)) \delta p_a $, so our perturbative symplectic quotient is given by the cotangent space of $\target$, modulo the flow generated by the vector field $v^a = \frac{\partial H}{\partial p_a}(q,p_0(q))$ (the velocity $\dot{q}^a$ of the background trajectory). We can make this more concrete by choosing a gauge condition $\chi(q,p)=0$ which selects a single point along each background trajectory, so that the physical phase space is identified as the cotangent bundle of the $(n-1)$-dimensional submanifold $\chi(q,p_0(q))=0$ of $\target$.

To implement this quantum mechanically, we start with a `background' semi-classical wavefunction $\psi_0(q) = A(q) e^{\frac{i}{\hbar}S(q)}$ which solves the Wheeler-DeWitt equation (so $S(q)$ solves the Hamilton-Jacobi equation and $A$ contains one-loop and higher order corrections). This encodes the $(n-1)$-dimensional family of classical trajectories described above, with $p_0(q) = \partial S(q)$. Then we take an ansatz for invariant wavefunctions of the form $\psi(q)A(q) e^{\frac{i}{\hbar}S(q)}$, where $\psi(q)$ is held fixed in the $\hbar\to 0$ limit. Requiring that this wavefunction solves the Wheeler-DeWitt equation to leading nontrivial order tells us that
$v^a \frac{\partial \psi}{\partial q^a}=0$, where $v^a = \frac{\partial H}{\partial p_a}(q,\partial S)$ is the tangent vector to gauge transformations as above (this is equivalent to $\delta S = -i\hbar\log \psi$ solving the Hamilton-Jacobi equation perturbed around the background solution $S$). So as expected from the above, our states are wavefunctions on $\mathcal{Q} = \target/\RR$, where $\RR$ is the group generated by the flow of $v^a$. And if $\chi(q,p)$ is a good gauge-fixing function then solutions to $v^a\partial_a\psi=0$ are in one-to-one correspondence with initial data giving $\psi$ on the $(n-1)$-dimensional surface $\chi(q,\partial S)=0$ (at least locally near $\chi=0$).

To check the identification of our Hilbert space with $L^2(\mathcal{Q})$, we also need to compare inner products, which we can compute as matrix elements of $\kappa_{\chi,\xi}$. Using \eqref{eq:kappaxiSC}, at one loop order we indeed get an inner product on the $(n-1)$-dimensional surface $\chi(q,\partial S)=0$ of the form $\int_{\chi=0} |\psi|^2$ (the choice of measure can be absorbed into a change of $A$ in the background state $\psi_0$, with a corresponding rescaling of $\psi$). Alternatively, we can treat $\kappa_{\chi,\xi}(\psi A e^{\frac{i}{\hbar}S})$ as a gauge-fixed representative of a co-invariant state (as outlined in section \ref{ssec:coinvGF}), and use group averaging to compute the inner product (dominated by short times or small lapse). We expect these to give the same results to all orders in perturbation theory if $\chi$ is a good gauge for our background trajectories (but group averaging should be taken as the correct physical prescription for going beyond perturbation theory).

Note, however, that this simple Hilbert space of functions with $L^2$ inner product applies only at this order. The formalism we have developed can be used to systematically compute corrections to the states and inner product. The loop corrections will typically contain higher and higher derivatives, indicating that summing corrections leads to an inner product which is non-local in configuration space.

\subsection{One-dimensional target examples}

 To illustrate these ideas more concretely, we return to the example of a one-dimensional target space (see section \ref{ssec:1Dtarget}) with $H=-\frac{p^2}{2}+V(q)$.

In particular, we mostly consider the case where the potential  $V$ approaches a positive constant $\frac{\omega^2}{2}$ for $q\to-\infty$ and monotonically decreases, becoming negative for $q>q_0$ where $q_0$ is the unique zero of $V$. Then there is a unique classical solution (solving the constraints, and up to gauge transformations), which comes in from $q\to-\infty$, bounces off the potential at $q=q_0$, and returns back to $q\to-\infty$. As already remarked in sections \ref{sssec:maxvol} and \ref{sssec:p0gauge}, this could be a simple model for an expanding and re-collapsing bang/crunch cosmology, or a model for different bulk time slices in an asymptotically AdS spacetime with the same asymptotic boundary time. With the latter in mind, many of our remarks will be closely analogous to the discussion in \cite{Witten:2022xxp}. The physics in this case is somewhat trivial (with a one-dimensional Hilbert space) but it is nonetheless sufficient to demonstrate many ideas that readily generalise, and we will remark on such generalisations along the way.

\subsubsection{Group averaging}

We begin by considering the group average inner product $\lc q|q'\rc$. Each trajectory contributes to the semi-classical inner product \eqref{eq:etaSC} as
\begin{equation}
    \lc q|q'\rc \sim \frac{1}{\sqrt{\dot{q}_i \dot{q}_f}} \exp\left({\frac{i}{\hbar} \int_{q'}^{q} p\,  dq}\right)+\cdots,
\end{equation}
where $p=\pm\sqrt{2V(q)}$ in the integrand, and the one-loop prefactor contains the  initial and final velocities $\dot{q}_{i,f}$ of the classical solution.\footnote{Some care is required to keep track of extra phases from the turning point $q=q_0$ where the solution reflects and $\dot{q}$ changes sign. For example, this leads to the $\frac{\pi}{4}$ phase shift familiar from WKB solutions.} For fixed $q'$, this is a WKB solution to $H|\psi\rangle=0$ as a function of $q$ (and the same holds with $q\leftrightarrow q'$).

Now, in the main case of interest where $q,q'<q_0$ both lie within the classically allowed region $V>0$ (and $q\neq q'$), there are in fact four zero-energy classical solutions which contribute. One trajectory goes directly from $q'$ to $q$ and one reflects from the potential barrier at $q_0$, and also we have the time-reverse of each of these. Summing the contributions from all four solutions gives a product of WKB scattering wavefunctions $\lc q|q'\rc = \psi_\mathrm{WKB}^*(q')\psi_\mathrm{WKB}(q)$, indicating that in this case the Hilbert space is one-dimensional. 
In contrast, if we had a potential which is everywhere positive there would be two distinct classical solutions (left-moving and right-moving) and the group average becomes a sum of two terms $\eta(q,q') = \psi_\mathrm{+}^*(q')\psi_\mathrm{+}(q)+\psi_\mathrm{-}^*(q')\psi_\mathrm{-}(q)$.

In section \ref{ssec:nonpert}, we will return to calculation of $\eta(q,q')$ in cases where there is no classical solution from $q'$ to $q$ (one endpoint lies in the classically forbidden region, or the endpoints are separated by a potential barrier).

\subsubsection{Invariant (scattering) wavefunctions}

In the case when $V(q)<0$ for $q>q_0$, there is a single solution to the Wheeler-DeWitt equation which does not blow up as $q\to\infty$. This is the WKB wavefunction $\psi_\mathrm{WKB}(q)$ which appeared in the calculation of $\eta$ above. Nonetheless, there are two natural basis vectors for this one-dimensional Hilbert space --- the scattering in state and out state --- and the relative phase contains physical information. Here we compute the  wavefunction of an in-state (an out-state is obtained by complex conjugation). This is defined as in \ref{eq:scatstate} by a long evolution of a momentum eigenstate $e^{-iH T}|p\rangle$ in the limit $T\to\infty$ (choosing $p=-\omega$ to solve the constraint in the asymptotic `free' region $q\to-\infty$, and normalising similarly to \eqref{eq:psiIn}). We could immediately write down the answer as a WKB wavefunction, but instead we will take the opportunity to illustrate methods which generalise more readily.

Anticipating that we will compute inner products of invariant states using the $p=0$ gauge-fixing, we will initially compute the wavefunction in the momentum basis, and restrict to evaluating at $p=0$. That is, we compute the matrix element $\langle p=0|e^{-iH T}|p=-\omega\rangle$ in the large $T$ limit. We can compute this by a (configuration space) path integral on the interval $t\in [-T,0]$ (holding the final time fixed to get a limiting classical solution $q_c(t)$ when we take $T\to\infty$ at fixed $t$). The initial and final states $|p\rangle$ are specified by Neumann boundary conditions $\dot{q}(-T) = \omega$, $\dot{q}(0) = 0$. The corresponding action is
\begin{equation}\label{eq:Sin}
    S = \int_{-T}^0 \left(-\frac{1}{2}\dot{q}^2 -V(q)\right)dt -q(-T)\omega,
\end{equation}
where we have added a boundary term at $t=-T$ to get a good variational problem with the specified boundary conditions.

A semi-classical calculation begins by looking for a classical solution $q_c(t)$, here with boundary conditions $\dot{q}_c(0)=0$ and $\dot{q}_c(-T)=\omega$. The solution for $T\to\infty$ has $q_c(0)$ at the turning point $q_0$ (with $V(q_0)=0$) to give zero energy, so the initial point $q(-T)\sim -\omega T$ in the far asymptotic region has velocity $\omega$. Explicitly, this solves $\frac{1}{2}\dot{q}_c^2=V(q_c)$ so $\int_{q_c(t)}^{q_0} \frac{dq}{\sqrt{2V(q)}} = |t|$. The on-shell action is $-\int_{q_c(-T)}^{q_0} \sqrt{2V(q)}dq -\omega q_c(-T)$, which in the $T\to\infty$ limit becomes
\begin{equation}\label{eq:Sinclass}
    S_\mathrm{in} = -\int_{-\infty}^{q_0} (\sqrt{2V(q)}-\omega)dq -q_0\omega.
\end{equation}
This is the WKB phase at the turning point of a zero energy scattering state.

Loop corrections are accounted for by expanding fluctuations around the classical solution, $q = q_\mathrm{c}+\delta$. At one loop we need only the quadratic action $-\frac{1}{2}\int_{-\infty}^0(\dot{\delta}^2 + V''(q_c(t)) \delta^2 )dt$, and we integrate over $\delta(t)$ with initial and final boundary conditions $\dot{\delta}=0$ (since we are working with states of fixed momentum). This action is that of harmonic oscillator with time-dependent frequency (and wrong sign kinetic term), with Hamiltonian $H_\mathrm{one-loop}=-\frac{p^2}{2}+\frac{1}{2}V''(q_c(t))\delta^2$. The corresponding time-dependent Schr\"odinger equation is solved by
\begin{equation}\label{eq:psioneloop}
    \psi_\mathrm{one-loop}(\delta,t) = \frac{1}{\sqrt{ f(t)}} \exp\left(\frac{1}{2i\hbar} \frac{\dot{f}(t)}{f(t)}\delta^2\right),
\end{equation}
where $\ddot{f}(t) = V''(q_c(t))f(t)$. By differentiating the classical equation of motion $\ddot{q}_c = V'(q_c)$, one solution is $f = \dot{q}_c$, and this gives the correct initial constant wavefunction (zero momentum for $\delta$, normalised as $\frac{1}{\sqrt{\omega}}$) as $t\to-\infty$. The $\delta'(0)=0$ boundary condition is implemented by taking the overlap with the final zero-momentum wavefunction at time $t=0$ (i.e., integrating over $\delta$ with a factor $(2\pi\hbar)^{-1/2}$), which gives us the result for the one-loop determinant $\frac{1}{\sqrt{i  \dot{f}(0)}} = \frac{1}{\sqrt{i  V'(q_0)}}$.
So, we get the one-loop result 
\begin{equation}\label{eq:psiinSC}
   \psi_\mathrm{in}(p=0) \sim  \frac{\exp(\frac{i}{\hbar}S_\mathrm{in})}{\sqrt{ i  V'(q_0)}}
\end{equation}
for the momentum-space wavefunction evaluated at $p=0$.

Our semi-classical calculation generalises straightforwardly to compute the momentum-space in-state wavefunction at different values of $p$ with $|p|<\omega$. The action \eqref{eq:Sin} gets an additional boundary term $-p q(0)$, and we look for classical solutions with $\dot{q}(0)=-p$. We use the same solution up to a time-translation to obey the boundary condition, so that $V(q(0))=\frac{p^2}{2}$. The on-shell action is similar to \eqref{eq:Sinclass} with upper integration limit given by the endpoint $q(0)$, and the addition of the boundary term $-p q(0)$. For $p>0$ we need to integrate from $-\infty$ to the turning point $q_0$, and then back to the endpoint after going to the other branch of the square root (with $\dot{q}<0$) to follow the reflected trajectory. For small momentum, the resulting action is 
\begin{equation}\label{eq:Ssmallp}
    S(p) = S_\mathrm{in} - p q_0 - \frac{p^3}{6V'(q_0)} + \cdots.
\end{equation}

We can check  this result by comparing with the WKB approximation, which gives the position space wavefunction
\begin{equation}\label{eq:psiinWKB}
    \psi_\mathrm{in}(q) \sim e^{\frac{i}{\hbar}S_\mathrm{in}+i\frac{\pi}{4}} \frac{2\cos\left(\frac{1}{\hbar}\int_q^{q_0} \sqrt{2V}-\frac{\pi}{4}\right)}{(2V(q))^\frac{1}{4}}
\end{equation}
for $q<q_0$. This is normalised to give $\psi_\mathrm{in}\sim \frac{1}{\sqrt{\omega}}e^{-\frac{i}{\hbar}\omega q} + \cdots$ at $q\to-\infty$, where $\cdots$ indicates the reflected part $\propto e^{+\frac{i}{\hbar}\omega q}$. Near the turning point $q=q_0$, this becomes the Airy function
\begin{equation}
    \psi_\mathrm{in}(q)\sim e^{\frac{i}{\hbar}S_\mathrm{in}} \sqrt{\frac{2\pi \hbar}{i V'(q_0)}} \lambda \Ai\left(\lambda \left(q-q_0\right)\right),
\end{equation}
where $\lambda^3 = \frac{-2V'(q_0)}{\hbar^2}$. This is precisely the Fourier transform of our momentum space result \eqref{eq:Ssmallp} at small $p$, including the correct phase and one-loop factor from \eqref{eq:psiinSC}.

We can also directly calculate the position-space wavefunction $\psi_\mathrm{in}(q) \propto \lim_{T\to\infty}\langle q|e^{-iH T}|p\rangle$ similarly, by imposing a Dirichlet boundary condition fixing $q(0)$. In the classically-allowed region $q<q_0$ there will be two classical solutions contributing: one coming directly from infinity and one reflected from the potential (again, these always come from the same underlying solution up to time-translation). These contribute the two exponentials in the cosine in \eqref{eq:psiinWKB}. The one-loop calculation is similar, except the different boundary conditions mean that we evaluate the one-loop wavefunction \eqref{eq:psioneloop} at $\delta=0$ to give the usual WKB denominator proportional to $(\dot{q})^{-1/2}=(\pm 2V(q))^{-1/4}$.

For $p^2>\omega^2$ or for $q>q_0$  there are no real classical solutions. A semi-classical analysis in that case requires complex solutions (which also can contribute important non-perturbative corrections to the calculations we have already considered). We discuss this in section \ref{ssec:nonpert}.

\subsubsection{Zero momentum gauge-fixing}

We now examine the sesquilinear form $\kappa_p$ obtained in section \ref{sssec:p0gauge} from the gauge-fixing fermion $\Psi=-i\hat{b}p$, which we expect to give us a physical inner product, interpreted as coming from imposing the gauge condition $p=0$. We found the explicit position-space expression (here restoring $\hbar$)
\begin{equation}\label{eq:kappapSC}
    \li \psi_2|\psi_1\ri_p = -\frac{1}{2\pi \hbar}\int dq_1 dq_2\,\psi_2^*(q_2)\frac{V(q_1)-V(q_2)}{q_1-q_2} \psi(q_1).
\end{equation}
This does not localise to the $p=0$ component of the states $\psi_{1,2}$ in any obvious way. Here we see how to nonetheless evaluate it from the $p=0$ value of the invariant states in a semi-classical limit, order-by-order perturbation theory.

The crucial observation is that \eqref{eq:kappapSC} will be dominated by the vicinity of the classical turning point $q_0$ where $V(q_0)=0$ in the semi-classical limit. For the classically allowed region $q<q_0$ the integrals over $q_{1,2}$ will be highly oscillatory, while for the classically forbidden region  $q>q_0$ the wavefunction will be exponentially small. Zooming in on this region, the factor $\frac{V(q_1)-V(q_2)}{q_1-q_2}$ in the vicinity of this turning point becomes simply a constant $V'(q_0)$  to leading order, and we get
\begin{equation}\label{eq:kappapLO}
    \li \psi_2|\psi_1\ri_p \sim (-V'(q_0)) \left(\int_{-\infty}^\infty \frac{dq_2}{\sqrt{2\pi\hbar}} \psi_2^*(q_2)\right)  \left(\int_{-\infty}^\infty \frac{dq_1}{\sqrt{2\pi\hbar}} \psi(q_1)\right).
\end{equation}
The integrals in the brackets give precisely the momentum-space wavefunctions  evaluated at $p=0$, which we write as $\psi_{1,2}(p=0)$. So this one-loop order result is given by evaluating the invariant wavefunctions on the gauge-fixing surface $p=0$, along with the classical value of the Faddeev-Popov determinant (with operator ordering ambiguities not arising at this order in $\hbar$). In a system with more degrees of freedom, we would also be left with an integral over the unfixed transverse variables.

We can use this to evaluate the norm of the in-state $\li \psi_\mathrm{in}|\psi_\mathrm{in}\ri_p$, using the result \eqref{eq:psiinSC} for the $p=0$ wavefunction: the measure factor of $-V'(q_0)>0$ cancels the one-loop determinant to get unit norm, agreeing with the result from group averaging as computed in section \ref{ssec:1Dtarget}. We can also compute the overlap between in-state and out-state  (with momentum-space wavefunction $\psi_\mathrm{out}(p)=\psi_\mathrm{in}^*(-p)$) obtaining $\li \psi_\mathrm{out}|\psi_\mathrm{in}\ri_p \sim i \exp(\frac{2i}{\hbar}S_\mathrm{in})$, which is the correct scattering phase to order $\hbar^0$.


\subsubsection{Higher orders in perturbation theory}

Now we would like to illustrate the strategy for extending this to higher order in perturbation theory. The philosophy here is that we are given only the zero momentum component $\psi^{p=0}$ of the wavefunctions in a semi-classical expansion (perhaps by extending the above analysis to higher loops), and we would like to compute the inner product only in terms of this data. For this, the first step is to Taylor expand $\frac{V(q_1)-V(q_2)}{q_1-q_2} $ around the turning point:
\begin{equation}\label{eq:VVexpansion}
    \frac{V(q_1)-V(q_2)}{q_1-q_2} \sim \sum_{n=1}^\infty \frac{V^{(n)}(q_0)}{n!} \frac{(q_1-q_0)^n-(q_2-q_0)^n}{q_1-q_2} = V'(q_0) + \frac{1}{2}V''(q_0) (q_1+q_2-2q_0)+\cdots \,.
\end{equation}
The $n$th term in this expansion will contribute at $n$-loop order in perturbation theory, with insertions of powers $(q-q_0)^{n-1}$ with $q=q_1$ or $q_2$. This gives us integrals of the form $\int (q-q_0)^k \psi(q) dq$, which is not determined only by the momentum wavefunction at $p=0$, but  by its $k$th derivative there $\psi^{(k)}(p=0)$. These derivatives are not part of the given data, but we can nonetheless obtain them by using the fact that our wavefunctions solve the Wheeler-DeWitt equation, perturbatively near $q=q_0$ (or at small $p$ in momentum space). Importantly, we need only local data (derivatives of the potential at $q=q_0$).

Let's see how to do this explicitly, choosing a solution normalised so that $\psi^{p=0}=\frac{1}{\sqrt{2\pi\hbar}}$ (so $\int_{-\infty}^\infty \psi(q) dq =1$). Expanding near the turning-point, the wavefunction will be given by an Airy function (or we can use the semi-classical momentum space result \eqref{eq:Ssmallp}):
\begin{equation}
    \psi(q) \sim \lambda \Ai(\lambda(q-q_0)), \qquad \psi(p)\sim \frac{1}{\sqrt{2\pi\hbar}} e^{i \frac{p^3}{3\lambda^3}-\frac{i}{\hbar} p q_0}
\end{equation}
where $\lambda^3=\frac{-2V'(q_0)}{\hbar^2}$. This suggests writing the Wheeler-DeWitt equation in terms of the variable $u=\lambda(q-q_0)$ and expanding for large $\lambda$ with $u$ held fixed, which gives
\begin{equation}
    \frac{d^2\psi}{du^2} - u\psi(u) = \sum_{n=2}^\infty \lambda^{1-n} \frac{V^{(n)}(q_0)}{n! V'(q_0)} u^n\psi(u) =  \frac{1}{\lambda} \frac{V''(q_0)}{2V'(q_0)} u^2\psi(u) + \cdots\,.
\end{equation}
Now if we make the perturbative ansatz $\psi(u) \sim \sum_{n=0}^\infty \lambda^{1-n} \psi_n(u)$ with $\psi_0(u) = \Ai(u)$, at fixed order in $\lambda$ we will have an  equation giving $\psi_n''(u)-u \psi_n(u) $ in terms of lower-order $\psi_k$ with $k<n$. This will have a unique solution once we fix $\int \psi_n(u)du=0$ (one coefficient in the two-dimensional space of solutions is fixed by decaying at $u\to\infty$, and the remaining coefficient of $\Ai(u)$ is fixed by the integral condition fixing the $p=0$ wavefunction\footnote{These integrals may not converge due to growing rapidly oscillating tails as $u\to-\infty$, but they are nonetheless well-defined in a distributional sense (defined by the Fourier transform evaluated at $p=0$, for example). No such issue appears if we work directly in momentum space.}). To first order, we get
\begin{equation}
    \psi_1(u) = \frac{V''(q_0)}{10V'(q_0)}\left(u^2\Ai'(u)-u\Ai(u)\right)= \frac{V''(q_0)}{10V'(q_0)}\left(\Ai^{(5)}(u)-5\Ai''(u)\right).
\end{equation}
Higher order solutions can all be similarly written as sums of derivatives of $\Ai(u)$. In fact, it is rather nicer to do this calculation in momentum space (and more directly compatible with semi-classical calculations); we do this systematically in appendix \ref{app:pertWDW}.

The upshot is that the required derivatives $\psi^{(k)}(p=0)$ are determined order-by-order in perturbation theory in terms of the original data $\psi(p=0)$ by solving the Wheeler-DeWitt equation, in terms of local data (derivatives of the potential at $q_0$). Using this  data in the expansion of \eqref{eq:VVexpansion} in turn determines the inner product $\li \psi_2|\psi_1\ri_{p}$, from only the data $\psi_{1,2}(p=0)$ and a finite number of derivatives of $V$ at the turning point at each order in perturbation theory. We explain how to systematically carry this out in appendix  \ref{app:pertWDW}, finding a first correction at order $\hbar^2$:
\begin{equation}
    \frac{\li \psi_2|\psi_1\ri_{p}}{\psi_2^*(p=0)\psi_1(p=0)} = -V'(q_0)\left(1 - \frac{3 V''(q_0)^3+V^{(4)}(q_0) V'(q_0)^2-4 V^{(3)}(q_0) V'(q_0) V''(q_0)}{12  V'(q_0)^4} \hbar^2 +\cdots \right).\nonumber
\end{equation}
We also give two examples in the appendix, including one (an exponential potential) with no perturbative corrections to the leading $-V'(q_0)$ result at any order in $\hbar$, but nonetheless with non-perturbative corrections.

\subsubsection{Gaussian average gauge fixing}

Above, we have used $p=0$ gauge-fixing to evaluate inner products of invariant states. We can also make use of this gauge fixing as the map $\kappa_p$ to construct gauge-fixed co-invariant representatives of a state (as discussed in sections \ref{ssec:GFintro}, \ref{ssec:coinvGF}). However, $\kappa_p$ does not result in a state which is well-localised in position space: it results in a wavefunction proportional to $\frac{V(q)}{q-q_0}$ (to leading order in $\hbar$). To get a state which is well-localised near the turning point in position space (as well as near $p=0$ in momentum space), we can instead use the Gaussian averaged version $\kappa_{p,\xi}$ of the gauge-fixing map.

We evaluate $\kappa_{p,\xi}$  acting on the in-state semi-classical wavefunction (in position space) using the result \eqref{eq:kappapxi}. This results in the integral
\begin{equation}
    \kappa_{p,\xi}\psi_\mathrm{in}(q)\sim -\frac{e^{\frac{i}{\hbar} S_\mathrm{in}+i\frac{\pi}{4}}}{2\pi\hbar}\int dq' \frac{V(q)-V(q')}{q-q'}\frac{1}{(-2 V(q'))^\frac{1}{4}}e^{-\frac{\xi}{2\hbar}(q-q')^2 - \frac{1}{\hbar}\int_{q_0}^{q'}\sqrt{-2V(x)}dx},
\end{equation}
where we have written the state so that the action is manifestly real in the classically forbidden region $q'>0$. For $q'<0$ we should include both possible branches $q'+i\epsilon$ (the incoming part) and $q'-i\epsilon$ (the reflected part).

 We now evaluate this by saddle-point, with the saddle point equation giving $\xi(q-q')=-\sqrt{-2V(q')}$. For $q$ close to $q_0$ (specifically, $|q-q'|\ll \xi^{-2}$), we find a saddle in the classically forbidden region at $q'-q_0\sim -\frac{\xi^2}{2 V'(q_0)} (q-q_0)^2>0$, with  action $-\frac{1}{2}\xi (q-q_0)^2$. So performing the one-loop integral around this saddle 
gives us a Gaussian wavefunction
\begin{equation}
    \kappa_{p,\xi}\psi_\mathrm{in}(q)\sim e^{\frac{i}{\hbar} S_\mathrm{in}+i\frac{\pi}{4}}  \sqrt{\frac{-V'(q_0)}{2\pi\hbar}}e^{-\frac{\xi}{2\hbar}(q-q_0)^2}
\end{equation}
close to its peak. To verify this, we can integrate against $\psi_\mathrm{in}^*$ (which is simplest if we first Fourier transform to momentum space so we avoid needing the Airy function) to recover the norm $\li \psi_\mathrm{in}|\psi_\mathrm{in}\ri$.

This does not agree with our general result in \eqref{eq:kappaxiSC}. The reason is that the semi-classical form of the position space wavefunction breaks down at the turning point $q_0$ where the gauge condition $p=0$ is satisfied. However, it does precisely agree if we instead work in momentum space.

\subsubsection{Non-perturbative effects}\label{ssec:nonpert}

We finish our discussion of the semi-classical approximation with examples of interesting non-perturbative effects in the group average inner product. These also serve as simple examples where our strategy of imposing constraints as operator in a quantum theory gives different results to the alternative of imposing constraints classically before quantising.

To illustrate the ideas in a simple setting, we will study the model with quadratic potential $V(q) = \frac{1}{2}\lambda^2(q^2-a^2)$. This has two distinct classically allowed regions $q<-a$ and $q>a$, separated by the barrier $|q|<a$ where $V(q)<0$. We will be interested in contributions to the group average inner product which come from tunnelling under this barrier.

First, we  compute the propagator $\langle q|e^{-i H t/\hbar}|q'\rangle$. For any endpoints $q,q'$ and any time $t$, there is a unique classical solution to $\ddot{q} =\lambda^2 q$ with $q(0)=q'$ and $q(t)=q$. The on-shell action of this solution is
\begin{equation}
	S(t) = \tfrac{1}{2}a^2\lambda^2 t-\lambda\frac{(q^2+q'^2)\cosh\left(\lambda t\right)-2q' q}{2\sinh\left(\lambda t\right)},
\end{equation}
and including a one-loop determinant gives us the semi-classical approximation for the propagator (which happens to be exact for the quadratic potential):
\begin{equation}
    \langle q|e^{-\frac{i}{\hbar} H t} |q'\rangle = \sqrt{\frac{\lambda}{-2\pi i \hbar \sinh(\lambda t)}} e^{\frac{i}{\hbar}S(t)}.
\end{equation}

However, these are not generally solutions of the full gravitational theory because they do not solve the constraint $H=0$. This is imposed when we group average, since the integral over $t$ is dominated semi-classically by  saddle-points where $S'(t)=-E=0$. The energy is
\begin{equation}
	E(t) =-S'(t) =  -\tfrac{1}{2}a^2\lambda^2 -\lambda^2\frac{(q^2+q'^2)-2 \cosh\left(\lambda t\right)q' q}{2\sinh^2\left(\lambda t\right)}.
\end{equation}

Now if $q,q'<-a$ (so both endpoints lie in the same classically allowed region), there are four real solutions to $E(t)=0$. If we write $q=-a \cosh(\lambda t_1)$ and $q'=-a \cosh(\lambda t_2)$ (so that $t_{1,2}$ are the times for $E=0$ solutions to get from $q,q'$ to the turning point), then these real solutions are $t=\pm t_1\pm t_2$, where the signs can be chosen independently.

However, in other cases there are no real solutions: for example, when $q=-a \cosh(\lambda t_L)<-a$ and $q'=a\cosh(\lambda t_R)>a$ (for $t_L,t_R>0$) so the endpoints are separated by the potential barrier. Nonetheless, we have the group average inner product written as an integral over real $t$, which correspond to off-shell Lorentzian geometries (off-shell because they do not solve the constraint, which is the $g_{tt}$ equation of motion from varying the lapse), and this integral is nonzero. We can compute it by deforming the real $t$ defining contour into the complex plane.

\begin{figure}
    \centering
    \includegraphics[width=0.5\linewidth]{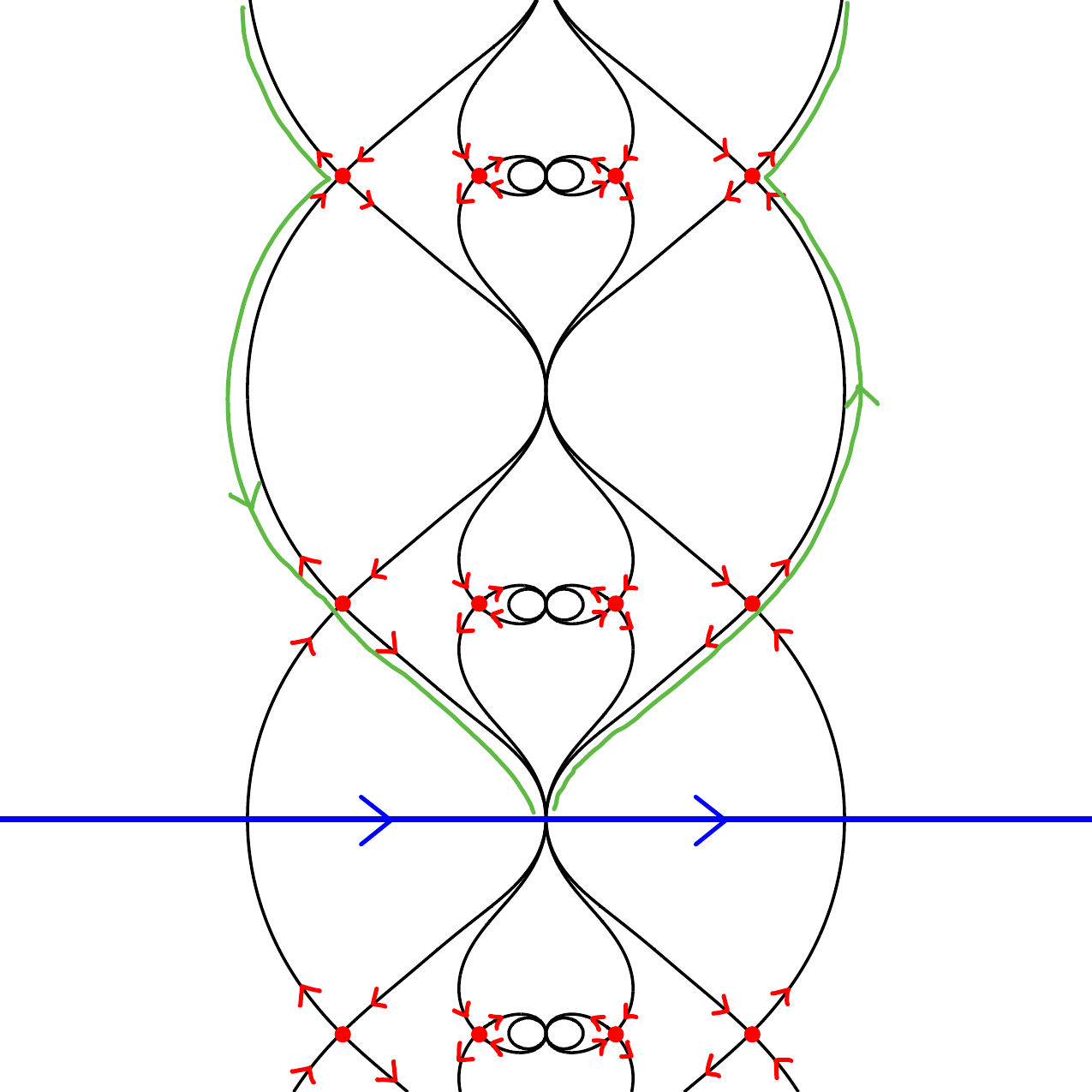}
    \caption{The complex $t$ plane for the group averaging integral computing $\lc q|q'\rc$, where $q,q'$ are separated by a potential barrier. Red dots indicate saddle points $E=0$. Steepest descent contours are shown in black, with the red arrows indicating the descent direction (i.e., the direction of increasing $\Im S$). The figure is periodically repeated in the imaginary direction. The original real $t$ contour of integrating is in blue. We can deform this to the green contour, since $\Im S$ is large and positive as $\Im t\to +\infty$.}
    \label{fig:SteepestDescent}
\end{figure}

In this parameterisation, the energy can be written as
\begin{equation}
    E(t) = -a^2\lambda^2 \frac{(\cosh(\lambda t)+\cosh(\lambda (t_L+t_R)))(\cosh(\lambda t)+\cosh(\lambda (t_L-t_R)))}{2\sinh^2\left(\lambda t\right)}.
\end{equation}
From this we see that there are four families of solutions to $E(t)=0$, namely $t=\pm t_L \pm t_R + \frac{i\pi}{\lambda}(1+2n)$, where the signs are chosen independently and $n$ is an integer. Importantly, the group averaging interpretation of our path integral means that we know the defining Lorentzian (real $t$) contour,\footnote{In our case, this integral over real lapse (after doing all other integrals) is convergent. In other cases (such as target spaces with sufficiently high dimension) this may not be true because of the singularity at $t=0$, and more care is required to identify the correct prescription. A closely related point is discussed in \cite{Banihashemi:2024aal}.} so we can identify the correct way to deform the integration contour to a path of steepest descent, and identify which of these complex saddle-points contribute (and with what sign or phase). The appropriate deformation is shown in figure \ref{fig:SteepestDescent}.

The integrand is suppressed (i.e., $S$ has positive imaginary part) in the upper half-plane $\Im t>0$, so we deform the contour in that direction and find that the appropriate contour passes through the saddle-points at $t=\pm(t_L + t_R) + \frac{i\pi}{\lambda}(1+2n)$ for $n\geq 0$. We can interpret the $n=0$ solutions as a trajectory which travels to the turning point in Lorentzian time (real lapse), tunnels under the barrier with a Euclidean geometry (imaginary lapse) taking imaginary time $i\frac{\pi}{\lambda}$, and then continues to the final point with another real Lorentzian evolution. Alternatively we can choose an equivalent complex path (for example with constant complex lapse) and deformations can be thought of as complex-valued diffeomorphisms: the invariant definition of a solution is a homotopy class of paths on the Riemann surface $H(q,p)=0$ with complex $q,p$.

The integral is dominated by the two saddle points at $t=\frac{i\pi}{\lambda}\pm (t_1+t_2)$, which gives us
\begin{equation}
    \lc q|q'\rc \sim \frac{2}{a\lambda\sqrt{\sinh(\lambda t_L)\sinh(\lambda t_R)}}e^{-\frac{1}{2\hbar} a^2\pi\lambda} \cos\left(\frac{a^2\lambda}{4\hbar}(\sinh(2\lambda t_L)-2\lambda t_L+\sinh(2\lambda t_R)-2\lambda t_R)\right).
\end{equation}
In particular, we do not have contributions from the saddle points at $t=\frac{i\pi}{\lambda}\pm (t_1-t_2)$, which are equally dominant.

Even in the case when there are real saddle-points, there can be additional non-perturbative contributions from complex saddle-points with important physical implications. In the case $q=-a \cosh(\lambda t_1)<-a$ and $q'=-a \cosh(\lambda t_2)<-a$, a similar analysis shows that in addition to four real saddle-points at $t=\pm t_1 \pm t_2$, the steepest descent contour passes through complex saddle-points at $t=\pm (t_1+t_2) + \frac{2\pi i}{\lambda}n$ for $n=1,2,3,\cdots$. Despite providing exponentially small corrections (scaling as $e^{-\frac{1}{\hbar} a^2\pi\lambda}$), they have a qualitatively important effect. The real saddles give a result which factorises as $\lc q|q'\rc\sim \psi(q)\psi^*(q')$ to all orders in perturbation theory, indicating that all the states $|q\rc$ for $q<a$ are physically equivalent up to normalisation. But the leading non-perturbative contributions violate this factorisation, indicating that the physical Hilbert space is in fact two-dimensional (since they describe mixing with the classical solution on the other side of the barrier).

An important and general feature is that we do not have any contribution from the paths with $t$ in the lower half-plane, which have $\Im S<0$ and hence would give exponentially large results. It is clear that such saddles can never lie on the contour of steepest descent, since the defining integral runs over a contour with oscillatory integrand, so the integral must be suppressed in the limit $\hbar\to 0$ (never enhanced).

An equivalent way to express the fact that the contributing saddles all lie in the upper half-plane $\Im t\geq 0$ is that the lapse has (or rather, can be chosen to have) non-negative imaginary part $\Im N\geq 0$. This is a very simple version of the Kontsevich-Segal-Witten criterion for allowable complex metrics \cite{Kontsevich:2021dmb,Witten:2021nzp}. This raises a question of whether such a result holds more generally: if we begin with a Lorentzian gravitational path integral and deform to its the steepest-descent contour, does any saddle-point along the path necessarily obey this criterion? This seems necessary for there to be a consistent perturbation theory around the relevant complex spacetimes.

Note that if we impose the constraints classically in this example, we eliminate these interesting non-perturbative effects entirely. The symplectic quotient phase space consists of two points, corresponding to the solutions on the left or right side of the barrier. Quantising this gives a two-dimensional Hilbert space, but there is no sense in which the states on the left and right can mix.

Non-perturbative effects can have a yet more dramatic effect, even changing the dimension of Hilbert space. For example, take a cubic potential $V(q)= -\lambda (q-q_1)(q-q_2)(q-q_3)$ with $\lambda>0$ and three real roots $q_1<q_2<q_3$. Now the classically allowed region consists of $q<q_1$, as well as the finite interval $q_2<q<q_3$. In the latter region, there is a classically bound orbit giving a second point in the constrained phase space. So constraining first would lead to a two-dimensional Hilbert space. But non-perturbative effects render this bound state metastable, and in fact the  Hilbert space has a single state. It would be an interesting exercise to understand the details of this from a semi-classical perspective.


More generally (especially in realistic non-renormalisable theories of gravity with a complicated perturbation theory), one might worry that exponentially small non-perturbative corrections are negligible and should not be taken seriously if we do not completely understand the high-order perturbative behaviour. The examples in this section illustrate a more general principle of why they are nonetheless informative: the interesting non-perturbative contributions are of a qualitatively different nature. In particular, perturbative corrections to the gravitational inner product are all local in superspace (and come from the small lapse/shift part of the group average integral). The non-perturbative corrections can be completely non-local, because they involve a finite amount of time evolution. This allows them to manifest phenomena which are invisible in perturbation theory, such as tunnelling in the examples above, or (as commented on in section \ref{ssec:null}) non-local contributions to the inner product in black hole interiors inducing null states

\section{Discussion}\label{sec:disc}

\subsection{Additional considerations in higher dimensions}

\subsubsection*{More constraints}

The most obvious complication in higher dimensions is that we must deal with many constraints: we now have a Hamiltonian constraint $\mathcal{H}(x)$ for every spatial point $x$, and we also have momentum constraints $\mathcal{P}_i(x)$. There are correspondingly many choices in how we account for the constraints in the quantum theory. We can define physical wavefunctions as invariants (solutions to the Wheeler-DeWitt equation), co-invariant equivalence classes, or a mixture. There is probably no universal best strategy: it is useful to have this freedom in mind to make whatever choice is most convenient or illuminating for a given problem of interest.

In the context of perturbation theory around a static background (such as Minkowski space), we can use a quantisation scheme which makes close contact with approaches familiar from gauge theory (or the string worldsheet). This involves dividing the constraints into (non-Hermitian) positive frequency parts  and their conjugate negative frequency parts with respect to time translation (under a conveniently gauge-fixed Hamiltonian). We then choose physical states which are annihilated by the positive frequency parts (invariance), and impose equivalence relations under negative frequency parts (co-invariance). Typically, these respectively remove unphysical transverse and longitudinal polarisations of photons/gluons/gravitons. In BRST language, this means choosing representatives of cohomology classes which are  annihilated by the positive frequency part of ghost operators. See appendix \ref{app:Maxwell} for details of this in abelian gauge theory. The subtleties of inner products that have been the main subject of this paper are absent in this case, since we impose invariance under non-Hermitian combinations of constraints as in the simple example in  section \ref{ssec:imN}.

It is reasonable to try to generalise such an approach to perturbation theory around general backgrounds, as discussed in \cite{Giddings:2022hba}. However, in time-dependent backgrounds (e.g., for cosmology or black holes) there may not be any natural or obvious choice of positive-frequency modes, so there is a great deal of arbitrariness in this quantisation, and other approaches my be useful. Even for a simple case of static closed universes, only modes of non-zero spatial momentum divide naturally into positive/negative frequency parts, and the zero-mode constraints must be dealt with separately (perhaps by the methods described in this paper). Such a hybrid description of the Hilbert space (positive/negative frequency split for some modes, co-invariant representation for some other modes) is likely to by the most practical in many cases. For example, for perturbation theory around homogeneous cosmological spacetimes it may be useful to use the methods of this paper to treat the zero modes of constraints, while finding a positive/negative frequency split for spatially inhomogeneous perturbations (e.g., describing gravitons).

\subsubsection*{Field-dependent structure constants}

The additional challenges of constraints in more realistic models of gravity is not primarily from their number, but rather from their nature. Their algebra is not only non-abelian, but also has field-dependent structure constants. Specifically, the Poisson bracket of two Hamiltonian constraints depends on the inverse spatial metric $\gamma^{ij}$:
\begin{equation}
	 \left\{ \int N_1 \mathcal{H},\int N_2 \mathcal{H} \right\} = \int \gamma^{ij}(N_1 \partial_j N_2-N_2 \partial_j N_1) \mathcal{P}_i \,.
\end{equation}
This leads to additional technical complications, for example when defining measures for group averaging. The usual Faddeev-Popov procedure does not immediately work because the field-dependence means that gauge transformations do not form an ordinary group.

Fortunately, the  BFV/BRST formalism takes care of these generalisations rather elegantly. For gravity in particular, the BRST charge continues to take the form familiar from gauge theories, namely $c$ times constraints plus a term with $bcc$ contracted into structure constants (for more general algebras there can be terms with more ghosts, $b^2c^3$ etc.~\cite{Henneaux:1985kr}):
\begin{equation}
    Q = \int d^dx\left[c \mathcal{H} + c^i_\perp \mathcal{P}_i - i b_i^\perp c_\perp^j\partial_j c_\perp^i+ib c_\perp^i\partial_i c + i \gamma^{ij} b_i^\perp c\partial_j c \right].
\end{equation}
This is valid classically (with $\{Q,Q\}=0$); for the quantum BRST charge we have to resolve operator ordering in a way which depends on details of the quantisation. With $Q$ in hand, we can continue to define physical states as the cohomology,  pure gauge operators as BRST exact commutators $[\Psi,Q]$, and so forth (adding also $\int(\hat{c}\Pi + \hat{c}_\perp^i \Pi^\perp_i)$  for the primary constraints in analogy to \eqref{eq:QBFV}, where $\Pi$ and $\Pi_i^\perp$ are the momenta conjugate to lapse $N$ and shift $N^i_\perp$ respectively). 

For example, to compute a group averaged inner product (in synchronous gauge $\dot{N}=\dot{N}^i_\perp=0$)  we might use the gauge-fixing fermion $\Psi = -i\int (bN+b_i^\perp N_\perp^i)$, generalising section \ref{sssec:GA}. With this, $\{Q,\Psi\}$ has bosonic part $-\int (N\mathcal{H}+ N_\perp^i\mathcal{P}_i)$, generating a diffeomorphism.\footnote{In the quantum theory, ordering constants required by Hermiticity of $Q$ might also give rise to shifts in $\mathcal{H},\mathcal{P}_i$. This is similar to the case of non-unimodular Lie groups, where the left-invariant and right-invariant Haar measures differ so the correct measure for the group average is not so obvious, and the Dirac/Wheeler-DeWitt condition for invariant states needs to be modified \cite{Giulini:1998kf}. This is automatically incorporated in the BRST formalism, where demanding Hermiticity of $Q$ requires a quantum shift in the constraint term \cite{Shvedov:2001ai}.}  But the ghost part will contain a term of the form $\gamma^{ij} N b_i^\perp \partial_j c $ from the last term in the BRST charge: this gives us a measure for the group average integral which depends on $\gamma^{ij}$.

Another example of the complications of field dependent structure constants arises in the maximal-volume gauge-fixing $\Tr K =0$. The corresponding pure gauge generator $\{Q, \int \hat{b} \Tr K\}$ contains a term proportional to $\int \gamma^{ij} \hat{b} b_i^\perp c\partial_j c$, quadratic in ghosts. Presumably this will be necessary to retain gauge-invariance in the ideas of \cite{Witten:2022xxp}  beyond one loop in perturbation theory.

\subsubsection*{Comparison with previous approaches}

Here we comment on the relation of our ideas with some recent work describing gravitational Hilbert spaces (and inner products in particular).

Precisely the BRST formalism we have discussed here was recently applied in \cite{Henneaux:2025ocw}. The problem in question was BMS symmetry in asymptotically flat space, though the formalism is much more generally applicable so this may serve as a useful reference to application of our ideas in Einstein gravity and beyond.

The Wheeler-DeWitt Hilbert space in perturbation theory around AdS was discussed in \cite{Chowdhury:2021nxw}. The basic idea was to split the perturbations $h$ of the spatial metric $\gamma$ into longitudinal ($h^L$), trace ($h^T$), and transverse-traceless ($h^{TT}$) components, and similarly for the conjugate momentum $\pi$. The first two are unphysical gauge modes (longitudinal and timelike polarisations of gravitons), while the final component is physical. An inner product was proposed which involves fixing $h^L$ and $\pi^T$ at any prescribed values, and integrating over $h^{TT}$. We can think of this as the $\kappa$ map induced by choosing $\chi=(h^L-h^L_0,\pi^T-\pi^T_0)$ (similar but not quite identical to the maximal-volume gauge, along with a transverse gauge to fix spatial coordinates). To leading order in perturbation theory, the momentum and Hamiltonian constraints are linear differential operators acting on $\pi^L$ and $h^T$ respectively: this means that the Faddeev-Popov factor arising from this gauge-fixing is a field-independent constant (coming from determinants of the differential operators in question). This explains why no measure factor was necessary to get a good inner product at this order, but we expect that modifications would be required at higher orders in perturbation theory. At this linearised order in constraints, gravity behaves much like the case of abelian gauge theory (see appendix \ref{app:Maxwell}).

\subsection{Backgrounds with symmetries, including de Sitter space}

We are often interested in backgrounds which preserve some symmetry group $G$, and perturbation theory around such a backgrounds comes with additional subtleties. Classically, there are linearisation stability constraints \cite{Moncrief:1975xtw}. This means that not every solution to the Einstein equation linearised around the background corresponds to a family of exact solutions: there is a constraint that perturbations should have vanishing charges which generate the group $G$.

In the quantum theory, the $G_N\to 0$ limit should not simply reduce to QFT on a fixed spacetime, but we should also gauge by $G$ (since the isometries are the diffeomorphism symmetries which are not spontaneously broken by the background). This naturally shows up when we calculate an inner product via group averaging, since the group average includes an integral over the isometries, and these cannot be accounted for by any gauge-fixing in the symmetric background. The result is that after taking gauge fixing into account, at $G_N\to 0$ we are still left with a group average integral over $G$.

A prominent example is de Sitter space, with a non-compact $SO(d+1,1)$ isometry group. Indeed, an early example of group averaging was the application to de Sitter at $G_N=0$ (where we have QFT on a fixed dS background and gauge only the isometries) \cite{Higuchi:1991tk,Higuchi:1994vc}. Further aspects of this $SO(d+1,1)$ group average were studied and applied in \cite{Marolf:2008hg,Chandrasekaran:2022cip}. A potentially confusing aspect is that the Bunch-Davies vacuum state of QFT on de Sitter (treated as a representative of a co-invariant state) is non-normalisable in the usual group average norm, because it is invariant under the isometries so the integral gives an infinite factor of the volume of $SO(d+1,1)$. States with few particles are similarly non-normalisable \cite{Marolf:2008hg}. On the other hand, if we treat the Bunch-Davies state as an invariant state (or similarly for suitable de Sitter invariant two-particle wavefunctions, for example), then we conclude that the state has vanishing norm with respect to any of our Klein-Gordon type $\kappa$ inner products: this comes from zero modes in the Faddeev-Popov measure (because no gauge-fixing condition can eliminate the isometries). Perhaps the best resolution is to treat such states separately, using a differently normalised inner product (which amounts to including a formally divergent or vanishing factor in the normalisation of such wavefunctions), with states requiring different inner products living in different superselection sectors. This idea is discussed in \cite{Marolf:2008hg} (see also \cite{Marolf:1995cn}).

We should also connect our perspective on gravitational Hilbert spaces to recent discussions of wavefunctions in de Sitter near asymptotic future infinity \cite{Chakraborty:2023yed} and in particular the inner product studied in \cite{Chakraborty:2023los}. These papers classified the solutions to the Wheeler-DeWitt equation (or invariant wavefunctions) in a limit of large conformal factor (taking spatial metrics $\gamma_{ij} = \Omega^2\hat{\gamma}_{ij}$ with $\Omega\to\infty$). These take the form $\Psi[\gamma] \sim e^{iS[\gamma]}Z[\gamma]$, where $S[\gamma]$ is a fixed phase (the same for every state) which becomes large in the $\Omega\to\infty$ limit (proportional to the volume of space $\int d^d x \sqrt{\gamma}$ at leading order), and $Z[\gamma]$ is any conformally-invariant functional (up to known anomaly terms). The conformal invariance emerges as a limit of the Hamiltonian constraint near asymptotic infinity. The inner product in \cite{Chakraborty:2023los} is evaluated by fixing a gauge, specifying conditions on $\gamma_{ij}$. This is a Klein-Gordon inner product: in our language, it is a $\kappa$ map obtained by a gauge-fixing $\chi(\gamma)$ depending only on configuration space variables $\gamma_{ij}$ (not momenta $\pi^{ij}$). From our general considerations, one should expect the  resulting inner product to include a measure factor which is linear in momenta $\pi^{ij}$, but in this limit there is a simplification: at large $\Omega$ the momentum is directly directly determined by the metric (thinking of $\pi^{ij}$ as $-i \frac{\delta}{\delta\gamma_{ij}}$, we need only keep the terms where it acts on the phase $e^{iS}$). The result is a positive-definite inner product, because the chosen gauge-fixing condition $\chi=0$ constitutes a good gauge for the large expanding universes under consideration.

Additionally, these papers relate this Wheeler-DeWitt `invariant' Hilbert space to the co-invariant Hilbert space and group averaging. Specifically, they consider wavefunctions constructed as a polynomial in metric perturbations (and matter fields) times the Hartle-Hawking state, which in our language can be thought of as representatives of co-invariant cosets. Group averaging (in this late time limit, integrating over spatial diffs and Weyl transformations) should then give back the invariant states above, and the group average inner product  $\eta$ should give identical results to the Klein-Gordon inner product in this regime. This was shown to hold for the $G_N\to 0$ limit in \cite{Chakraborty:2023yed,Chakraborty:2023los}, in which the gauge symmetry reduces to the $SO(d+1,1)$ isometry group.

 The particular issue of the norm of the Hartle-Hawking state was studied in \cite{Cotler:2025gui}. In our language, they were interested in the norm of the state $|\mathrm{HH}\ri$ in the invariant (Wheeler-DeWitt) Hilbert space. They approached this by constructing a resolution of the identity constructed from co-invariant states $|\gamma\rc$ (represented by delta-function wavefunctions of fixed spatial metric $\gamma$) for universes at late time. A resolution of the identity is inverse to the inner product, and the inner product $\lc\gamma'|\gamma\rc$ they use on co-invariants is equivalent to the group averaging $\eta$. Concretely, this resolution of the identity requires gauge-fixing the diffeomorphisms (which once again become spatial diffs and Weyl transformations at late time), just as for the $\kappa$ maps in the language of this paper. So, their identity $\hat{\mathbb{1}}$ is precisely an example of what we would call $\kappa$, and their justification (e.g., their equation (3.30)) corresponds to verifying the generalised inverse identity $\eta=\eta\kappa\eta$. For the Hartle-Hawking state, this leaves a residual unfixed gauge group of the de Sitter isometries, leading to zero norm. This reproduces the earlier results mentioned above, but now explicitly including one-loop gravitational fluctuations.

\subsection{Singularities: constraints are not essentially self-adjoint}

There is one feature of the gravitational constraints (appearing even for mini-superspace models) that we have not addressed at all so far: they typically fail to be essentially self-adjoint. In simple terms, they act as Hermitian operators on sufficiently nice (e.g., smooth, compactly supported) functions, but do not have a unique definition on a larger domain, usually because some boundary conditions need to be specified. While this may seem to be mostly a technical issue (preventing operators from being exponentiated to finite gauge transformations, for example), it points at an important physical property of gravity.

In fact, this quantum phenomenon is a symptom of a classical illness. The Cauchy problem in gravity does not allow us to evolve for an arbitrary amount of time (in a given gauge) for general initial data. Instead, we often encounter a singularity in finite time. These can be curvature singularities, or they can be coordinate singularities (occurring even in Minkowski space for a poor choice of initial Cauchy surface and/or gauge).

In the classical context, one usually deals with this simply by terminating spacetime and its evolution at such a singularity, professing ignorance of what might lie beyond. In the symplectic quotient construction of the classical phase space, we declare equivalence between two sets of initial data only if they correspond to data on a pair of Cauchy surfaces in a common singularity-free (but not necessarily geodesically complete) spacetime. One could guess at additional equivalences that occur because of evolution through a singularity to some new classical spacetime, but that would be extremely speculative. Equally, we do not exclude a point from the phase space if its evolution leads to a singularity either in the future (such as a black hole) or in the past (such as a big bang).
 
We advocate that this conservative approach can guide an analogous prescription for the quantum theory, at least semi-classically. In the path integral (for example, computing inner products by group averaging), this would simply means integrating over smooth spacetimes, which in the semi-classical limit amounts to allowing only  smooth metrics as saddle-points (though making this precise may lead to yet more subtleties when determining whether a given complex saddle-point contributes). Note that this restriction only applies to the part of spacetime (in a finite band of time) relevant to the saddle-point in question; singularities in an extension the classical solution further into the future or past are allowed (excluding these, as in the similar prescription of \cite{Cotler:2023eza}, would project out certain states which we regard as physical). In operator terms, the upshot is that we do not choose a self-adjoint extension; instead we accept that gauge transformations no longer generate unitaries, roughly because any part of the wavefunction that evolves to a singularity gets discarded. We spell out the details in the example of JT gravity \cite{Held:2024rmg}. In this simple solvable 2D model, this approach is perfectly satisfactory even non-perturbatively, giving a well-defined Hilbert space whose classical limit agrees with the standard phase space.

We can contrast our approach with an alternative suggested in DeWitt's treatise \cite{DeWitt:1967yk}, which postulates a boundary condition that invariant wavefunctions should vanish in limits of singular metrics. But this amounts to a reflecting boundary condition (generalising Dirichlet boundary conditions for a non-relativistic particle on a half-line), asserting that gravitational collapse to a singularity precedes a `bounce' back to a re-expanding classical geometry (at least locally; presumably spatial gradients make the actual dynamics rather complicated). For example, in DeWitt's discussion of a Friedmann cosmology this boundary condition turns a classical bang-crunch spacetime into a cyclic `bouncing' cosmology, where every crunch is followed by a new big bang. We regard this idea that evolution continues through a singularity and re-emerges to a classical spacetime as highly speculative.

\subsection{Null states}\label{ssec:null}

One of our main motivations for revisiting canonical methods in quantum gravity arose from the phenomenon of `null states' induced by non-perturbative gravitational effects. It has long been appreciated that the perturbative Hilbert space of gravity (around a fixed classical background) can appear to contain too many states, in extreme kinematic regimes. The archetypal example is the interior of an old black hole (after time of order $(\hbar G_N)^{-1}$), which can allow more than $\exp(\frac{A}{4\hbar G_N})$ states, in tension with a microscopic state-counting interpretation of Bekenstein-Hawking entropy. A possible resolution is that the states constructed in perturbation theory are not in fact linearly independent, so that various states are in fact `null' --- they have zero norm, and the physical states are taken to be equivalence classes under such wavefunctions. This requires modifications to the physical inner product which are non-perturbatively small (i.e., exponential in $(\hbar G_N)^{-1}$)  for individual matrix elements, but finely tuned. Various  recent results from the path integral \cite{Penington:2019kki,Balasubramanian:2022gmo,Maxfield:2023mdj,Iliesiu:2024cnh} give indirect and direct evidence that non-perturbative gravitational effects can supply precisely the needed corrections to the inner product.\footnote{\label{foot:factorization}At least, gravitational effects can provide some of these modifications, with sums over geometries being sufficient to make gravity compatible with the Bekenstein-Hawking entropy. But there remains the outstanding issue of the factorization problem \cite{Maldacena:2004rf,Marolf:2020xie}, whereby exchange of closed `baby' universes leads to superselection sectors for long-distance physics. As yet, there does not seem to be any evidence that gravity alone is sufficient to solve this without deliberate fine-tuning, so perhaps stringy or other effects are necessary.}

However, from this path integral perspective these null states seem rather miraculous: for a matrix of inner products to have a large kernel (but no negative eigenvalues), its entries must be finely tuned in a carefully coordinated way. Such a coincidence demands an explanation. We propose that the underlying mechanism is gauge symmetry, with null states arising from the usual redundancy under gauge transformations. But we are not invoking some new gauge symmetry: instead, we attribute the null states to the usual diffeomorphism symmetry of gravity.\footnote{We may need to allow a mild extension of the usual diffeomorphism, for example to permit jump discontinuities to account for topology-changing effects (see \cite{Maxfield:2023mdj} and the next subsection). More speculatively, perhaps diffeomorphisms are part of an extended (perhaps stringy) gauge symmetry which is restored in the UV, which may be behind additional null states such as those required for factorization (see footnote \ref{foot:factorization}).} More specifically, we refer to the non-perturbative part of diffeomorphisms which are not close to the identity.

We can make this more concrete in the language of co-invariants. A perturbative Hilbert space might be constructed by choosing a gauge-fixing which is good to all orders in perturbation theory. For example, if we have a gauge condition which selects a unique Cauchy surface in the dominant semi-classical spacetime, we can use this to make an ansatz for states as sketched in section \ref{ssec:coinvGF} (perhaps using a $\kappa$ map to construct the ansatz and compute an inner product order-by-order in perturbation theory). But even after this gauge-fixing, there could be additional equivalences arising from a residual group of unfixed diffeomorphisms which are far from the identity. Geometrically, this might happen when there are spacetimes contributing to the inner product (perhaps encoding non-perturbative effects as in section \ref{ssec:nonpert}) which contain several Cauchy surfaces satisfying the gauge condition.



This idea can be made completely precise in a simple toy model \cite{Maxfield:2023mdj} where the non-perturbative part of diffeomorphism symmetry is the symmetric group $\operatorname{Sym}(n)$ permuting several components of space. We hope that the ideas of this paper might be useful to implement a similar idea in more realistic theories of gravity. Additionally, this canonical perspective (which does not rely on any asymptotic boundary or AdS/CFT-type duality) may help to understand other circumstances under which similar effects are relevant, for example in cosmological spacetimes.

\subsection{Topology change}

 Having motivated the interest of non-perturbative gravitational effects, we should point out that the most interesting examples involve topology change. Such effects are very natural in the context of path integrals where the relevant spacetimes have Euclidean signature. However, topology change is incompatible with ordinary Lorentzian geometry, since any two Cauchy surfaces (in a globally hyperbolic spacetime) must have the same topology. Given this, one might reasonably believe  that these effects are impossible to capture in a canonical formalism like we have explored in this paper, since this typically leads to an interpretation in terms of Lorentzian spacetimes.

Fortunately, this is too pessimistic since topology change is possible if we allow a mild relaxation of a Lorentzian geometry, where the metric is permitted to have certain degenerations along surfaces of codimension two (or higher). One example is the Louko-Sorkin `crotch singularity' \cite{Louko:1995jw}, which locally looks like an $n$-fold cover of Minkowski space ramified over a codimension-2 branch locus. Such singularities appears in replica wormhole calculations \cite{Almheiri:2019qdq,Penington:2019kki} from the Lorentzian perspective emphasised in \cite{Dong:2016hjy,Marolf:2021ghr,Colin-Ellerin:2020mva}, and similar configurations have been used to calculate the entropy of black holes \cite{Marolf:2022ybi} and de Sitter space \cite{Dittrich:2024awu} from Lorentzian path integrals.

But while these Lorentzian path integrals make closer contact with the perspective of this paper, we still lack a complete understanding of topology change from an \emph{ab initio} canonical perspective. This might look like `interaction terms' in the Hamiltonian (perhaps required for a consistent self-adjoint completion of the constraints) which couple sectors of different spatial topology. We believe that filling this gap is currently the most pressing problem in canonical quantum gravity.


\paragraph{Acknowledgements}

We would like to thank Shoaib Akhtar, Steve Carlip, Kristan Jensen, Don Marolf and Mukund Rangamani for many helpful discussions and comments on drafts. JH is supported by U.S. Department of Defense through the National Defense Science and Engineering Graduate (NDSEG) Fellowship Program as well as NSF grant PHY-2408110 and funds from the University of California. HM is supported by DOE grant DE-SC0021085 and a Bloch fellowship from Q-FARM. This research was supported in part by grant NSF PHY-2309135 to the Kavli Institute for Theoretical Physics (KITP).

\appendix

\section{Proof of Klein-Gordon causal propagator identity}\label{app:KGid}

Let $\psi$ be a solution to the constraint $H\psi=0$ with compactly supported initial data, and $f$ a source function with compact support in time. Write $\psi_+ = \mathcal{G}_+f$ for the retarded solution to $H\psi_+ = f$. We have the identity
\begin{equation}
    \tfrac{1}{2}\nabla\cdot\left[\psi_+\nabla\psi-\psi\nabla\psi_+\right] = f\psi ,
\end{equation}
using $H = -\frac{1}{2}\nabla^2 + V$. Now, since $f$ is compactly supported in time we can choose a Cauchy surface $\Sigma$ lying to the future of the support of $f$. Let $\target_-$ denote the region to the past of $\Sigma$. Since the right hand side of our identity is supported only on $\target_-$, we can evaluate $\int_\target f\psi$  by integrating the left hand side over $\target_-$. This integral evaluates to a boundary term on $\Sigma$ only (there is no boundary term in the past, because $\psi_+$ is the retarded solution and vanishes there):
\begin{equation}
    \begin{aligned}
        \int_\target f \psi &= \tfrac{1}{2}\int_\Sigma (\psi_+\nabla_N\psi-\psi\nabla_N\psi_+) \\
        &= \Omega(\psi,\Delta f),
    \end{aligned}
\end{equation}
where for the second line we use the fact that $\psi_+ = \Delta f$ on $\Sigma$, since it lies to the future of the support of $f$.

\section{Static target spaces}\label{app:staticTarget}

Here we consider the group-averaging and Klein-Gordon inner products in the special case of a static target space. This means that $\target$ is symmetric under translations and reversals of a `time' coordinate $T= q^0$, with the remaining coordinates $q^i$ denoted by $X^i$; we have $G_{00} = -f(X)$ for a positive function $f$, $G_{0,i} =0$, and $G_{ij}(X)$ a Euclidean metric on a manifold $\Sigma$. For simplicity of exposition we will additionally assume that $\Sigma$ is compact, and take the potential $V(X)$ to be positive (for example, a constant $\frac{m^2}{2}$ with positive nonzero `mass' $m$). Then the Hamiltonian is
\begin{equation}
    H = -\tfrac{1}{2}f(X)p_T^2 + \mathcal{D},
\end{equation}
where $\mathcal{D}$ is a strictly positive Hermitian operator acting on  $X$ variables only  (namely, the potential $V$ minus half the Laplacian on $\target$ acting on functions independent of $T$). 

Now, we can decompose a general wavefunction on $\target$ into simultaneous eigenfunctions of $H$ and $p_T$,
\begin{gather}
    f(T,X) = \frac{1}{\sqrt{2\pi}}\sum_n \int d\omega \hat{f}_n(\omega) e^{-i T \omega} \phi_n^\omega(X),\\
    (\mathcal{D}-\tfrac{1}{2}\omega^2 f(X)) \phi_n^\omega(X) = E_n(\omega)  \phi_n^\omega(X).
\end{gather}
Here, $\phi_n^\omega$ are a complete set of eigenfunctions for the operator $\mathcal{D}-\tfrac{1}{2}\omega^2 f$, which (for any fixed $\omega$) has a discrete spectrum by our assumption that $\Sigma$ is compact (relaxing this, we could replace $n$ by continuous labels, such as `spatial momenta' $k$ in the case that $\target$ is Minkowski space with constant $V=\frac{m^2}{2}$). We normalise them such that $\int_\Sigma dX \sqrt{-G} \,  (\phi_m^\omega(X))^* \phi_n^\omega(X) = \delta_{mn}$.

The energy eigenvalues $E_n(\omega)$ are even functions of $\omega$ (since they depend only on $\omega^2$ with $E_n(\omega)>0$, and strictly decreasing for $\omega>0$. This last property follows from `time-independent perturbation theory', which gives $E_n'$ as an `expectation value' of $f$:
\begin{equation}\label{eq:Enp}
    E_n'(\omega) = -\omega \int_\Sigma dX \sqrt{-G} \,  f(X)|\phi_n^\omega(X)|^2.
\end{equation}
In particular, from this it follows that $E_n(\omega)=0$ for exactly one positive frequency $\omega = \omega_n>0$ and its negative $\omega = - \omega_n$ (there must be a zero for every $n$ because \eqref{eq:Enp} implies an upper bound in terms of the minimum of $f$ on $\Sigma$, $E_n(\omega)<E_n(0) - \tfrac{1}{2}\omega^2 \min f$, which is negative for sufficiently large $\omega$).

It is now straightforward to identify invariant and co-invariant states. Invariants are linear combinations of the positive and negative frequency modes solving the constraints,
\begin{equation}
    \psi(T,X) = \frac{1}{\sqrt{2\pi}}\sum_n (\psi_n^+ e^{- i T \omega_n}\phi_n^{\omega_n}(X) + \psi_n^- e^{+ i T \omega_n}\phi_n^{-\omega_n}(X)),
\end{equation}
with complex coefficients $\psi_n^\pm$. Co-invariants are given by smooth functions $f_n(\omega)$, with two functions $f\sim g$ identified as the same state if $f_n(\pm\omega_n)=g_n(\pm\omega_n)$ for all $n$.\footnote{We should also include technical conditions: $f_n(\omega)$ should be analytic of exponential type  to ensure compactness in time (though a weaker condition suffices), and we should have growth conditions on $\psi^\pm_n$ and $f_n$ at large $n$. There is some freedom in the precise conditions, though they are replaced with weaker and more rigid requirements once we have an inner product and pass to the Hilbert space completion.} The natural pairing between the two is
\begin{equation}
    \lc f| \psi \ri = \sum_n (f_n^*(\omega_n) \psi^+_n + f_n^*(-\omega_n) \psi^-_n).
\end{equation}

The decomposition into `Fourier modes'  $\hat{f}_n(\omega)$ also allows us to easily write the action of the rigging maps $\eta$ (either group average or Klein-Gordon). For the group average, we have
\begin{align}
    \delta(H) f(T,X) &= \frac{1}{\sqrt{2\pi}}\sum_n \int d\omega \, \delta(E_n(\omega))\hat{f}_n(\omega) e^{-i T \omega} \phi_n^\omega(X) \\
    &= \frac{1}{\sqrt{2\pi}} \sum_n   \frac{1}{|E_n'(\omega_n)|}(\hat{f}_n(\omega_n) e^{-i T \omega_n}\phi_n^{\omega_n}(X) + \hat{f}_n(-\omega_n) e^{+ i T \omega_n}\phi_n^{-\omega_n}(X))  . \nonumber
    \end{align}
So, we see that $\eta$ projects the wavefunction onto the modes satisfying the constraint $E=0$, with the positive factor $|E_n'(\omega_n)|^{-1}$ (\eqref{eq:Enp} ensures that $E_n'(\omega_n)$ is nonzero). The resulting inner products on co-invariants and invariants give us
\begin{gather}
    \lc g|f\rc := \langle g|\delta(H)|f\rangle =  \sum_n  \frac{\hat{f}_n(\omega_n)\hat{g}_n^*(\omega_n) + \hat{f}_n(-\omega_n)\hat{g}_n^*(-\omega_n)}{|E_n'(\omega)|} , \\
    \li \psi|\psi \ri =\langle\psi|\kappa|\psi\rangle =   \sum_n |E_n'(\omega)|(|\psi_n^+|^2 + |\psi_n^-|^2 ),
\end{gather}
which is manifestly positive definite.

We can compare this to the Klein-Gordon form, which corresponds to $\eta_{KG} = \frac{1}{2\pi i} \Delta$ given by the causal propagator $\Delta = \mathcal{G}_+-\mathcal{G}_-$. For this, we can first write the retarded and advanced solutions $ \mathcal{G}_\pm f$ to $H\psi = f$ as
\begin{equation}
    \mathcal{G}_\pm f(T,X) = \frac{1}{\sqrt{2\pi}}\sum_n \int d\omega \frac{1}{E_n(\omega) \pm i \epsilon E_n'(\omega)} \hat{f}_n(\omega) e^{-i T \omega} \phi_n^\omega(X).
\end{equation}
These are manifestly solutions, and the $i\epsilon$ prescription ensures that the poles in $\omega$ are located in the upper or lower half-plane for retarded and advanced propagators respectively, yeilding the appropriate boundary conditions. The difference between these gives a $\delta$-function, but with a sign depending on the $i\epsilon$:
\begin{equation}
\begin{aligned}
\Delta_n(\omega) &= \frac{1}{E_n(\omega) + i \epsilon E_n'(\omega)}-\frac{1}{E_n(\omega) - i \epsilon E_n'(\omega)} \\
&= -2\pi i \sgn(E_n'(\omega))\delta(E_n(\omega)).
\end{aligned}
\end{equation}
The resulting Klein-Gordon form on co-invariants is
\begin{equation}
\begin{aligned}
    \lc g|f\rc_{KG} := \frac{1}{2\pi i}\langle g|\Delta|f\rangle &=  -\sum_n  \frac{\hat{f}_n(\omega_n)\hat{g}_n^*(\omega_n) + \hat{f}_n(-\omega_n)\hat{g}_n^*(-\omega_n)}{E_n'(\omega)} \\
    &= \sum_n  \frac{\hat{f}_n(\omega_n)\hat{g}_n^*(\omega_n) - \hat{f}_n(-\omega_n)\hat{g}_n^*(-\omega_n)}{|E_n'(\omega)|},
    \end{aligned}
\end{equation}
where in the last line we use the fact that $E_n$ is decreasing for positive $\omega$. This is almost identical to co-invariants, up to a crucial sign. It gives us a non-degenerate but indefinite Hermitian form: a positive norm for the positive-frequency modes, but negative frequency modes have negative norm.

We can give the sign appearing in the Klein-Gordon form a classical interpretation, by noting it comes from the sign of $E_n'(\omega)$ in the $i\epsilon$ prescription. This derivative of energy with respect to $\omega$ is classically equal to the derivative of target space time $T$ with respect to worldline proper time $\tau$, $\frac{dT}{d\tau} = \{T,H\}  = -\frac{\partial H}{\partial \omega}$ (using the fact that $-\omega$ is the momentum conjugate to $T$). Positive frequency then corresponds to  $\frac{dT}{d\tau}>0$ and negative frequency $\frac{dT}{d\tau}<0$. Later we will see how to generalise this idea, interpreting the sign of the norm as the direction in which the solution crosses the slice $\Sigma$ in phase space, which will allow us to write more general expressions for both group average and Klein-Gordon type inner products on invariants.

\section{More on BRST calculations}\label{app:BRST}
 
 Our discussion of inner products in the language of BFV or non-minimal BRST in section \ref{sec:BFVIP} relied heavily on exponentiating the anti-commutator of the BRST charge with some gauge fixing fermion $\Psi$. Here we will briefly elucidate some of the finer details which led to our results in the hope that it may aid with future calculations done by other authors. We first present an identity  which is often useful for exponentiating $[Q,\Psi]_+$ in relatively simple cases. 

Let $\epsilon$ be an operator which squares to zero $\epsilon^2=0$ and commutes with operators $X,Y$ (for our applications, $\epsilon$ is a product of ghosts such as $c\hat{b}$). Then
\begin{equation}
    e^{X+ \epsilon Y} = e^X + \epsilon \int_0^1 ds e^{(1-s)X} Y e^{s X}.
    \label{eq:expIdentity}
\end{equation}
We can think of this as taking the linear term in $\epsilon$ and averaging over all the possible orderings between $e^X$ and $Y$.

To prove this, write the exponentials on both sides as power series and compare terms. For the left hand side we use
\begin{equation}
    (X+\epsilon Y)^n = X^n + \epsilon\sum_{k=1}^n X^{k-1}Y X^{n-k},
\end{equation}
so $\epsilon X^p Y X^q$ appears in the $n$th term of the exponential with $n=p+q+1$, giving a coefficient $\frac{1}{(p+q+1)!}$.
For the right the same term appears with coefficient
\begin{equation}
    \int_0^1 ds \frac{(1-s)^p }{p!} \frac{s^q}{q!},
\end{equation}
which evaluates to the same. Use of this particular identity is useful in cases where $[Q,\Psi]_+$ has no terms involving conjugate pairs of ghosts and ghost momenta (i.e. $cb$ or $\hat{c}\hat{b}$).

When computing the inner products between invariants, we  take $\Psi = -i\hat{b} \chi$, where $\chi$ is some `gauge-fixing' operator on the matter Hilbert space. We get $[Q,\Psi]_+= -i\Pi \chi -i [H,\chi]  c\hat{b}$. Using the identity proven above we find
\begin{equation}
    e^{[Q,\chi]_+} = e^{-i\Pi \chi} -i c\hat{b} \int ds e^{-i(1-s)\Pi \chi}[H,\chi] e^{-i s \Pi \chi}.
\end{equation}
In the main text we also make liberal use of the fact that,
\begin{equation}
    [H, e^{-i \Pi \chi}] = -i\Pi \int ds e^{-i(1-s)\Pi \chi}[H,\chi] e^{-i s \Pi \chi}.
\end{equation}
This can similarly be proven by Taylor expanding the exponential on either side. On the left hand side we get,

\begin{equation}
    \begin{split}
        [H, e^{-i \Pi \chi}]&=\sum_{n=0}^\infty \frac{(-i\Pi)^n}{n!}[H,\chi^n]=\sum_{p,q=0}^{\infty}\frac{(-i\Pi)^{p+q+1}}{(p+q+1)!}\chi^p [H,\chi]\chi^q
        \\&= -i\Pi \int_0^1ds\,e^{-i(1-s)\Pi\chi}\,[H,\chi]\,e^{-is\Pi\chi},
   \end{split}
\end{equation}
Which is precisely the right hand side.

We can additionally generalize the identity \eqref{eq:expIdentity} slightly to the case where the operator in the exponential (the anti-commutator of the BRST charge with the gauge fixing fermion) does not just have a nilpotent term proportional to $c\hat{b}$ or $\hat{c}b$ but also has a term like $cb$ or $\hat{c}\hat{b}$ (e.g., for the discussion above \eqref{eq:BRSTrescaled}). The cases we'd like to consider are when the operator we're exponentiating takes one of the two following forms,
\begin{equation}
    \begin{split}
        \mathcal{O}& = bcX+cbY+\hat{c}bZ\\
        \,\,&\quad\quad\quad\text{or}\\
        \tilde{\mathcal{O}}&=\hat{b}\hat{c}\tilde{X}+\hat{c}\hat{b}\tilde{Y} + c\hat{b}\tilde{Z},
    \end{split}
\end{equation}
Where $X,Y,Z$ and their tilded variants are all taken to be Hermitian operators on the part of the factor of the Hilbert space without the ghosts. These operators arise naturally in contexts where we want to compute inner products between invariants using a gauge fixing fermion which does not act as the identity operator on the lapse degrees of freedom, or when we want to compute inner products between coinvariants which does not act as the identity operator on unconstrained Hilbert space.

Note that the presence of non-nilpotent combinations of ghosts means that we cannot straightforwardly apply equation \eqref{eq:expIdentity}. However, for the purpose of the inner product calculations for which this problem is relevant the important quantities will be the exponentiation of these operators evaluated in the $|\uparrow\downarrow\rangle$ and $|\downarrow\uparrow\rangle$ states (corresponding to coinvariants and invariants) respectively. As such we will specialize to these cases. As before, we will write the matrix exponentials in a series form

\begin{equation}
    \begin{split}
        \langle\uparrow\downarrow|e^{\mathcal{O}}|\uparrow\downarrow\rangle &=\sum_{n}\frac{1}{n!}\langle\uparrow\downarrow|\big(bcX+cbY+\hat{c}bZ\big)^n |\uparrow\downarrow\rangle\\
        \langle\downarrow\uparrow|e^{\tilde{\mathcal{O}}}|\downarrow\uparrow\rangle &=\sum_{n}\frac{1}{n!}\langle\downarrow\uparrow|\big(\hat{b}\hat{c}\tilde{X}+\hat{c}\hat{b}\tilde{Y}+c\hat{b}\tilde{Z}\big)^n |\downarrow\uparrow\rangle.
    \end{split}
\end{equation}
Now we note that since $cb|\uparrow\downarrow\rangle=|\uparrow\downarrow\rangle$, $\hat{c}b|\uparrow\downarrow\rangle=bc|\downarrow\uparrow\rangle$, and $\hat{c}b|\downarrow\uparrow\rangle=bc|\uparrow\downarrow\rangle=cb|\downarrow\uparrow\rangle =0$ we find that 
\begin{equation}
 \langle\uparrow\downarrow|\big(bcX+cbY+\hat{c}bZ\big)^n|\uparrow\downarrow\rangle=\langle\uparrow\downarrow|\downarrow\uparrow\rangle\times\sum_{i+j+1=n}Y^i Z X^j,
\end{equation}
 and by similar reasoning, we find that 
\begin{equation}
    \langle\downarrow\uparrow|\big(\hat{b}\hat{c}\tilde{X}+\hat{c}\hat{b}\tilde{Y}+c\hat{b}\tilde{Z}\big)^n|\downarrow\uparrow\rangle=\langle\downarrow\uparrow|\uparrow\downarrow\rangle\times\sum_{i+j+1=n}\tilde{Y}^i \tilde{Z} \tilde{X}^j.
\end{equation}
Using these results, we arrive at the conclusion that
\begin{equation}
    \begin{split}
        \langle\uparrow\downarrow|e^{\mathcal{O}}|\uparrow\downarrow\rangle &=\langle\uparrow\downarrow|\downarrow\uparrow\rangle\times\sum_{i,j}\frac{Y^i Z X^{j}}{(i+j+1)!} = \langle\uparrow\downarrow|\downarrow\uparrow\rangle\times\int_0^1 ds\,e^{(1-s)Y}Ze^{sX}
        \\
        \langle\downarrow\uparrow|e^{\tilde{\mathcal{O}}}|\downarrow\uparrow\rangle &=\langle\downarrow\uparrow|\uparrow\downarrow\rangle\times\sum_{i,j} \frac{\tilde{Y}^{i}\tilde{Z}\tilde{X}^{j}}{(i+j+1)!}=\langle\downarrow\uparrow|\uparrow\downarrow\rangle\times\int_0^{1}ds\,e^{(1-s)\tilde{Y}}\tilde{Z}e^{s\tilde{X}}.
    \end{split}
\end{equation}

\section{Semi-classical calculations}

\subsection{Perturbative solution of WDW and inner product}\label{app:pertWDW}

In the $u$ variable, we have
\begin{equation}
    \psi''(u)-u\psi(u) = \sum_{n=2}^\infty \frac{V^{(n)}(q_0)}{n! V'(q_0)} \frac{u^n}{\lambda^{n-1}} \psi(u).
\end{equation}
Go to momentum space by writing an Ansatz
\begin{equation}
    \psi(u) = \int \frac{dv}{2\pi} e^{i v u +i v^3/3}f(v),
\end{equation}
which gives us
\begin{equation}
    f'(v) = \sum_{n=2}^\infty \frac{V^{(n)}(q_0)}{n! V'(q_0)} \frac{i^{n+1}}{\lambda^{n-1}} e^{-i v^3/3} 
    \partial_v^n\left(e^{i v^3/3}f(v)\right).
\end{equation}
Now we simply expand $f(v) = \sum_{n=0}^\infty \lambda^{1-n}f_n(v)$ with $f_0(v)=1$, and to $n$th order in $\lambda$ we only get finitely many terms:
\begin{equation}
    f_n'(v) = \sum_{k=2}^{n+1} \frac{V^{(k)}(q_0)}{k! V'(q_0)} \frac{i^{k+1}}{\lambda^{k-1}} e^{-i v^3/3} 
    \partial_v^k\left(e^{i v^3/3}f_{n-k+1}(v)\right).
\end{equation}
At each order, the right hand side is a polynomial  which we can easily integrate with $f_n(0)=0$ for $n\geq 1$ to find $f_n$ (a polynomial in $v$ of degree $5n$). Furthermore, only coefficients of $q^{5n-3k}$ for integer $k$ are nonzero.

Now let's compute the inner product. We can expand it as
\begin{equation}
    \frac{V(q_1)-V(q_2)}{q_1-q_2} = \sum_{k_1,k_2=0}^\infty \frac{V^{(k1+k_2+1)}(q_0)}{(k1+k_2+1)!}  q_1^{k_1}q_2^{k_2},
\end{equation}
and calculate terms from $\int q^k \psi(q) dq = \left. \frac{i^k}{\lambda^{k+1}}\partial_v^k\left(e^{i v^3/3}f(v)\right)\right|_{v=0}$. Acting on $f_n$, only derivatives where $k-n$ is a multiple of 3 give a nonzero result, so this has an expansion in powers of $\lambda^{-3}$, or $\hbar^2$. We find
\begin{equation}
    \Vert \psi\Vert^2 = (-V'(q_0))|\tilde{\psi}(p=0)|^2 \left(1+ \frac{3 V''(q_0)^3+V^{(4)}(q_0) V'(q_0)^2-4 V^{(3)}(q_0) V'(q_0) V''(q_0)}{6  V'(q_0)^3} \lambda^{-3} +\cdots \right).
\end{equation}

\subsection{Examples}

Our first example is an exponential potential,
\begin{equation}
    V(q)=\frac{\omega^2}{2}(1- e^{\mu q}).
\end{equation}
The normalisable  solution to the Wheeler-DeWitt equation is
\begin{equation}
    \psi(q) = K_{\frac{2i\omega}{\hbar \mu}}\left(\tfrac{2\omega}{\hbar\mu}e^{\frac{\mu q}{2}}\right),
\end{equation}
which we have not normalised. The $q\to -\infty$ asymptotics are
\begin{equation}
    \psi(q) \sim \tfrac{1}{2} e^{\tfrac{2i\omega}{\hbar\mu}\log \tfrac{\omega}{\hbar\mu}} \Gamma(\tfrac{-2i \omega}{\hbar\mu}) e^{\frac{i\omega q}{\hbar}} + \mathrm{c.c.}= \sqrt{\frac{\pi \mu \hbar}{2\omega \sinh(\frac{2\pi \omega}{\hbar\mu})}} \cos(\tfrac{\omega q}{\hbar}  +\theta),
\end{equation}
so from this we can read off the group-averaged norm of the state (using section \ref{ssec:1Dtarget}) to be
\begin{equation}
    \Vert\psi\Vert^2 = \frac{\pi \mu\hbar^2}{8\sinh(\frac{2\pi \omega}{\hbar\mu})}.
\end{equation}

The momentum space wavefunction is given by the Fourier transform
\begin{equation}
    \psi(p) = \int \frac{dq}{\sqrt{2\pi\hbar}}\psi(q)e^{-ipq/\hbar}  = \frac{1}{2 \mu \sqrt{2\pi\hbar}}\left(\frac{\mu  \hbar }{\omega }\right)^{-\frac{2 i p}{\mu  \hbar }} \Gamma \left(\frac{p-\omega }{i\mu  \hbar }\right) \Gamma \left(\frac{p+\omega}{i\mu  \hbar }\right),
\end{equation}
so in particular we have
\begin{equation}
    \psi(p=0) = \sqrt{\frac{\pi\hbar}{2}} \frac{1}{2\omega\sinh(\frac{\pi \omega}{\hbar\mu})}.
\end{equation}
Now, if we normalise this to get an in-state with unit norm we have
\begin{equation}
    \psi_\mathrm{in}(p=0) = \sqrt{\frac{2\pi}{\omega}} \frac{1}{2\omega\sinh(\frac{\pi \omega}{\hbar\mu})} \frac{e^{\tfrac{2i\omega}{\hbar\mu}\log \tfrac{\omega}{\hbar\mu}}}{ \Gamma(\tfrac{2i \omega}{\hbar\mu})} \sim \frac{e^{\frac{i}{\hbar}\frac{2\omega}{\mu}(1-\log 2)}}{\sqrt{-i\hbar \tfrac{1}{2}\omega^2\mu}}(1+O(\hbar)),
\end{equation}
in precise agreement with \eqref{eq:psiinSC}. Note that as well as perturbative corrections (from the $\Gamma$ function), expanding the $\sinh$ in the denominator gives us non-perturbative corrections suppressed by factors of $e^{-\frac{2\pi \omega}{\hbar\mu}}$. These can be accounted for by complex saddle points with paths ending at $q=\frac{2\pi i n}{\mu}$ where $V'(q)=0$.

Now comparing the exact norm with the leading order result from the $p=0$ gauge-fixed norm, we find
\begin{equation}
    \frac{\hbar(-V'(0))|\psi(p=0)|^2}{\Vert\psi\Vert^2} = \frac{1}{\tanh\left(\frac{\pi \omega}{\hbar\mu}\right)},
\end{equation}
which is one plus non-perturbatively small corrections. So for this potential, the leading order expression for the $p=0$ gauge-fixed inner product is correct to all orders in perturbation theory, but not exact.

Another example is useful to show that the absence of perturbative corrections is special to this example. Take
\begin{equation}
    V(q) = \frac{\omega^2}{2}\left(1-\frac{a^2}{q^2}\right),
\end{equation}
where we restrict to $q<0$ (and demand $\psi(q)\to0$ at $q\to 0$ as the unique  boundary conditions to get a self-adjoint Hamiltonian). The solution to WDW is
\begin{equation}
    \psi(q) = \frac{1}{\hbar} \sqrt{-2\pi q} J_n\left(-\tfrac{\omega}{\hbar}q\right),
\end{equation} where $n^2 = \frac{a^2\omega^2}{\hbar^2}+\frac{1}{4}$, and we have normalised so that $\Vert\psi\Vert =1$ using group-averaging (by looking at the $q\to -\infty$ tails). The zero momentum component (which we obtained from integrating $(-q)^{-k} \psi(q)$ for $k>0$ and taking $k\to 0$) is
\begin{equation}
    \psi(p=0) = \sqrt{\frac{2}{\omega^3}}\frac{\Gamma\left(\tfrac{n}{2}+\tfrac{3}{4}\right)}{\Gamma\left(\tfrac{n}{2}+\tfrac{1}{4}\right)} \sim \sqrt{\frac{a}{\hbar\omega^2}}\left(1+\frac{\hbar^2}{8a^2\omega^2} +\cdots \right).
\end{equation}
This gives us 
\begin{equation}
    \Vert\psi\Vert^2  =(-\hbar V'(a))|\psi(p=0)|^2 \left( 1-\frac{\hbar^2}{4a^2\omega^2} +\cdots \right),
\end{equation}
with the expansion matching the perturbative corrections computed in the main text.

\section{String worldsheet analogy}\label{app:string}

While many of the ingredients we are discussing may seem unconventional, any reader familiar with the basics of string theory will already have encountered are version of all of them in the quantisation of the string worldsheet (see \cite{Polchinski:1998rq}, for example). Since this analogy may be a helpful guide in a more familiar context, we sketch the main ideas here.

Specifically, consider the quantisation of a closed bosonic string in conformal gauge (similar comments apply to open strings and superstrings). We are left with the Hilbert space of a worldsheet CFT (e.g., 26 free bosons), but must still impose the constraints from the residual conformal gauge transformations generated by $L_n,\bar{L}_n$. The inner product on the space of physical string states is not usually discussed (since in this context the more relevant interpretation is in terms of the target space, rather than physics from the perspective of the worldsheet), but our considerations are nonetheless closely related to certain  familiar amplitudes. Using `old covariant quantisation' to identify the physical Hilbert space, for non-zero $n$ we impose `invariance' $L_n|\psi\rangle=0$ on the positive-frequency modes $n>0$, and `co-invariance' equivalence relations on the negative-frequency modes $n<0$.\footnote{This arises in BRST quantisation by choosing a state for the corresponding ghosts which is annihilated by $b_n$ and $c_n$ for $n<0$.} We will not say more about these constraints, concentrating on the remaining zero modes $H=L_0+\bar{L}_0-2$ and $P=L_0-\bar{L}_0$. The momentum $P$ generates the compact $U(1)$ group of translations of the string, so we can simply project onto the $P=0$ subspace (or average over the finite-volume group) and this does not lead to any subtleties with constructing a gauge-invariant inner product on the space of physical states. The only tricky part concerns the Hamiltonian constraint $H$ (and this works  essentially the same as the single constraint $H$ in the one-dimensional models that this paper concentrates on).

The first (and most standard) way to define physical states imposes `invariance' for the zero modes also, $H|\psi\rangle=P|\psi\rangle=0$. This gives rise to the usual string states corresponding to marginal ($h=\bar{h}=1$) vertex operators of the CFT.  But in fact we also encounter states of the string which we might try to interpret as co-invariants with respect to $H$: these are boundary states, representing a string emitted from a D-brane. These states are annihilated by $P$, but we do not require that they are annihilated by $H$ (and the same applies for all modes $L_n+\bar{L}_n$ of the Hamiltonian constraints).\footnote{Keeping track of the ghosts, the vertex operator states are annihilated by $b_0,\bar{b}_0$ (Siegel gauge), as appropriate for invariants. In the path integral, fixing the coordinate location of vertex operators leads to a ghost operator insertion $c\tilde{c}$, which under the state operator correspondence leads to the correct state. Boundary states are annihilated by $b_n-\tilde{b}_n$, which are the $b$-ghosts associated with momentum constraints, but by $c_n+\tilde{c}_n$ corresponding to Hamiltonian constraints.}

Now, even for non-interacting strings (at zero string coupling) we would expect to be able to construct an inner product (or at least  a gauge-invariant amplitude) from a pair of such physical states. The simplest case is when we take one of each: a vertex operator state $|\mathcal{V}\rangle$ and  boundary state $|\mathcal{B}\rangle$ have a natural finite gauge-invariant pairing given by the disk one-point function. This amplitude is particularly simple since it has no moduli to integrate over, and no residual gauge symmetry. It is analogous to the canonical pairing $\lc\mathcal{B}|\mathcal{V}\ri$ between invariants and co-invariants, and the simplicity of the amplitude reflects that this pairing is natural, requiring no further choices to be made.

The pairing of two vertex operators $\mathcal{V}_1$ and $\mathcal{V}_2$ gives us the sphere two-point amplitude, but this is more complicated because there is an infinite-volume residual symmetry generated by $H$ (as well as the finite-volume rotation generated by $P$). This comes about because vertex operator states are not normalisable due to the continuous spectrum of $H$. For a string in $D$ non-compact spacetime dimensions ($D=26$ in the simplest case), the na\"ive inner product is proportional to $\delta^{(D)}(k_\mu-k'_\mu)$, a delta-function in the full spacetime. The physically sensible result includes only a delta-function $\delta^{(D-1)}(\vec{k}-\vec{k}')$ in the spatial directions, with the last equality automatically following from the mass-shell condition $p^2+m^2=0$. To get a sensible finite result requires some gauge-fixing condition (such as $X^0=0$ at a chosen point on the worldsheet) \cite{Erbin:2019uiz} (which is associated with a $c$-ghost insertion in the BRST formalism). This is completely analogous to the Klein-Gordon-like $\kappa$ map.

Finally, for two boundary states $\mathcal{B}_1,\mathcal{B}_2$ the obvious pairing requires the cylinder amplitude (or annulus). In this case there is a modulus (the length of the cylinder for fixed radius), a single-parameter family of gauge-inequivalent worldsheet conformal structures to integrate over (associated with a $b$-ghost insertion). We get an amplitude of the form $\int dt \langle\mathcal{B}_2|e^{-i H t}|\mathcal{B}_1\rangle$, which should be compared to the group-averaging integral in \eqref{eq:GAIP}. There is an important difference for the usual string amplitude: we integrate over imaginary $t$ (Euclidean worldsheets), and only over a half-line rather than all real $t$. This is because the string amplitude has a target space interpretation, and is not intended as an inner product of worldsheet states (see comments in \cite{Casali:2021ewu}).

\begin{table}
    \centering
    \begin{tabular}{cc}
 \textbf{Gravitational Hilbert space} & \textbf{String worldsheet Hilbert space} \\ \hline
        Invariant states  $|\psi\ri$   & Vertex operators \\
       Co-invariant states $|\psi\rc$  & Boundary states (D-branes) \\
       Pairing $\li\psi_2|\psi_1\ri$ between invariants  & Sphere two-point amplitude  \\
       Pairing $\li\psi_2|\psi_1\rc$ between invariants and co-invariants  & Disk one-point amplitude \\
        Pairing $\lc\psi_2|\psi_1\rc$ between co-invariants & Cylinder/annulus amplitude \\
         & \\
         & \\
    \end{tabular}
    \caption{Analogies between the gravitational Hilbert space and that of the string worldsheet.}
    \label{tab:my_label}
\end{table}

\section{Maxwell theory}\label{app:Maxwell}

To illustrate some of the ideas in a more familiar setting, we here discuss time evolution and inner products in the BFV formalism for electrodynamics.

In terms of the (Lorentz non-covariant) electric scalar potential $\phi$ and magnetic vector potential $\vec{A}$, the Lagrangian of the theory is
\begin{equation}
\begin{gathered}
       L  = \tfrac{1}{2}\int d^3x (E^2-B^2), \\
       \vec{B} = \nabla\times \vec{A}, \quad \vec{E} = -\frac{\partial \vec{A}}{\partial t} -\vec{\nabla} \phi.
\end{gathered}
\end{equation}
From this, the electric field $\vec{E}$ is minus the canonical momentum for $\vec{A}$ (and we will not introduce separate notation for this), while the momentum $\Pi$ conjugate to $\phi$ vanishes: we have the primary constraint $\Pi=0$. The canonical construction of the Hamiltonian gives (after an integration by parts)
\begin{equation}
    H = \int d^3 x \left[ \frac{1}{2}(E^2+B^2) - \phi \nabla\cdot E \right] \qquad (\vec{B}=\nabla\times \vec{A}).
\end{equation}
Demanding that this commutes with $\Pi$, we get the secondary constraint $\nabla\cdot E=0$ (Gauss's law), and $\phi$ acts as the Lagrange multiplier imposing this constraint.

Since the two constraints $\Pi$, $\nabla\cdot E$ commute, the BRST charge is simply
\begin{equation}
    Q = \int d^3 x \left[ c\, \nabla\cdot E + \hat{c}\, \Pi \right].
\end{equation}
This BRST charge does not commute with $H$ as given above, signalling the need for additional ghost terms. But $Q$ does commute with the first term, and so we can write the most general Hamiltonian as $H_\Psi = \frac{1}{2}\int (E^2+B^2) + [Q,\Psi]_+$ for some choice of gauge-fixing fermion $\Psi$. A nice family of choices are the Gaussian-averaged Lorenz gauges (parameterised by a real number $\xi$):
\begin{equation}
    \Psi = -\int d^3 x \left(b \phi +\hat{b} \nabla\cdot A + \tfrac{1}{2}\xi \hat{b} \Pi\right).
\end{equation}
This gives us a Hamiltonian $H_\Psi = H_\xi  + H_\mathrm{ghost}$, with
\begin{equation}
\begin{gathered}
      H_\xi = \int d^3 x \left[ \tfrac{1}{2}(E^2+B^2) - \phi \nabla\cdot E - \Pi \nabla\cdot A - \tfrac{1}{2}\xi \Pi^2  \right]\\
      H_\mathrm{ghost}= i\int d^3 x\left(\hat{c} b + c \nabla^2 \hat{b}\right).
\end{gathered}
\end{equation}
The correspondence to Lorenz gauge is visible from the Hamiltonian equation of motion for $\phi$ (obtained from varying $\Pi$), which is $\dot{\phi} = \frac{\delta H}{\delta \Pi} = -\nabla\cdot A - \xi \Pi$, which we can write as $\xi \Pi = -\partial_\mu A^\mu$. For $\xi=0$ this is the Lorenz gauge condition $\partial_\mu A^\mu=0$ directly, while for $\xi\neq 0$ we get the Lorenz gauge condition after imposing the constraint $\Pi=0$. More generally, the bosonic part $H_\xi$ can also be obtained by adding a term $-\frac{1}{2\xi}(\partial_\mu A^\mu)^2$ to the original Lagrangian. Landau gauge is $\xi=0$, and Feynman gauge is $\xi=1$.

\subsection{Wavefunctional representation}

Let's understand the action of the time-evolution operator $e^{-i H_\xi t}$ in this gauge (since $H_\mathrm{ghost}$ is independent of the original fields we'll concentrate on the bosonic part for now, coming back to the ghost contribution later). This will be slightly easier in momentum space, defining $\phi(x) = \int \frac{d^3 x}{(2\pi)^{3/2}} e^{i k\cdot x} \phi(k)$ (with $\phi(k)^\dag = \phi(-k)$) and similarly for other fields, to get
\begin{equation}
    H_\xi = \int d^3 k \left[ \tfrac{1}{2} |E(k)|^2 + \tfrac{1}{2}(k^2 |A(k)|^2 - |k\cdot A(k)|^2) - \tfrac{1}{2}\xi |\Pi(k)|^2  - i \phi(-k)  k \cdot E(k) - i\Pi(-k)k\cdot A(k)  \right].
\end{equation}
Now we can separate the physical transverse modes of the gauge field and electric field from the nonphysical longitudinal components: $\vec{A}(k) = \vec{A}_\perp(k) + \frac{\vec{k}}{|k|} A_\parallel(k)$  where $A_\parallel(k) = \frac{1}{|k|}\vec{k}\cdot \vec{A}(k)$ is the component of $\vec{A}$ in the direction $\vec{k}$, and similarly for $E_\parallel$. Note that $A_\parallel(k)^\dag = -A_\parallel(-k)$. The $B^2$ `potential' term for $\vec{A}$ projects onto the physical transverse part, so we can write
\begin{equation}
\begin{gathered}
    H_\xi =H_\perp + H_\mathrm{gauge}, \\
    H_\perp =  \int d^3 k \left[ \tfrac{1}{2} |E_\perp(k)|^2 + \tfrac{1}{2}k^2 |A_\perp(k)|^2 \right],\\
    H_\mathrm{gauge} = \int d^3 k\left[\tfrac{1}{2}|E_\parallel(k)|^2  - \tfrac{1}{2}\xi |\Pi(k)|^2  - i |k|  \phi(-k)  E_\parallel (k) - i |k| \Pi(-k) A_\parallel(k)  \right].
\end{gathered}
\end{equation}
The transverse piece is simply the free Hamiltonian for the physical components of the fields (with massless relativistic dispersion relation $\omega^2=k^2$) and is unaffected by the constraints, so we can concentrate on the (commuting) `gauge' piece, acting on the longitudinal modes and $\phi,\Pi$.

Now, since different values of momentum $k$ are all independent (except that modes with $-k$ and $k$ are related by complex conjugation) we can treat them all separately, treating each $\vec{k}$ (together with $-\vec{k}$) as an ordinary quantum mechanics with two complex coordinates $A_\parallel$ and $\phi$, with conjugate momenta $-E_\parallel$ and $\Pi$. To recover the full results we need only restore $k$-dependence and take appropriate integrals over $k$ in the end. With this we can write the momentum $\pm \vec{k}$ contribution to $H_\mathrm{gauge}$ as
\begin{equation}
    H_\mathrm{gauge}(k) = |E_\parallel|^2  - \xi |\Pi|^2  - i |k|  (\bar{\phi}  E_\parallel-\phi  \bar{E}_\parallel + \bar{\Pi} A_\parallel-\Pi \bar{A}_\parallel).
\end{equation}
 For these complex coordinates the nontrivial commutation relations are $[\phi,\bar{\Pi}] =[\bar{\phi},\Pi] =i$, and $[A_\parallel,\bar{E}_\parallel] = [\bar{A}_\parallel,E_\parallel] = -i$.

Since $H_\mathrm{gauge}$ is linear in $A_\parallel$ and $\phi$, it is simplest to solve the Schr\"odinger equation in the momentum space representation (with wavefunctions depending on $E_\parallel$, $\Pi$), so $\phi = i \partial_{\bar{\Pi}} $ and so forth. Then, $H_\mathrm{gauge}$ is a first-order differential operator,
\begin{equation}
H_\mathrm{gauge} = |E_\parallel|^2  - \xi |\Pi|^2  + k   \left(E_\parallel  \frac{\partial}{\partial \Pi} - 
\bar{E}_\parallel  \frac{\partial}{\partial \bar{\Pi}} + \Pi \frac{\partial}{\partial E_\parallel} - \bar{\Pi} \frac{\partial}{\partial \bar{E}_\parallel}\right)
\end{equation}
so the Schr\"odinger equation can then be solved by the method of characteristics. The Schr\"odinger evolution for time $t$ from initial wavefunction $\psi_0$ then gives
\begin{gather}
       \psi_t(E_\parallel,\Pi) = e^{i \vartheta}\psi_0(E_\parallel \cos(k t)-i \Pi \sin(k t),\Pi\cos(k t)-i E_\parallel \sin(k t)), \\
      \vartheta = \tfrac{\xi-1}{2}t \left(|E_\parallel|^2+|\Pi|^2\right)
       +\tfrac{\xi+1}{2}\tfrac{\sin(kt)}{k}\left(\cos(k t)(|\Pi|^2-|E_\parallel|^2)+i \sin(k t)(\bar{E}_\parallel \Pi-E_\parallel\bar{\Pi})\right).\nonumber
\end{gather}

Now we have to choose how we want to represent our initial and final states: invariants, co-invariants, or a mixture of the two. For co-invariants, we choose arbitrary $E_\parallel$ dependence, but impose invariance under the primary constraint $\Pi=0$: so the initial wavefunction is  $\psi(E_\parallel)\delta(\Pi)$ (the notation is not supposed to imply that $\psi$ is holomorphic, and the $\delta$-function is two-dimensional, defined for a complex value). Evolving this gives us a state
\begin{equation}
    \psi_t(E_\parallel,\Pi) =  \exp\left[i\left(\tfrac{\xi-1}{2}t \sec^2(k t)
       -\tfrac{\xi+1}{2}\tfrac{\tan(kt)}{k}\right)|E_\parallel|^2\right] \delta(\Pi \cos(k t)-i E_\parallel \sin(k t))\psi(\sec(k t)E_\parallel) 
\end{equation}
where we have assumed that $\cos(k t)$ is nonzero to write the answer in this form (otherwise the $\delta$-function sets $E_\parallel$ to zero and the remaining dependence must be written in terms of $\Pi$).
Taking the inner product with a similar $\Pi=0$ state means that we evaluate at the final wavefunction at $\Pi=0$. This gives simply
\begin{equation}
    \psi_t(E_\parallel(k),0) =\left(\frac{\sin(k t)}{k} \right)^{-2}\delta(k E_\parallel) \psi(0) ,
\end{equation}
where the prefactor is a Jacobian coming from extracting a factor from the delta-function $\delta(-i \sin(kt)E_\parallel)$. This factor should be cancelled by the contribution of the ghosts. Now $k E_\parallel$ is simply a momentum space representation of the Gauss law constraint $\nabla\cdot E$, so our result gives us matrix elements of $\eta \propto \delta[\nabla\cdot E]$, the group averaged inner product projecting onto solutions of the constraints.

The Jacobian should be cancelled by a contribution from the ghosts. Following the same momentum space representation, we can write the momentum $k$ ghost Hamiltonian as
\begin{equation}
    H_\mathrm{ghost}(k) = i \left(\hat{c}^\dag b  - k^2 c^\dag \hat{b}) + i(\hat{c} b^\dag - k^2 c \hat{b}^\dag \right),
\end{equation}
with $[c,b^\dag]_+ =[c^\dag,b]_+ =1$ and similarly for hatted ghosts. The two terms we have written commute, which is useful for calculations.
For the co-invariant inner product, we want to calculate the matrix elements of $e^{-i t  H_\mathrm{ghost}(k)}$ between states $|\uparrow\uparrow\downarrow\downarrow\rangle$ annihilated by $c,c^\dag,\hat{b},\hat{b}^\dag$. To do this, simply use the Taylor expansion for the exponentials and note that the only nonzero contribution comes from odd powers with alternating terms (meaning products like $(\hat{c}^\dag bc^\dag \hat{b})^{2n} \hat{c}^\dag b$). These sum to give
\begin{equation}
    \langle\uparrow\uparrow\downarrow\downarrow| e^{-i t  H_\mathrm{ghost}(k)}|\uparrow\uparrow\downarrow\downarrow\rangle = \left(\frac{\sin(k t)}{k} \right)^{2} \langle\uparrow\uparrow\downarrow\downarrow|\downarrow\downarrow\uparrow\uparrow\rangle,
\end{equation}
precisely cancelling the Jacobian factor from the bosonic part.

If we instead use invariant states, then we have initial and final wavefunctions $\delta(k E_\parallel)\psi(\Pi)$ for arbitrary $\psi(\Pi)$. Evolving this gives
\begin{gather}
       \psi_t(E_\parallel,\Pi) =\exp\left[i \left(\tfrac{\xi-1}{2}t \sec^2(k t)
       + \tfrac{\xi+1}{2}\tfrac{\tan(kt)}{k}\right)|\Pi|^2 \right]\delta(E_\parallel k \cos(k t)-i \Pi k \sin(k t))\psi\left(\frac{\Pi}{\cos(kt)}\right).
\end{gather}
Taking the overlap with $\delta(k E_\parallel)$, we get
\begin{gather}
       \frac{1}{k}\psi_t(0,\Pi) = (k \sin(k t))^{-2} \delta(\Pi)\psi\left(0\right).
\end{gather}
Similarly to above, the ghost contribution is $(k \sin(k t))^{2}$, cancelling the Jacobian factor as expected.

\subsection{Oscillator representation and Gupta-Bleuler}

To explain the relation to the more familiar Lorenz gauge Gupta-Bleuler quantisation, we write the fields in terms of oscillators. Define annihilation operators $\varphi_k$ and $\vec{a}_k$ (which is vector-valued) and their conjugate creation operators as follows:
\begin{equation}
    \begin{aligned}
        \phi(x) &= \int \frac{d^3k}{(2\pi)^{3/2}} \frac{e^{i k\cdot x}}{\sqrt{2|k|}} \varphi_k + \text{h.c.}\\
        \vec{A}(x) &= \int \frac{d^3k}{(2\pi)^{3/2}} \frac{e^{i k\cdot x}}{\sqrt{2|k|}} \vec{a}_k + \text{h.c.} \\
         \Pi(x) &=\frac{2i}{\xi+1} \int \frac{d^3k}{(2\pi)^{3/2}} e^{i k\cdot x}\sqrt{\frac{|k|}{2}}(\varphi_k-a^\parallel_k)+ \text{h.c.} \\
         \vec{E}(x) &= i\int \frac{d^3k}{(2\pi)^{3/2}} e^{i k\cdot x}\sqrt{\frac{|k|}{2}} \left(\vec{a}^\perp_k-\frac{2}{\xi+1} (\varphi_k-a^\parallel_k) \frac{\vec{k}}{|k|}\right) + \text{h.c.}
    \end{aligned}
\end{equation}
Here $a_k^\parallel := \frac{\vec{k}\cdot \vec{a}_k}{|k|}$ is the longitudinal component of $ \vec{a}_k$, and $\vec{a}^\perp_k = \vec{a} - a_k^\parallel\frac{\vec{k}}{|k|}$ the transverse part. This decomposition is chosen so that the explicitly written pieces are `positive frequency': Heisenberg evolution with $H_\xi$ gives the time-dependence
\begin{equation}
    \begin{aligned}
        \vec{a}_k(t) &= e^{-i|k| t}\left(\vec{a}_k-i\frac{\xi-1}{\xi+1}(\varphi_k - a_k^\parallel) \vec{k} t\right), \\
        \varphi_k(t) &= e^{-i|k| t}\left(\varphi_{k}-i\frac{\xi-1}{\xi+1}(\varphi_k - a_k^\parallel) |k| t\right),
    \end{aligned}
\end{equation}
which is almost positive frequency oscillation (slightly spoiled by the terms linear in $t$). Here we see that Feynman gauge $\xi=1$ simplifies things significantly, so we specialise to that case. We have non-trivial commutators
\begin{equation}\label{eq:MaxwellOscComm}
    [\varphi_k,\varphi^\dag_{k'}] = - \delta(k-k'), \quad [a^i_k,(a^j_{k'})^\dag] = \delta_{ij} \delta(k-k'),
\end{equation}
so that $\vec{a}_k$ are ordinary annihilation operators, while the extra sign in the first relation means that (for the algebra at least) $\varphi_k$ behaves as a creation operator. We can combine these back into a four-vector $a^\mu_k$ with $\varphi_k = a^0_k$, with covariant commutation relations $ [a^\mu_k,(a^\nu_{k'})^\dag] = \eta^{\mu\nu} \delta(k-k')$. In terms of the oscillators, the Hamiltonian (up to a constant from operator ordering) is
\begin{equation}
    H_{\xi=1} = \int d^3 k |k| \left(\vec{a}^\dag_k\cdot\vec{a}_k - \varphi^\dag_k\varphi_k\right).
\end{equation}

Now, a quantisation of the oscillator commutation relations with a unitary representation on the Hilbert space must treat $\varphi_k^\dag$ as an annihilation operator, so there is a `vacuum' state annihilated by $a^i_k$ and $\varphi_k^\dag$ and other states are built by acting with $(a^i_k)^\dag$ and $\varphi_k$. Then $H_\xi$ would be unbounded from below (since $\varphi_k$ lowers the energy). However, Gupta-Bleuler quantisation uses instead a non-unitary representation built on a Lorentz-invariant vacuum state $|\Omega\rangle$ annihilated by $\varphi_k$ and $a^i_k$ (and acting with $(a^i_k)^\dag$ and $\varphi_k^\dag$ creates other states, always increasing energy by $|k|$). Before imposing constraints, we have a `Hilbert space' with a non-degenerate but indefinite inner product (for example, a single photon state with timelike polarisation  $\varphi_k^\dag|\Omega\rangle$ has negative norm). The negative norm states are not immediately a problem, but they should be eliminated when we impose constraints and pass to the physical Hilbert space.

We can make a similar mode decomposition of the ghosts, with complex ghost modes $b_k$ and $c_k$ defined as follows:
\begin{equation}
    \begin{aligned}
        c(x) &= \int \frac{d^3k}{(2\pi)^{3/2}} \frac{e^{i k\cdot x}}{\sqrt{2|k|}} c_k + \text{h.c.}\\
        b(x) &= \int \frac{d^3k}{(2\pi)^{3/2}} e^{i k\cdot x}\sqrt{\frac{|k|}{2}} b_k  + \text{h.c.} \\
         \hat{c}(x) &= i\int \frac{d^3k}{(2\pi)^{3/2}} e^{i k\cdot x}\sqrt{\frac{|k|}{2}} c_k  + \text{h.c.} \\
         \hat{b}(x) &= i\int \frac{d^3k}{(2\pi)^{3/2}} \frac{e^{i k\cdot x}}{\sqrt{2|k|}} b_k + \text{h.c.}
    \end{aligned}
\end{equation}
These definitions are chosen so that the Heisenberg evolution of $b_k$ and $c_k$ simply produces a factor of $e^{-i |k|t}$, a positive-frequency oscillation. The only nontrivial anti-commutation relations following from these definitions are $[b_k^\dag,c_{k'}]_+ = \delta(k-k')$ (and the conjugate $[b_k,c_{k'}^\dag]_+ = \delta(k-k')$). The BRST charge and the ghost Hamiltonian are given in terms of oscillators by
\begin{align}
    Q &= \int d^3 k \, |k| \left(c_k^\dag(\varphi_k-a^\parallel_k) + c_k(\varphi_k-a^\parallel_k)^\dag \right) \, ,\\
     H_\mathrm{ghost} &= \int d^3 k \, |k| \left(c_k^\dag b_k + b_k^\dag  c_k \right) \,.
\end{align}
This means that $b_k$ and $c_k$ both lower the energy by $|k|$, and  $b_k^\dag$ and $c_k^\dag$ both raise it.

To recover the standard Gupta-Bleuler quantisation we place the ghosts in their ground state, annihilated by all the positive-frequency modes $b_k$ and $c_k$. To get a BRST-closed state, we then require that the matter part of the state is annihilated by $\varphi_k-a^\parallel_k$: this is the Gupta-Bleuler condition that the state is annihilated by the positive-frequency part of $\partial_\mu A^\mu$ (or equivalently, by the positive-frequency part of $\Pi$ or of $\nabla\cdot E$).  This roughly means that every unphysical timelike-polarized photon is accompanied by another unphysical longitudinal-polarized photon. The BRST-exact states of this form, given by  $Q$ acting on a state with a single $b_k^\dag$ ghost excitation, give the image of $(\varphi_k-a^\parallel_k)^\dag$ in the matter sector. Imposing equivalence under addition of such a state, we learn that we can remove a pair of timelike and longitudinal photons and get the same physical state. The upshot is that the physical content of the BRST cohomology leaves only the physical transverse polarizations.

In this case, we consistently get an inner product in which the ghost vacuum (annihilated by $b_k,c_k$) has non-zero norm, and the matter oscillator vacuum has finite norm. Additionally, the physical state Gupta-Bleuler condition eliminates negative norm states and the quotient by BRST exact states leaves a sensible positive-definite inner product for the physical Hilbert space. So, there is no requirement to insert a non-trivial gauge transformation in order to define a physical inner product, as we did for one-dimensional gravity. 

However, this would not be quite so simple if we changed things slightly to have a zero mode: for example, if we put the theory on a compact space.  In that case, the BRST charge gets a zero-mode contribution of the form $c_0 Q_m +\hat{c}_0 \Pi_0$, where $Q_m$ is the total electric charge of matter coupled to our gauge field (which we haven't considered until now). The relevant zero-mode piece of the Hamiltonian $H_\xi$ is $-Q_m \phi_0-\frac{\xi}{2}\Pi^2 + i \hat{c}_0 b_0$. Now the situation is extremely similar to that considered in the text, where we constructed the group-averaged inner product. We can take our states to be co-invariant with respect to the zero-mode of the Gauss-law constraint, namely the vanishing of the total charge~$Q_m=0$; this means taking the ghost state annihilated by $c_0$ and $\hat{b}_0$ and a wavefunction independent of the zero mode of the electric potential $\phi_0$ (annihilated by $\Pi_0$), but a state of arbitrary charge. The norms of such states take the indefinite form $0\times \infty$ from the ghost and $\phi_0$ sectors (for a non-compact gauge group). But any amount of time-evolution by $H_\xi$ resolves this, with the zero-modes contributing a factor $t \int \frac{d\phi_0}{2\pi} e^{i t Q_m \phi_0} = \sgn(t) \delta(Q_m)$ (for any $\xi$). This projects onto the zero-charge component of the state (with a delta-function as appropriate for continuous charge under a non-compact gauge group), by averaging over the group of zero-mode gauge transformations (constant gauge parameter). For a compact gauge group and quantised charge we would change this range of integration; the BRST formalism does not directly address such global properties of the gauge group.

\bibliography{biblio.bib}

\end{document}

%% file: Figures/nkn.tex
\begin{tikzpicture}
    \draw[ultra thick] (0,2) ellipse (1 and 0.25);
    \draw[ultra thick,red] (0,0) ellipse (1 and 0.25); 
    \draw[ultra thick] (0,-2) ellipse (1 and 0.25);
    \draw[ultra thick,variable=\y,domain=0:2,samples=80] plot ({(-0.05*sin(2*\y*360/2)+0.025*sin(\y*4*360/2)-1},{\y});
    \draw[ultra thick,variable=\y,domain=0:2,samples=80] plot ({(-0.05*sin(2*\y*360/2)+0.025*sin(\y*4*360/2)+1},{\y});
    \draw[ultra thick,variable=\y,domain=-2:0,samples=80] plot ({(0.1*sin(2*\y*360/2)-0.02*sin(\y*4*360/2)-1},{\y});
    \draw[ultra thick,variable=\y,domain=-2:0,samples=80] plot ({(0.1*sin(2*\y*360/2)-0.02*sin(\y*4*360/2)+1},{\y});

    \draw[red] node[right] (kappa) at (1,0) {$\kappa$};
    \draw node[below] (psiIL) at (0,-2.25) {$|\psi\rrangle$};
    \draw node[above] (psiFL) at (0,2.25) {$\llangle \psi'|$};
    \draw node[right] (eta1) at (1,-1) {$\eta$};
    \draw node[right] (eta2) at (1,1) {$\eta$};

    \draw node[] (g1) at (0,-1) {$g_1$};
    \draw node[] (g2) at (0,1) {$g_2$};

    \draw node at (2.5,0) {\Large=};

    \begin{scope}[shift={(4.5,0)}]
        \draw[ultra thick] (0,2) ellipse (1 and 0.25); 
        \draw[ultra thick] (0,-2) ellipse (1 and 0.25);
        \draw[ultra thick,variable=\y,domain=-2:2,samples=80] plot ({(-0.075*sin(\y*360/2)-1)},{\y});
        \draw[ultra thick,variable=\y,domain=-2:2,samples=80] plot ({(-0.075*sin(\y*360/2)+1)},{\y});

        \draw node[below] (psiIR) at (0,-2.25) {$|\psi\rrangle$};
        \draw node[above] (psiFR) at (0,2.25) {$\llangle \psi'|$};
        \draw node[right] (eta3) at (1,0) {$\eta$};

         \draw node[] (g3) at (0,0) {$g$};
    \end{scope}
\end{tikzpicture}

%% file: Figures/pairings.tex
\begin{tikzpicture}[scale=0.9]
\begin{scope}
  
    \draw[ultra thick, variable=\y,domain=-4.5:0.5,samples=80] plot ({-0.1*(1-sin(2*360*(\y-0.5)/5+90))},{\y});
    \draw node[right] at (0,-2) {$\eta$};
    \draw node[above] at (0,0.5) {$\lc\psi'|$};
    \draw node[below] at (0,-4.5) {$|\psi\rc$};

    \node at (0,0.5) (ccf) {};
    \node at (0,-4.5) (cci) {};

    \fill[black] (ccf) circle(2pt);
    \fill[black] (cci) circle(2pt);
\end{scope}

\begin{scope}[shift={(3,0)}]

  \draw[dotted, ultra thick] (0,0)--(0,0.5); 
    \draw[ultra thick, variable=\y,domain=-4:0,samples=80] plot ({0.1*(1-sin(2*360*\y/4+90))},{\y});
    \draw[dotted, ultra thick] (0,-4)--(0,-4.5);
    \draw node[red, right] at (0,-2) {$\kappa$};
    \node at (0,-2) (iikappa){};
    \fill[red] (iikappa) circle(2pt);
    \draw node[above] at (0,0.5) {$\li\psi'|$};
    \draw node[below] at (0,-4.5) {$|\psi\ri$};
\end{scope}

\begin{scope}[shift={(6,0)}]

\draw[dotted, ultra thick] (0,0)--(0,0.5);
     \draw[ultra thick, variable=\y,domain=-4.5:0,samples=80] plot ({0.1*(1-sin(2*360*(\y)/4.5+90))},{\y});
    \draw node[above] at (0,0.5) {$\li\psi'|$};
    \draw node[below] at (0,-4.5) {$|\psi\rc$};

    \node at (0,-4.5) (ici) {};
    \fill[black] (ici) circle(2pt);
    
\end{scope}
\end{tikzpicture}

%% file: biblio.bib
@article{Almheiri:2019qdq,
    author = "Almheiri, Ahmed and Hartman, Thomas and Maldacena, Juan and Shaghoulian, Edgar and Tajdini, Amirhossein",
    title = "{Replica Wormholes and the Entropy of Hawking Radiation}",
    eprint = "1911.12333",
    archivePrefix = "arXiv",
    primaryClass = "hep-th",
    doi = "10.1007/JHEP05(2020)013",
    journal = "JHEP",
    volume = "05",
    pages = "013",
    year = "2020"
}

@article{Chakraborty:2023yed,
    author = "Chakraborty, Tuneer and Chakravarty, Joydeep and Godet, Victor and Paul, Priyadarshi and Raju, Suvrat",
    title = "{The Hilbert space of de Sitter quantum gravity}",
    eprint = "2303.16315",
    archivePrefix = "arXiv",
    primaryClass = "hep-th",
    doi = "10.1007/JHEP01(2024)132",
    journal = "JHEP",
    volume = "01",
    pages = "132",
    year = "2024"
}

@article{Giddings:2022hba,
    author = "Giddings, Steven B. and Perkins, Julie",
    title = "{Perturbative quantum evolution of the gravitational state and dressing in general backgrounds}",
    eprint = "2209.06836",
    archivePrefix = "arXiv",
    primaryClass = "hep-th",
    doi = "10.1103/PhysRevD.110.026012",
    journal = "Phys. Rev. D",
    volume = "110",
    number = "2",
    pages = "026012",
    year = "2024"
}

@article{Cotler:2025gui,
    author = "Cotler, Jordan and Jensen, Kristan",
    title = "{Norm of the no-boundary state}",
    eprint = "2506.20547",
    archivePrefix = "arXiv",
    primaryClass = "hep-th",
    month = "6",
    year = "2025"
}

@article{Henneaux:2025ocw,
    author = "Henneaux, Marc",
    title = "{Wheeler-DeWitt Equation and Bondi-Metzner-Sachs (BMS) Symmetry}",
    eprint = "2506.02240",
    archivePrefix = "arXiv",
    primaryClass = "hep-th",
    doi = "10.1103/29w3-3mmc",
    journal = "Phys. Rev. Lett.",
    volume = "135",
    number = "6",
    pages = "061501",
    year = "2025"
}

@article{Chowdhury:2021nxw,
    author = "Chowdhury, Chandramouli and Godet, Victor and Papadoulaki, Olga and Raju, Suvrat",
    title = "{Holography from the Wheeler-DeWitt equation}",
    eprint = "2107.14802",
    archivePrefix = "arXiv",
    primaryClass = "hep-th",
    doi = "10.1007/JHEP03(2022)019",
    journal = "JHEP",
    volume = "03",
    pages = "019",
    year = "2022"
}

@article{Chakraborty:2023los,
    author = "Chakraborty, Tuneer and Chakravarty, Joydeep and Godet, Victor and Paul, Priyadarshi and Raju, Suvrat",
    title = "{Holography of information in de Sitter space}",
    eprint = "2303.16316",
    archivePrefix = "arXiv",
    primaryClass = "hep-th",
    doi = "10.1007/JHEP12(2023)120",
    journal = "JHEP",
    volume = "12",
    pages = "120",
    year = "2023"
}

@article{Penington:2019kki,
    author = "Penington, Geoff and Shenker, Stephen H. and Stanford, Douglas and Yang, Zhenbin",
    title = "{Replica wormholes and the black hole interior}",
    eprint = "1911.11977",
    archivePrefix = "arXiv",
    primaryClass = "hep-th",
    doi = "10.1007/JHEP03(2022)205",
    journal = "JHEP",
    volume = "03",
    pages = "205",
    year = "2022"
}

@article{Saad:2019lba,
    author = "Saad, Phil and Shenker, Stephen H. and Stanford, Douglas",
    title = "{JT gravity as a matrix integral}",
    eprint = "1903.11115",
    archivePrefix = "arXiv",
    primaryClass = "hep-th",
    month = "3",
    year = "2019"
}

@article{Harlow:2018tqv,
    author = "Harlow, Daniel and Jafferis, Daniel",
    title = "{The Factorization Problem in Jackiw-Teitelboim Gravity}",
    eprint = "1804.01081",
    archivePrefix = "arXiv",
    primaryClass = "hep-th",
    doi = "10.1007/JHEP02(2020)177",
    journal = "JHEP",
    volume = "02",
    pages = "177",
    year = "2020"
}

@article{Marolf:2008hg,
    author = "Marolf, Donald and Morrison, Ian A.",
    title = "{Group Averaging for de Sitter free fields}",
    eprint = "0810.5163",
    archivePrefix = "arXiv",
    primaryClass = "gr-qc",
    doi = "10.1088/0264-9381/26/23/235003",
    journal = "Class. Quant. Grav.",
    volume = "26",
    pages = "235003",
    year = "2009"
}

@article{Moncrief:1975xtw,
    author = "Moncrief, Vincent",
    title = "{Spacetime symmetries and linearization stability of the Einstein equations. I}",
    doi = "10.1063/1.522572",
    journal = "J. Math. Phys.",
    volume = "16",
    number = "3",
    pages = "493",
    year = "1975"
}

@article{Balasubramanian:2022gmo,
    author = "Balasubramanian, Vijay and Lawrence, Albion and Magan, Javier M. and Sasieta, Martin",
    title = "{Microscopic Origin of the Entropy of Black Holes in General Relativity}",
    eprint = "2212.02447",
    archivePrefix = "arXiv",
    primaryClass = "hep-th",
    doi = "10.1103/PhysRevX.14.011024",
    journal = "Phys. Rev. X",
    volume = "14",
    number = "1",
    pages = "011024",
    year = "2024"
}

@article{Marolf:2020xie,
    author = "Marolf, Donald and Maxfield, Henry",
    title = "{Transcending the ensemble: baby universes, spacetime wormholes, and the order and disorder of black hole information}",
    eprint = "2002.08950",
    archivePrefix = "arXiv",
    primaryClass = "hep-th",
    doi = "10.1007/JHEP08(2020)044",
    journal = "JHEP",
    volume = "08",
    pages = "044",
    year = "2020"
}

@article{Maldacena:2004rf,
    author = "Maldacena, Juan Martin and Maoz, Liat",
    title = "{Wormholes in AdS}",
    eprint = "hep-th/0401024",
    archivePrefix = "arXiv",
    reportNumber = "ITFA-2003-57",
    doi = "10.1088/1126-6708/2004/02/053",
    journal = "JHEP",
    volume = "02",
    pages = "053",
    year = "2004"
}

@article{Giulini:1998rk,
    author = "Giulini, Domenico and Marolf, Donald",
    title = "{On the generality of refined algebraic quantization}",
    eprint = "gr-qc/9812024",
    archivePrefix = "arXiv",
    reportNumber = "ZU-TH-98-20, SU-GP-98-10-1",
    doi = "10.1088/0264-9381/16/7/321",
    journal = "Class. Quant. Grav.",
    volume = "16",
    pages = "2479--2488",
    year = "1999"
}

@article{Marolf:2021ghr,
    author = "Marolf, Donald and Maxfield, Henry",
    title = "{The page curve and baby universes}",
    eprint = "2105.12211",
    archivePrefix = "arXiv",
    primaryClass = "hep-th",
    doi = "10.1142/S021827182142027X",
    journal = "Int. J. Mod. Phys. D",
    volume = "30",
    number = "14",
    pages = "2142027",
    year = "2021"
}

@article{Iliesiu:2022kny,
    author = "Iliesiu, Luca V. and Murthy, Sameer and Turiaci, Gustavo J.",
    title = "{Black hole microstate counting from the gravitational path integral}",
    eprint = "2209.13602",
    archivePrefix = "arXiv",
    primaryClass = "hep-th",
    month = "9",
    year = "2022"
}

@article{Dittrich:2024awu,
    author = {Dittrich, Bianca and Jacobson, Ted and Padua-Arg{\"u}elles, Jos{\'e}},
    title = "{de Sitter horizon entropy from a simplicial Lorentzian path integral}",
    eprint = "2403.02119",
    archivePrefix = "arXiv",
    primaryClass = "gr-qc",
    doi = "10.1103/PhysRevD.110.046006",
    journal = "Phys. Rev. D",
    volume = "110",
    number = "4",
    pages = "046006",
    year = "2024"
}

@article{Dong:2016hjy,
    author = "Dong, Xi and Lewkowycz, Aitor and Rangamani, Mukund",
    title = "{Deriving covariant holographic entanglement}",
    eprint = "1607.07506",
    archivePrefix = "arXiv",
    primaryClass = "hep-th",
    doi = "10.1007/JHEP11(2016)028",
    journal = "JHEP",
    volume = "11",
    pages = "028",
    year = "2016"
}

@article{Colin-Ellerin:2020mva,
    author = "Colin-Ellerin, Sean and Dong, Xi and Marolf, Donald and Rangamani, Mukund and Wang, Zhencheng",
    title = "{Real-time gravitational replicas: Formalism and a variational principle}",
    eprint = "2012.00828",
    archivePrefix = "arXiv",
    primaryClass = "hep-th",
    doi = "10.1007/JHEP05(2021)117",
    journal = "JHEP",
    volume = "05",
    pages = "117",
    year = "2021"
}

@inproceedings{Arisue:1981qs,
    author = "Arisue, H. and Fujiwara, T. and Inoue, T. and Ogawa, K.",
    title = "{GENERALIZED SCHRODINGER REPRESENTATIONS AND ITS APPLICATION TO GAUGE THEORIES.}",
    booktitle = "{1981 INS Symposium on Quark and Lepton Physics}",
    year = "1981"
}

@article{Batalin:1977pb,
    author = "Batalin, I. A. and Vilkovisky, G. A.",
    title = "{Relativistic S Matrix of Dynamical Systems with Boson and Fermion Constraints}",
    doi = "10.1016/0370-2693(77)90553-6",
    journal = "Phys. Lett. B",
    volume = "69",
    pages = "309--312",
    year = "1977"
}

@article{Cotler:2023eza,
    author = "Cotler, Jordan and Jensen, Kristan",
    title = "{Isometric Evolution in de Sitter Quantum Gravity}",
    eprint = "2302.06603",
    archivePrefix = "arXiv",
    primaryClass = "hep-th",
    doi = "10.1103/PhysRevLett.131.211601",
    journal = "Phys. Rev. Lett.",
    volume = "131",
    number = "21",
    pages = "211601",
    year = "2023"
}

@article{Maxfield:2023mdj,
    author = "Maxfield, Henry",
    title = "{Counting states in a model of replica wormholes}",
    eprint = "2311.05703",
    archivePrefix = "arXiv",
    primaryClass = "hep-th",
    month = "11",
    year = "2023"
}

@article{Iliesiu:2024cnh,
    author = "Iliesiu, Luca V. and Levine, Adam and Lin, Henry W. and Maxfield, Henry and Mezei, M{\'a}rk",
    title = "{On the non-perturbative bulk Hilbert space of JT gravity}",
    eprint = "2403.08696",
    archivePrefix = "arXiv",
    primaryClass = "hep-th",
    doi = "10.1007/JHEP10(2024)220",
    journal = "JHEP",
    volume = "10",
    pages = "220",
    year = "2024"
}

@article{Fradkin:1975cq,
    author = "Fradkin, E. S. and Vilkovisky, G. A.",
    title = "{QUANTIZATION OF RELATIVISTIC SYSTEMS WITH CONSTRAINTS}",
    doi = "10.1016/0370-2693(75)90448-7",
    journal = "Phys. Lett. B",
    volume = "55",
    pages = "224--226",
    year = "1975"
}

@article{Louko:1995jw,
    author = "Louko, Jorma and Sorkin, Rafael D.",
    title = "{Complex actions in two-dimensional topology change}",
    eprint = "gr-qc/9511023",
    archivePrefix = "arXiv",
    reportNumber = "SU-GP-95-5-1, WISC-MILW-95-TH-16, MDDP-PP-96-40",
    doi = "10.1088/0264-9381/14/1/018",
    journal = "Class. Quant. Grav.",
    volume = "14",
    pages = "179--204",
    year = "1997"
}

@article{Banihashemi:2024aal,
    author = "Banihashemi, Batoul and Jacobson, Ted",
    title = "{On the lapse contour in the gravitational path integral}",
    eprint = "2405.10307",
    archivePrefix = "arXiv",
    primaryClass = "hep-th",
    doi = "10.1103/PhysRevD.111.066014",
    journal = "Phys. Rev. D",
    volume = "111",
    number = "6",
    pages = "066014",
    year = "2025"
}

@article{DeWitt:1962cg,
    author = "DeWitt, Bryce S.",
    title = "{The Quantization of geometry}",
    year = "1962"
}

@article{Giddings:2005id,
    author = "Giddings, Steven B. and Marolf, Donald and Hartle, James B.",
    title = "{Observables in effective gravity}",
    eprint = "hep-th/0512200",
    archivePrefix = "arXiv",
    doi = "10.1103/PhysRevD.74.064018",
    journal = "Phys. Rev. D",
    volume = "74",
    pages = "064018",
    year = "2006"
}

@article{Marolf:2022ybi,
    author = "Marolf, Donald",
    title = "{Gravitational thermodynamics without the conformal factor problem: partition functions and Euclidean saddles from Lorentzian path integrals}",
    eprint = "2203.07421",
    archivePrefix = "arXiv",
    primaryClass = "hep-th",
    doi = "10.1007/JHEP07(2022)108",
    journal = "JHEP",
    volume = "07",
    pages = "108",
    year = "2022"
}

@article{Louko:1988fi,
    author = "Louko, Jorma",
    title = "{A Feynman Prescription for the Hartle-hawking Proposal}",
    reportNumber = "DAMTP/R-88/8",
    doi = "10.1088/0264-9381/5/11/001",
    journal = "Class. Quant. Grav.",
    volume = "5",
    pages = "L181",
    year = "1988"
}

@article{Dasgupta:2001ue,
    author = "Dasgupta, A. and Loll, R.",
    title = "{A Proper time cure for the conformal sickness in quantum gravity}",
    eprint = "hep-th/0103186",
    archivePrefix = "arXiv",
    reportNumber = "AEI-2001-020",
    doi = "10.1016/S0550-3213(01)00227-9",
    journal = "Nucl. Phys. B",
    volume = "606",
    pages = "357--379",
    year = "2001"
}

@article{DeWitt:1967yk,
    author = "DeWitt, Bryce S.",
    editor = "Fang, Li-Zhi and Ruffini, R.",
    title = "{Quantum Theory of Gravity. 1. The Canonical Theory}",
    doi = "10.1103/PhysRev.160.1113",
    journal = "Phys. Rev.",
    volume = "160",
    pages = "1113--1148",
    year = "1967"
}

@Book{ChaosBook,
  title =     {Chaos: Classical and Quantum},
  publisher = {Niels Bohr Inst.},
  year =      {2016},
  author =    {P. Cvitanovi{\'c} and R. Artuso and R. Mainieri and G. Tanner and G. Vattay},
  address =   {Copenhagen},
  url =       {http://ChaosBook.org/}
}

@article{Gibbons:1978ac,
    author = "Gibbons, G. W. and Hawking, S. W. and Perry, M. J.",
    title = "{Path Integrals and the Indefiniteness of the Gravitational Action}",
    reportNumber = "PRINT-78-0375 (CAMBRIDGE)",
    doi = "10.1016/0550-3213(78)90161-X",
    journal = "Nucl. Phys. B",
    volume = "138",
    pages = "141--150",
    year = "1978"
}

@article{Kontsevich:2021dmb,
    author = "Kontsevich, Maxim and Segal, Graeme",
    title = "{Wick Rotation and the Positivity of Energy in Quantum Field Theory}",
    eprint = "2105.10161",
    archivePrefix = "arXiv",
    primaryClass = "hep-th",
    doi = "10.1093/qmath/haab027",
    journal = "Quart. J. Math. Oxford Ser.",
    volume = "72",
    number = "1-2",
    pages = "673--699",
    year = "2021"
}

@article{Witten:2021nzp,
    author = "Witten, Edward",
    title = "{A Note On Complex Spacetime Metrics}",
    eprint = "2111.06514",
    archivePrefix = "arXiv",
    primaryClass = "hep-th",
    month = "11",
    year = "2021"
}

@book{Polchinski:1998rq,
    author = "Polchinski, J.",
    title = "{String theory. Vol. 1: An introduction to the bosonic string}",
    doi = "10.1017/CBO9780511816079",
    isbn = "978-0-511-25227-3, 978-0-521-67227-6, 978-0-521-63303-1",
    publisher = "Cambridge University Press",
    series = "Cambridge Monographs on Mathematical Physics",
    month = "12",
    year = "2007"
}

@article{Hartnoll:2022snh,
    author = "Hartnoll, Sean A.",
    title = "{Wheeler-DeWitt states of the AdS-Schwarzschild interior}",
    eprint = "2208.04348",
    archivePrefix = "arXiv",
    primaryClass = "hep-th",
    doi = "10.1007/JHEP01(2023)066",
    journal = "JHEP",
    volume = "01",
    pages = "066",
    year = "2023"
}

@article{Cohen:1985sm,
    author = "Cohen, Andrew G. and Moore, Gregory W. and Nelson, Philip C. and Polchinski, Joseph",
    title = "{An Off-Shell Propagator for String Theory}",
    reportNumber = "HUTP-85/A058, UTTG-16-85",
    doi = "10.1016/0550-3213(86)90148-3",
    journal = "Nucl. Phys. B",
    volume = "267",
    pages = "143--157",
    year = "1986"
}

@article{Gribov:1977wm,
    author = "Gribov, V. N.",
    editor = "Nyiri, J.",
    title = "{Quantization of Nonabelian Gauge Theories}",
    reportNumber = "LENINGRAD-77-367",
    doi = "10.1016/0550-3213(78)90175-X",
    journal = "Nucl. Phys. B",
    volume = "139",
    pages = "1",
    year = "1978"
}

@article{Held:2024rmg,
    author = "Held, Jesse and Maxfield, Henry",
    title = "{The Hilbert space of de Sitter JT: a case study for canonical methods in quantum gravity}",
    eprint = "2410.14824",
    archivePrefix = "arXiv",
    primaryClass = "hep-th",
    month = "10",
    year = "2024"
}

@article{Woodard:1989ac,
    author = "Woodard, R. P.",
    title = "{Enforcing the Wheeler-de Witt Constraint the Easy Way}",
    reportNumber = "BROWN-HET-723",
    doi = "10.1088/0264-9381/10/3/008",
    journal = "Class. Quant. Grav.",
    volume = "10",
    pages = "483--496",
    year = "1993"
}

@article{Dirac:1950pj,
    author = "Dirac, Paul A. M.",
    title = "{Generalized Hamiltonian dynamics}",
    doi = "10.4153/CJM-1950-012-1",
    journal = "Can. J. Math.",
    volume = "2",
    pages = "129--148",
    year = "1950"
}

@article{Halliwell:1990qr,
    author = "Halliwell, Jonathan J. and Hartle, James B.",
    title = "{Wave functions constructed from an invariant sum over histories satisfy constraints}",
    reportNumber = "NSF-ITP-90-97",
    doi = "10.1103/PhysRevD.43.1170",
    journal = "Phys. Rev. D",
    volume = "43",
    pages = "1170--1194",
    year = "1991"
}

@article{Erbin:2019uiz,
    author = "Erbin, Harold and Maldacena, Juan and Skliros, Dimitri",
    title = "{Two-Point String Amplitudes}",
    eprint = "1906.06051",
    archivePrefix = "arXiv",
    primaryClass = "hep-th",
    doi = "10.1007/JHEP07(2019)139",
    journal = "JHEP",
    volume = "07",
    pages = "139",
    year = "2019"
}

@article{bar2010linear,
    title={Linear wave equations on Lorentzian manifolds},
    author={B{\"a}r, Christian},
    eprint = "1006.2354",
    archivePrefix = "arXiv",
    primaryClass = "math",
    year={2010}
}

@article{Casali:2021ewu,
    author = "Casali, Eduardo and Marolf, Donald and Maxfield, Henry and Rangamani, Mukund",
    title = "{Baby universes and worldline field theories}",
    eprint = "2101.12221",
    archivePrefix = "arXiv",
    primaryClass = "hep-th",
    doi = "10.1088/1361-6382/ac37cd",
    journal = "Class. Quant. Grav.",
    volume = "39",
    number = "13",
    pages = "134004",
    year = "2022"
}

@article{Hollands:2014eia,
    author = "Hollands, Stefan and Wald, Robert M.",
    title = "{Quantum fields in curved spacetime}",
    eprint = "1401.2026",
    archivePrefix = "arXiv",
    primaryClass = "gr-qc",
    doi = "10.1016/j.physrep.2015.02.001",
    journal = "Phys. Rept.",
    volume = "574",
    pages = "1--35",
    year = "2015"
}

@book{Wald:1995yp,
    author = "Wald, Robert M.",
    title = "{Quantum Field Theory in Curved Space-Time and Black Hole Thermodynamics}",
    isbn = "978-0-226-87027-4",
    publisher = "University of Chicago Press",
    address = "Chicago, IL",
    series = "Chicago Lectures in Physics",
    year = "1995"
}

@article{Hartle:1997dc,
    author = "Hartle, James B. and Marolf, Donald",
    title = "{Comparing formulations of generalized quantum mechanics for reparametrization - invariant systems}",
    eprint = "gr-qc/9703021",
    archivePrefix = "arXiv",
    reportNumber = "UCSBTH-96-15, NSF-ITP-96-59, UCSBTH-96-15, NSF-ITP-96-59",
    doi = "10.1103/PhysRevD.56.6247",
    journal = "Phys. Rev. D",
    volume = "56",
    pages = "6247--6257",
    year = "1997"
}

@article{Henneaux:1985kr,
    author = "Henneaux, M.",
    title = "{Hamiltonian Form of the Path Integral for Theories with a Gauge Freedom}",
    doi = "10.1016/0370-1573(85)90103-6",
    journal = "Phys. Rept.",
    volume = "126",
    pages = "1--66",
    year = "1985"
}

@book{Henneaux:1994lbw,
    author = "Henneaux, Marc and Teitelboim, Claudio",
    title = "{Quantization of Gauge Systems}",
    isbn = "978-0-691-03769-1, 978-0-691-21386-6",
    publisher = "Princeton University Press",
    month = "8",
    year = "1994"
}

@inproceedings{Marolf:2000iq,
    author = "Marolf, Donald",
    title = "{Group averaging and refined algebraic quantization: Where are we now?}",
    booktitle = "{9th Marcel Grossmann Meeting on Recent Developments in Theoretical and Experimental General Relativity, Gravitation and Relativistic Field Theories (MG 9)}",
    eprint = "gr-qc/0011112",
    archivePrefix = "arXiv",
    reportNumber = "SU-GP-11-2",
    month = "7",
    year = "2000"
}

@article{Witten:2022xxp,
    author = "Witten, Edward",
    title = "{A note on the canonical formalism for gravity}",
    eprint = "2212.08270",
    archivePrefix = "arXiv",
    primaryClass = "hep-th",
    doi = "10.4310/ATMP.2023.v27.n1.a6",
    journal = "Adv. Theor. Math. Phys.",
    volume = "27",
    number = "1",
    pages = "311--380",
    year = "2023"
}

@article{Araujo-Regado:2022gvw,
    author = "Araujo-Regado, Goncalo and Khan, Rifath and Wall, Aron C.",
    title = "{Cauchy slice holography: a new AdS/CFT dictionary}",
    eprint = "2204.00591",
    archivePrefix = "arXiv",
    primaryClass = "hep-th",
    doi = "10.1007/JHEP03(2023)026",
    journal = "JHEP",
    volume = "03",
    pages = "026",
    year = "2023"
}

@article{Higuchi:1991tk,
    author = "Higuchi, A.",
    title = "{Quantum linearization instabilities of de Sitter space-time. 1}",
    doi = "10.1088/0264-9381/8/11/009",
    journal = "Class. Quant. Grav.",
    volume = "8",
    pages = "1961--1981",
    year = "1991"
}

@article{Ashtekar:1995zh,
    author = "Ashtekar, Abhay and Lewandowski, Jerzy and Marolf, Donald and Mourao, Jose and Thiemann, Thomas",
    title = "{Quantization of diffeomorphism invariant theories of connections with local degrees of freedom}",
    eprint = "gr-qc/9504018",
    archivePrefix = "arXiv",
    reportNumber = "UCSBTH-95-7",
    doi = "10.1063/1.531252",
    journal = "J. Math. Phys.",
    volume = "36",
    pages = "6456--6493",
    year = "1995"
}

@article{Shvedov:2001ai,
    author = "Shvedov, Oleg Yu.",
    title = "{On correspondence of BRST-BFV, Dirac and refined algebraic quantizations of constrained systems}",
    eprint = "hep-th/0111270",
    archivePrefix = "arXiv",
    doi = "10.1006/aphy.2002.6305",
    journal = "Annals Phys.",
    volume = "302",
    pages = "2--21",
    year = "2002"
}

@article{Batalin:1994rd,
    author = "Batalin, Igor and Marnelius, Robert",
    title = "{Solving general gauge theories on inner product spaces}",
    eprint = "hep-th/9501004",
    archivePrefix = "arXiv",
    reportNumber = "GOTEBORG-94-32, ITP-G\textbackslash{}''OTEBORG-94-32, ITP-G\textbackslash{}''oteborg 94-32",
    doi = "10.1016/0550-3213(95)00141-E",
    journal = "Nucl. Phys. B",
    volume = "442",
    pages = "669--696",
    year = "1995"
}

@article{Giulini:1998kf,
    author = "Giulini, Domenico and Marolf, Donald",
    title = "{A Uniqueness theorem for constraint quantization}",
    eprint = "gr-qc/9902045",
    archivePrefix = "arXiv",
    reportNumber = "ZU-TH-99-2, SU-GP-99-1-1, NSF-ITP-99-08",
    doi = "10.1088/0264-9381/16/7/322",
    journal = "Class. Quant. Grav.",
    volume = "16",
    pages = "2489--2505",
    year = "1999"
}

@article{Halliwell:1988wc,
    author = "Halliwell, Jonathan J.",
    title = "{Derivation of the Wheeler-De Witt Equation from a Path Integral for Minisuperspace Models}",
    reportNumber = "NSF-ITP-88-25",
    doi = "10.1103/PhysRevD.38.2468",
    journal = "Phys. Rev. D",
    volume = "38",
    pages = "2468",
    year = "1988"
}

@article{Marolf:1996gb,
    author = "Marolf, Donald",
    title = "{Path integrals and instantons in quantum gravity: Minisuperspace models}",
    eprint = "gr-qc/9602019",
    archivePrefix = "arXiv",
    reportNumber = "UCSBTH-96-01",
    doi = "10.1103/PhysRevD.53.6979",
    journal = "Phys. Rev. D",
    volume = "53",
    pages = "6979--6990",
    year = "1996"
}

@article{Chandrasekaran:2022cip,
    author = "Chandrasekaran, Venkatesa and Longo, Roberto and Penington, Geoff and Witten, Edward",
    title = "{An algebra of observables for de Sitter space}",
    eprint = "2206.10780",
    archivePrefix = "arXiv",
    primaryClass = "hep-th",
    doi = "10.1007/JHEP02(2023)082",
    journal = "JHEP",
    volume = "02",
    pages = "082",
    year = "2023"
}

@article{Marolf:1995cn,
    author = "Marolf, Donald",
    title = "{Refined algebraic quantization: Systems with a single constraint}",
    eprint = "gr-qc/9508015",
    archivePrefix = "arXiv",
    reportNumber = "UCSBTH-95-16",
    month = "8",
    year = "1995"
}

@article{Marolf:1994wh,
    author = "Marolf, Donald",
    title = "{Quantum observables and recollapsing dynamics}",
    eprint = "gr-qc/9404053",
    archivePrefix = "arXiv",
    reportNumber = "CGPG-94-4-5",
    doi = "10.1088/0264-9381/12/5/011",
    journal = "Class. Quant. Grav.",
    volume = "12",
    pages = "1199--1220",
    year = "1995"
}

@article{Marnelius:1990eq,
    author = "Marnelius, Robert and Ogren, Mats",
    title = "{Symmetric inner products for physical states in BRST quantization}",
    reportNumber = "GOTEBORG-90-14",
    doi = "10.1016/0550-3213(91)90098-I",
    journal = "Nucl. Phys. B",
    volume = "351",
    pages = "474--490",
    year = "1991"
}

@article{Moss:1988wk,
    author = "Moss, Ian",
    title = "{Quantum Cosmology and the Selfobserving Universe}",
    reportNumber = "Print-88-0292 (NEWCASTLE)",
    journal = "Ann. Inst. H. Poincare Phys. Theor.",
    volume = "49",
    pages = "341--349",
    year = "1988"
}

@inproceedings{Marolf:1994ae,
    author = "Marolf, Donald",
    title = "{The Spectral analysis inner product for quantum gravity}",
    booktitle = "{7th Marcel Grossmann Meeting on General Relativity (MG 7)}",
    eprint = "gr-qc/9409036",
    archivePrefix = "arXiv",
    reportNumber = "UCSBTH-94-38, CGPG-94-9-1",
    pages = "851--853",
    month = "9",
    year = "1994"
}

@article{Wald:1993kj,
    author = "Wald, Robert M.",
    title = "{A Proposal for solving the 'problem of time' in canonical quantum gravity}",
    eprint = "gr-qc/9305024",
    archivePrefix = "arXiv",
    reportNumber = "PRINT-93-0424 (EFI-CHICAGO)",
    doi = "10.1103/PhysRevD.48.R2377",
    journal = "Phys. Rev. D",
    volume = "48",
    pages = "R2377--R2381",
    year = "1993"
}

@article{Higuchi:1994vc,
    author = "Higuchi, Atsushi and Wald, Robert M.",
    title = "{Applications of a new proposal for solving the 'problem of time' to some simple quantum cosmological models}",
    eprint = "gr-qc/9407038",
    archivePrefix = "arXiv",
    reportNumber = "BUTP-94-16",
    doi = "10.1103/PhysRevD.51.544",
    journal = "Phys. Rev. D",
    volume = "51",
    pages = "544--561",
    year = "1995"
}
